%% file: thesis.tex
\documentclass[12pt,twoside]{report}
\usepackage{theorem}
\usepackage{maxim}
\input diagrams   % commutative diagrams
\diagramstyle[centerdisplay,PostScript=dvips]
\usepackage{amssymb}
\usepackage{epsfig}
\usepackage{amsmath}
\usepackage{mathrsfs}
\usepackage{amsfonts}
\usepackage[colorlinks=true,ps2pdf=true,letterpaper=true,breaklinks=true]{hyperref}
\usepackage[bf,bf,outermarks]{titlesec}

\renewcommand{\thechapter}{\arabic{chapter}}
\titleformat{\chapter}[display]
{\bfseries\Large}
{\filleft\MakeUppercase{\chaptertitlename} \Huge\thechapter}
{4ex}
{\titlerule \titlerule
\vspace{2ex}%
\Huge\filright}
[\vspace{2ex}%
\titlerule]

\newpagestyle{basic}[]{
\headrule
\sethead[\usepage][][Chapter \thechapter: \chaptertitle]
{Chapter \thechapter: \chaptertitle}{}{\usepage}}

\newpagestyle{pref}[]{
\headrule
\sethead[\usepage][][\chaptertitle]
{\chaptertitle}{}{\usepage}}

\newpagestyle{main}[]{
\headrule
\sethead[\usepage][][Chapter \thechapter: \chaptertitle]
{\thesection. \sectiontitle}{}{\usepage}}

\newpagestyle{appa}[]{
\headrule
\sethead[\usepage][][Appendix A: Mathematical Background]
{\thesection. \sectiontitle}{}{\usepage}}

\begin{document}

\author{Raginsky, Max}
\sloppy
\input{title}
\pagenumbering{roman}
\setcounter{page}{2}
\input{copyright}
\chapter*{ABSTRACT}
\addcontentsline{toc}{chapter}{ABSTRACT}
\input{abstract}
\chapter*{Acknowledgments}
\addcontentsline{toc}{chapter}{Acknowledgments}
\input{acks}

\newpage
\pagestyle{pref}
\tableofcontents
\listoffigures

\newpage
\pagestyle{basic}

\input{chap1}

\newpage
\pagestyle{main}
\input{chap2}
\input{chap3}
\input{chap4}

\input{chap5}

\newpage
\pagestyle{basic}
\input{chap6}

\newpage
\pagestyle{appa}
\input{app}
\pagestyle{pref}

\input{refs}
\input{vita}

\end{document}

%% file: title.tex
\begin{titlepage}
\renewcommand{\baselinestretch}{1.5}

\begin{center} 

\href{http://www.northwestern.edu}{NORTHWESTERN UNIVERSITY}
%NORTHWESTERN UNIVERSITY
\vspace{1.0in}

{\Large { \bf Dynamical Aspects of Information Storage\\
in Quantum-Mechanical Systems}}

\vspace{.67in}
A DISSERTATION\\
\vspace{0.25in} 
SUBMITTED TO THE GRADUATE SCHOOL \\
IN PARTIAL FULFILLMENT OF THE REQUIREMENTS 
 
\vspace{0.5in} 
for the degree\\
\vspace{0.5in}
DOCTOR OF PHILOSOPHY\\

\vspace{0.25in} 
Field of Electrical and Computer Engineering \\
\vspace{.7in}

By \\
\vspace{0.25in} 

{\bf Maxim Raginsky} \\

\vspace{.5in}

EVANSTON, ILLINOIS \\

\vspace{.25in}
June 2002
\end{center}
\end{titlepage}

%% file: copyright.tex
\vspace*{2.5in}
\begin{center}

\baselineskip 16pt

\copyright \,\, Copyright by Raginsky, Maxim 2002
\break
All Rights Reserved

\end{center}

%% file: abstract.tex
\begin{center}

{\bf Dynamical Aspects of Information Storage\\
in Quantum-Mechanical Systems} \\
\vspace{0.2in} 
Maxim Raginsky
\end{center}

\vspace{0.15in}

We study information storage in noisy quantum registers and computers
using the methods of statistical dynamics.  We develop the concept
of a strictly contractive quantum channel in order to construct
mathematical models of physically realizable, i.e., nonideal, quantum
registers and computers.  Strictly contractive channels are simple
enough, yet exhibit very interesting features, which are meaningful
from the physical point of view.  In particular, they allow
us to incorporate the crucial assumption of finite
precision of all experimentally realizable operations.
Strict contractivity also helps us gain insight into the
thermodynamics of noisy quantum evolutions (approach to
equilibrium). Our investigation into thermodynamics focuses on the
entropy-energy balance in quantum registers and computers under the
influence of strictly contractive noise.  Using entropy-energy
methods, we are able to appraise the thermodynamical resources needed
to maintain reliable operation of the computer.  We also obtain
estimates of the largest tolerable error rate.  Finally, we explore
the possibility of going beyond the standard circuit model of error
correction, namely constructing quantum memory devices on the basis of
interacting particle systems at low temperatures.

%% file: acks.tex
%\vskip 1.1in

Following the hallowed tradition, first I would like to thank my
thesis advisor and mentor, Prof. Horace P. Yuen, who not only
profoundly influenced my thinking and the course of my career as a
graduate student, but also impressed upon me this very important
lesson:  in scientific research, one should never blindly defer to
\qt{authority,} but rather work everything out for oneself.  Next I
would like to thank Profs. Prem Kumar and Selim M. Shahriar for
serving on my final examination committee, Dr. Giacomo M. D'Ariano
(University of Pavia, Italy) for serving on my qualifying examination
committee and for a careful reading of this dissertation, which
resulted in several improvements, Dr. Viacheslav Belavkin (University
of Nottingham, United Kingdom) for interesting discussions, and
Dr. Masanao Ozawa (Tohoku University, Japan) for valuable comments on
my papers. I also gratefully acknowledge the support of the U.S. Army
Research Office for funding my research through the MURI grant
DAAD19-00-1-0177.

Most of my research was conceived and done in the many coffeehouses of
Evanston and Urbana-Champaign.  Therefore some credit is due the
following fine establishments: in Evanston, the Potion Liquid Lounge
(now unfortunately defunct), Unicorn Caf\'e, and Kafein; in
Urbana-Champaign, the Green Street Coffeehouse and Caf\'e Kopi (which
also serves alcohol).

During the three years I have spent at Northwestern as a grad student,
I have had a chance to meet some interesting characters, with whom it was a
real pleasure to discuss the Meaning of Life and other, less
substantial, matters, often over a pint or two of Guinness.  These
people are:  Jeff Browning, Eric Corndorf, Yiftie Eisenberg, Vadim
Moroz, Ranjith Nair, Boris Rubinstein, Jay Sharping, Brian Taylor, and
Laura Tiefenbruck.  Did I forget anyone? It is also a pleasure to
thank my friends outside Northwestern, for believing in me and for
being there.  This one goes out to the high-school crew:  Mark
Friedgan, Alex Rozenblat, Mike Sandler, Ilya Sutin, and Arthur Tretyak.

I owe a great deal of gratitude to my parents, Margarita and Anatoly
Raginsky, who always encouraged my interest in science and
mathematics, and to my brother Alex, with whom I made a bet that I
would earn my doctorate by the time he graduated from high school.
Fork over the fifty bucks, dude!  And, last but not least, I would
like to thank my parents-in-law, Rosa and Vladimir Lazebnik, and my
sister-in-law, Masha, for their support.

Finally, I must admit that above all I cherish and value the love of
my wonderful wife Lana. I dedicate this dissertation to her.

\newpage
\vspace*{2.5in}
\begin{center}

To Lana
\end{center}

%% file: chap1.tex
\chapter{Introduction}
\label{ch:intro}
\pagenumbering{arabic}

%\vskip 1in

Quantum memory will be a key ingredient in any viable implementation
of a quantum information-processing system (computer).  However, because any
quantum computer realized in a laboratory will necessarily be subject
to the combined influence of environmental noise and unavoidable
imprecisions in the preparation, manipulation, and measurement of
quantum-mechanical states, reliable storage of quantum information
will prove to be a daunting challenge.  Indeed, some authors
\cite{pse, unr} found that circuit-based quantum computation (i.e., a
temporal sequence of local unitary transformations, or quantum gates
\cite{bar}) is extremely vulnerable to noisy perturbations.  The same
noisy perturbations will also adversely affect information stored in
quantum registers (e.g., between successive stages of a computation).

Therefore, since it was first realized that maintaining
reliable operation of a large-scale quantum computer would pose a
formidable obstacle to any experimental realization thereof, many
researchers have expended a considerable amount of effort devising
various schemes for \qt{stabilization of quantum information.} These
schemes include, e.g., quantum error-correcting codes \cite{kl},
noiseless quantum codes \cite{zr}, decoherence-free subspaces
\cite{lcw}, and noiseless subsystems \cite{klv}.  (The last three of
these schemes boil down to essentially the same thing, but are arrived
at by different means.) Each of these schemes relies for its efficacy
upon explicit assumptions about the nature of the error
mechanism. Quantum error-correcting codes \cite{kl}, for instance,
perform best when different qubits in the computer are affected by
independent errors.  On the other hand, stabilization strategies that
are designed to handle collective errors \cite{klv,lcw,zr} make
extensive use of symmetry arguments in order to demonstrate existence
of \qt{noiseless subsystems} that are effectively decoupled from the
environment, even though the computer as a whole certainly remains
affected by errors.

In a recent publication \cite{zan}, Zanardi gave a unified description
of all of the above-mentioned schemes via a common algebraic
framework, thereby reducing the conditions for efficient stabilization
of quantum information to those based on symmetry considerations.  The
validity of this framework will ultimately be decided by experiment,
but it is also quite important to test its applicability in a
theoretical setting that would require minimal assumptions about the
exact nature of the error mechanism, and yet would serve as an
abstract embodiment of the concept of a physically realizable (i.e.,
nonideal) quantum computer.

In this respect, the assumption of finite precision of all physically
realizable state preparation, manipulation, and registration
procedures is particularly important, and can even be treated as an
empirical given.  This premise is general enough to subsume (a)
fundamental limitations imposed by the laws of quantum physics (e.g.,
impossibility of reliable discrimination between any two density
operators with nonorthogonal supports), (b) practical constraints
imposed by the specific experimental setting (e.g., impossibility of
synthesizing any quantum state or any quantum operation with arbitrary
precision), and (c) environment-induced noise.

As a rule, imprecisions in preparation and measurement procedures
will give rise to imprecisions in the building blocks of the
computer (quantum gates) because the precision of any experimental
characterization of these gates will always be affected by the
precision of preparation and measurement steps involved in any such
characterization.  Conversely, the precision of quantum gates will
affect the precision of measurements because the closeness of
conditional probability measures, conditioned on the gate used, is
bounded above by the closeness of the two quantum gates \cite{bv}.

Incorporation of the finite-precision assumption into the mathematical
model of noisy quantum memories and computers has to proceed in two
directions.  On the one hand, we must characterize the sensitivity of
quantum information-processing devices to small perturbations of both
states and operations.  This is important for the following reasons.
First, any unitary operation required for a particular computational
task must be {\em approximated} by several unitary operations taken
from the set of universal quantum gates \cite{bar}.  Since any quantum
computation is a long sequence of unitary operations, approximation
errors will propagate in time, and the resultant state at the end of
the computation will differ from the one that would be generated by
the \qt{ideal} computer.  This issue was addressed by Bernstein and
Vazirani \cite{bv} who found that if a sequence of gates
$G_1,G_2,\ldots,G_n$ is approximated by the sequence $G'_1,
G'_2,\ldots,G'_n$, where the $i$th approximating gate $G'_i$ differs
from the \qt{true} gate $G_i$ by $\epsilon_i$, then the corresponding
resultant states will differ by at most $\epsilon_1 + \epsilon_2 +
\ldots + \epsilon_n$.  Secondly, in the case of noisy computation,
each gate will be perturbed by noise, thus resulting in additional
error.  This situation was handled by Kitaev \cite{kit}, with the same
conclusion:  errors accumulate at most linearly.  Therefore, if we
approximate the gates sufficiently closely, and if the noise is
sufficiently weak, then we can hope that the resulting error in the
output state will be small.  The same reasoning can be applied to
perturbations of initial states:  if two states differ by $\epsilon$,
then the corresponding output states will also differ by at most
$\epsilon$.  However, these conclusions are hardly surprising; they
are, in fact, simple consequences of the continuity of quantum
channels and expectation values.

\begin{figure}
\includegraphics{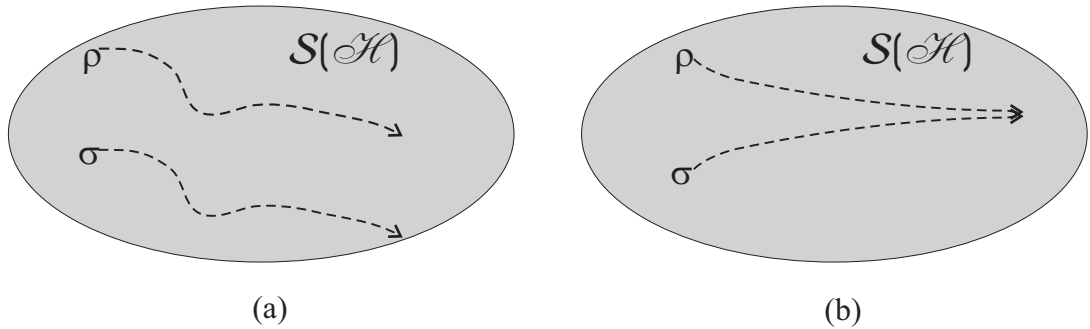}
\caption{Orbits defined by input density operators $\rho$ and $\sigma$
  in the state space $\cS(\sH)$ of the quantum system with  the
  Hilbert space $\sH$ in the case of (a) noiseless (reversible,
  unitary) channel; and (b) noisy (irreversible, non-unitary)
  channel.}
\label{fig:orbits}
\end{figure}

There is, on the other hand, another aspect of the noiseless/noisy
dichotomy, which has been so far largely overlooked.  Assuming for
simplicity that all operations in the quantum system (register or
computer) take place at integer times, each initial state (density
operator) $\rho_0$ defines an {\em orbit} in the state space of the
system, i.e., a sequence $\set{\rho_n}_{n \in \bbn}$ where $\rho_n$ is
the state of the system at time $n$.  According to the circuit model
of a quantum computer, each time step of the computation is a
unitary channel.  Now consider a pair of initial states $\rho_0,
\sigma_0$.  Then, by unitary invariance of the trace norm (cf. Section
\ref{ssec:trnorm}), we will have
$$
\trnorm{\rho_n - \sigma_n} = \trnorm{\rho_0 - \sigma_0},\qquad \forall
n \in \bbn.
$$
In other words, {\em the output states of a noiseless quantum system
  are distinguishable from one another exactly to the same extent as
  the corresponding input states.}  However, this is not the case for
general (non-unitary) channels.  Such channels are described by
trace-preserving completely positive maps (cf. Ch.~\ref{ch:prelim1})
and, for any such map $T$ on density operators, we have
$\trnorm{T(\rho) - T(\sigma)} \le \trnorm{\rho-\sigma}$ \cite{rus}.  A
noisy quantum system can be modeled by replacing a unitary channel at
each time step with a general completely positive trace-preserving
map. In this case, we will have
$$
\trnorm{\rho_{n+1} - \sigma_{n+1}} \le \trnorm{\rho_n -
  \sigma_n},\qquad \forall n \in \bbn,
$$
whence we see that, {\em for the case of a noisy quantum system, the
  output states are generally less distinguishable from one another than the
  corresponding input states.  Furthermore, distinguishability can
  only decrease with each time step.}  In other words, the distance
between two disjoint orbits in the state space of the system will
remain constant in the absence of noise, and shrink when noise is
present.  Both situations are depicted in Fig.~\ref{fig:orbits}.

The discussion above suggests that, apart from insensitivity to small
perturbations of states and operations, we should also pay attention
to {\em insensitivity to initial conditions}, i.e., the situation
where two markedly different input density operators will, over time,
evolve into effectively indistinguishable output density operators due
to the rapid shrinking of the distance between the corresponding
orbits in the state space.  One of the central goals of this
dissertation is to investigate noisy channels with the property that
{\em any two orbits} get uniformly exponentially close to each other
with time.  In a sense, this is the \qt{worst} kind of noise because
it renders the result of any sufficiently lengthy computation
essentially useless, as it cannot be distinguished reliably from the
result due to any other input state.  Noisy channels with this
property will be referred to as {\em strictly contractive}.

Why do we choose to focus on this seemingly extreme noise model?
First of all, as we will show later on, any noisy channel can be
approximated arbitrarily closely by a strictly contractive channel,
such that the two cannot be distinguished by any experimental means.
Essentially this implies that if a given noiseless channel is
perturbed to a noisy one, we may as well assume that the latter is
strictly contractive.  Secondly, we wish to incorporate the
finite-precision assumption into our mathematical framework.  In
particular, we want our model to be such that, in the presence of
noise, there is always a {\em nonzero} probability of making an error
when attempting to distinguish between any two quantum states, even
when these states are, in principle, maximally distinguishable.  As we
will demonstrate, this desideratum is fulfilled by strictly
contractive channels.  Finally, the strictly contractive model
provides a tool for investigations into the {\em statistical dynamics}
of noisy quantum channels.  In particular, the model already
accommodates two important ingredients for a theory of approach to
equilibrium, namely ergodicity and mixing (cf. \cite{kry} or
\cite[pp. 54-60, 237-243]{rs}).

Let us quickly recall these notions and outline the way they relate to
noisy quantum systems. The content of the so-called {\em ergodic
  hypothesis} of statistical mechanics can be succinctly stated as the
equivalence of statistical averages and time averages.  Physicists
usually take the pragmatic approach, assuming that the ergodic
hypothesis holds in any physically meaningful situation (cf., e.g.,
\cite[p. 4]{ll}).  Rigorous proofs of ergodicity have been obtained
only for very few cases (see, e.g., Sinai and Chernov \cite{sc}), none
of which are particularly interesting. The sad fate of the ergodic
programme in classical Hamiltonian mechanics had been sealed further
by the famous Kolmogorov-Arnold-Moser (KAM) theorem
\cite[p. 155]{scheck}, which states that the majority of Hamiltonian
evolutions do not satisfy the ergodic
hypothesis.\footnote{Incidentally, it has recently been noted by
  Novikov \cite{nov} that the results of Kolmogorov, Arnold, and Moser
  {\em have not been fully proved}. One can only wonder whether this
  will revive the research into the ergodic hypothesis for classical
  Hamiltonian systems.\\}  Quantum systems (spin systems in
particular), however, still serve as fruitful soil for various
investigations into ergodic theory \cite[Ch. 7]{af}.  Discrete-time
quantum channels are especially amenable to such studies; a general
quantum channel $T$ on density operators is termed ergodic\footnote{An
  alternative (and, in many respects, more natural) definition of
  ergodicity can be formulated for transformations of observables, i.e., for
  the Heisenberg picture of quantum dynamics.\\} if there exists a
unique density operator $\rho_T$ such that $T(\rho_T) = \rho_T$.  If
$\set{\rho_n}$ is an orbit generated by an ergodic channel $T$, then
it can be shown that, for any observable $A$, the time average
$\frac{1}{N+1} \sum^N_{n=0}\ave{A}_n$, where $\ave{A}_n \defeq \tr{(A
  \rho_n)}$, converges to the fixed-point average $\ave{A}_T \defeq
\tr{(A\rho_T)}$ as $N \rightarrow \infty$.

There exists also a stronger property, called {\em mixing}. In simple
terms, a channel $T$ is mixing if, for any observable $A$, we have
$\ave{A}_n \rightarrow \ave{A}_T$ as $n \rightarrow \infty$.  Mixing
obviously implies ergodicity, but the converse is not necessarily
true.  It turns out that strictly contractive channels are mixing, and
hence ergodic.  One of the most original thinkers on the subject of
statistical physics, Nikolai Krylov, believed \cite{kry} that mixing,
rather than ergodicity, should play central role in the theory of
approach to equilibrium.  In particular, he emphasized the importance
of the so-called {\em relaxation time}, i.e., the time after which the
system will be found, with very high probability, in a state very
close to equilibrium.  He showed that mixing, and not ergodicity, is
necessary for obtaining correct estimates of the relaxation time.
Qualitatively we can say that approach to equilibrium should be
exponentially fast, as confirmed by experimental evidence, and this is
precisely the feature that strictly contractive channels will be shown
to possess.

One of our central results is the following:  errors modeled by
strictly contractive channels cannot be corrected perfectly.  This
result, while of a negative nature, does not come as a complete
surprise:  in a nonideal setting, impossibility of perfect error
correction can only be expected.  We will, however, present an
argument that some form of \qt{approximate} error correction will
still be useful in many circumstances.  In particular, we will discuss
the possibility of either (a) going beyond the circuit model of
quantum computation, or (b) finding ways to introduce enough
parallelism into our quantum information processing so as to finish
any job we need to do before the effect of errors becomes appreciable.

In this respect we will mention an intriguing possibility of realizing
quantum information processing in massively parallel arrays of
interacting parcitles (quantum cellular automata \cite{ric}).  One
advantage furnished by such systems is the possibility of a phase
transition, i.e., a marked change in macroscopic behavior that occurs
when the values of suitable parameters cross some critical threshold.
In the classical case, the stereotypical example is provided by the
two-dimensional Ising ferromagnet which, at sufficiently low
temperatures, can \qt{remember} the direction of an applied magnetic
field even after the field is turned off.  This phenomenon is, of
course, at the basis of magnetic storage devices.  The concept of a
quantum phase transition \cite{sac} is tied to the ground-state
behavior of perturbed quantum spin systems on a lattice and refers to
an abrupt change in the macroscopic nature of the ground state as the
perturbation strength is varied.  We will discuss this concept in
greater detail later on; here we only mention that existence of a
quantm phase transition can be exploited fruitfully for reliable
storage of quantum information in the subcritical region at low
temperatures (assuming that the ground state carries sufficient
degeneracy, so as to accommodate the necessary amount of information).

The dissertation is organized as follows.  In Chapter \ref{ch:prelim1}
we give a quick introduction to the mathematical formalism of quantum
information theory.  Then, in Chapter \ref{ch:scc}, we discuss
strictly contractive quantum channels.  Chapter \ref{ch:enterg} is
devoted to the the study of noisy quantum registers and computers in
terms of the entropy-energy balance.  In particular, we give an
entropic interpretation of strict contractivity for bistochastic
strictly contractive channels.  In Chapter \ref{ch:spinsys} we briefly
comment on the possibility of reliable storage of quantum information
in spin systems on a lattice.  Concluding remarks are given in Chapter
\ref{ch:concl}.  The necessary mathematical background is collected in
Appendix A; Appendix B contains the list of symbols used throughout
the dissertation.

%% file: chap2.tex
\pagestyle{main}
\chapter{Basic notions of quantum
  information theory}
\label{ch:prelim1}

%\vskip 1in

In this chapter we introduce the abstract formalism of quantum
information theory.  But, before we proceed, it is pertinent to ask:
what exactly {\em is} quantum information?  Here is a definition taken
from an excellent survey article of Werner \cite{wer1}.

\begin{center}
\begin{minipage}{10cm}
{\em Quantum information is that kind of information which is carried
  by quantum systems from the preparation device to the measuring
  apparatus in a quantum-mechanical experiment.}
\end{minipage}
\end{center}

Of course, this definition is somewhat vague about the general notion
of \qt{information,} but we can take the pragmatic approach and say
that the information about a given physical system includes the
specification of the initial state of the system, as well as any other
knowledge that can be used to {\em predict} the state of the system at
some later time.  Note that we are not talking about any quantitative
measures of \qt{information content.}  For this reason, such notions
as channel capacity will be conspicuously absent form our
presentation.  For a lucid account of quantum channel capacity, the
reader is referred to the surveys of Bennett and Shor \cite{bs} and
Werner \cite{wer1}.
%add reference to M. Keyl

\section{Classical systems vs. quantum systems}
\label{sec:systems}

Classical systems are distinguished from their quantum counterparts
through such characteristics as size (macroscopic vs. microscopic) or
the nature of their energy spectrum (continuous vs. discrete). For
example, an electromagnetic pulse sent through an optical fiber can be
thought of as classical, whereas a single photon sent through the
fiber is regarded as quantum.  The most conspicuous differences,
however, are revealed through the statistics of experiments performed
upon these systems.  For instance, the joint probability distribution
of a number of classical random variables always has the form of a
limit of convex combinations of product probability distributions, but
this is generally not so in the quantum case.

In this section we introduce the mathematical formalism necessary for
capturing the essential features of classical and quantum systems.
Our exposition closely follows Werner's survey
\cite{wer1}. The requisite background on operator algebras is
collected in Appendix A.

\subsection{Algebras of observables}
\label{ssec:algebras}

For each physical system we need an abstract description that would
account not only for the classical/quantum distinction, but also for
such features as the structure of the set of all possible
configurations of the system.  Such a description is possible through
defining the {\em algebra of observables} of the system.  In order to
cover both classical and quantum systems, we will require from the
outset that their algebras of observables be C*-algebras with
identity.  For the moment, we do not elaborate on the reasons for this
choice, hoping that they will become clear as we go along.

Anyone who has taken an introductory course in quantum mechanics knows
that the presence of noncommuting observables is the most salient
feature of the quantum formalism.  Therefore we take for granted that
the algebra of observables of a quantum system must be noncommutative,
whereas the algebra of observables of a classical system must be
commutative (abelian).  Thus, without loss of generality, the algebra
of observables of a quantum system is the algebra $\cB(\sH)$ of
bounded operators on some Hilbert space $\sH$, whereas the
corresponding algebra for a classical system is the algebra $\cC(\sX)$
of continuous complex-valued functions on a compact set
$\sX$.\footnote{We have allowed ourselves a simplification which
  consists in requiring that the set $\sX$ be compact; this is not the
  case for a general abelian C*-algebra, where the set $\sX$ can be
  merely a locally compact space, but, because we have assumed that any
  algebra of observables must have an identity, the set $\sX$ will, in
  fact, be compact.\\}

Let us illustrate this high-level statement with some concrete
examples.  First we treat the simplest classical case, namely the
classical bit.  Here the set $\sX$ is the two-element set $\set{0,1}$,
and the corresponding algebra of observables is the set of all
complex-valued functions on $\set{0,1}$.  We can think of an element
of this algebra of observables as a random variable defined on the
two-element sample space $\sX$.  The simplest example of a quantum
system, the quantum bit (or {\em qubit}) is furnished by considering a
two-dimensional complex Hilbert space $\sH \simeq \bbc^2$, and the
algebra of observables is nothing but the set $\cM_2$ of $2 \times 2$
complex matrices.

In general, the structure of the configuration space of the system is
reflected in the set $\sX$ (in the classical case) or the Hilbert
space $\sH$ (in the quantum case).  Thus a set $\sX$ with $\abs{\sX} =
n$ would be associated to a classical system with an $n$-element
configuration space; similarly, the underlying Hilbert space of a
spin-$S$ quantum object would be $(2S+1)$-dimensional.  We can also
describe systems with countably infinite or uncountable configuration
spaces, e.g., the classical Heisenberg spin with the set $\sX$ being
$\bbs^2$ (the unit sphere in $\bbr^3$), or a single mode of an
electromagnetic field with the Hilbert space isomorphic to the space
$\ell^2$ of square-summable infinite sequences of complex numbers.
For simplicity let us suppose that, from now on, the algebras of
observables with which we deal are finite-dimensional.  This implies
that, if we are dealing with the algebra $\cC(\sX)$, then the set
$\sX$ is finite; similarly, given the algebra $\cB(\sH)$, the Hilbert
space $\sH$ must be finite-dimensional.

For the purposes of calculations it is often convenient to expand
elements of an algebra in a basis.  A canonical basis for $\cC(\sX)$
is the set of functions $e_x, x \in \sX$, defined by
\begin{equation}
e_x(y) = \left \{
	\begin{array}{ll}
		1 & {\rm if}\, x = y \\
	      0 & {\rm if}\, x \neq y
	\end{array}
	\right .,
\label{eq:functbasis}
\end{equation}
so that any function $f \in \cC(\sX)$ can be expanded as $f = \sum_{x
  \in \sX} f(x) e_x$.  A basis for $\cB(\sH)$ is constructed by
picking any orthonormal basis $\set{e_i}$ for $\sH$ and defining the
\qt{standard matrix units} $e_{ij} \defeq \ketbra{e_i}{e_j}$.  Thus
for any $X \in \cB(\sH)$ we have $X = \sum_{i,j} X_{ij}e_{ij}$ with
$X_{ij} \in \bbc$.

In order to describe composite systems, i.e., systems built up from
several subsystems, we need a way of combining algebras to form new
algebras.  Let us consider bipartite systems first, starting with the
classical case. Suppose we are given two classical systems, $\Sigma_1$
and $\Sigma_2$, with configuration spaces $\sX$ and $\sY$
respectively.  Then the configuration of the joint system, $\Sigma_1 +
\Sigma_2$, is characterized by giving an ordered pair $(x \in \sX,y
\in \sY)$.  Thus the configuration space of the joint system is simply
the Cartesian product $\sX \times \sY$, i.e., the set of all ordered
pairs of the kind described above.  The corresponding algebra of observables is
$\cC(\sX \times \sY)$, i.e., the algebra of functions $\map{f}{\sX
  \times \sY}{\bbc}$.  Any element $f$ of this algebra can be written
in the form
\begin{equation}
f = \sum_{x \in \sX, y \in \sY} f(x,y) e_{xy},
\label{eq:cartp}
\end{equation}
where the basis functions $e_{xy}$ are defined in the manner similar
to Eq.~(\ref{eq:functbasis}).  Furthermore, for any $x' \in \sX$ and
$y' \in \sY$ we have $e_{xy}(x',y') = e_x(x') e_y(y')$.  On the other
hand, a general element of the tensor product
$\tp{\cC(\sX)}{\cC(\sY)}$ has the form
\begin{equation}
f = \sum_{x \in \sX, y \in \sY} f(x,y) \tp{e_x}{e_y}.
\label{eq:tensp}
\end{equation}
Directly comparing Eqs.~(\ref{eq:cartp}) and (\ref{eq:tensp}), we see
that $\tp{\cC(\sX)}{\cC(\sY)} \simeq \cC(\sX \times \sY)$.

In the quantum case we start by taking the tensor product of the
Hilbert spaces of the subsystems.  Consider two quantum systems with
the Hilbert spaces $\sH$ and $\sK$.  Let $\set{e_i}$ and $\set{e_\mu}$
be orthonormal bases of $\sH$ and $\sK$ respectively.  Then the
set $\set{\tp{e_i}{e_\mu}}$ is the corresponding orthonormal basis of
the tensor product space $\tp{\sH}{\sK}$.  A typical element of the
algebra $\cB(\tp{\sH}{\sK})$ has the form
$$
X = \sum_{i,j,\mu,\nu} X_{ij,\mu\nu} \tp{e_{ij}}{e_{\mu \nu}},
$$
and a typical element of the product algebra $\tp{\cB(\sH)}{\cB(\sK)}$
has a similar form.  Thus we conclude that $\cB(\tp{\sH}{\sK}) \simeq
\tp{\cB(\sH)}{\cB(\sK)}$.

In both the classical case and the quantum case we see that the
algebra of observables of the bipartite system, whose subsystems are
assigned algebras $\cA$ and $\cB$, has the form $\tp{\cA}{\cB}$.
Algebras of observables for multipartite systems can now be
constructed inductively.  Using the tensor product, it is possible to
define algebras of observables for hybrid systems, i.e., systems with
both classical and quantum subsystems.  This is not necessary for our
purposes, and therefore we will not dwell on this.  An interested
reader is referred to Werner's survey \cite{wer1} for details.

\subsection{Pure and mixed states}
\label{ssec:states}

Our next step is to describe the statistics of both classical and
quantum systems in a unified fashion.  This is accomplished by
introducing states over the algebra of observables of the system.
Recall that a state over a C*-algebra $\cA$ is a positive normalized
linear functional on $\cA$, i.e., a mapping $\map{\omega}{\cA}{\bbc}$
that maps all positive elements of $\cA$ to nonnegative real numbers,
and for which we have $\omega(\idty) = 1$, where $\idty$ is the
identity element of $\cA$.  The number $\omega(A)$ then gives the
expected value of the observable $A$ measured on the system in the
state $\omega$.

The positive elements of a C*-algebra $\cA$ are precisely those
elements that can be written in the form $B^*B$ for some $B \in \cA$.
In the case of the algebra $\cC(\sX)$, a function $f$ is a positive
element if and only if $f(x) \ge 0$ for all $x \in \sX$ or,
equivalently, if and only if $f(x) = \abs{g(x)}^2$ for some $g \in
\cC(\sX)$.  In the case of $\cB(\sH)$, an operator $X$ is positive if
and only if $\braket{\psi}{X \psi} \ge 0$ for all $\psi \in \sH$ or,
equivalently, if and only if $X = Y^* Y$ for some $Y \in \cB(\sH)$.

Of especial importance to the statistical framework of quantum
information theory is  the subset of $\cA$ consisting of those
elements $F$ for which $F \ge 0$ and $\idty - F \ge 0$ (this is
written as a double inequality $0 \le F \le \idty$).  These
observables are referred to as {\em effects}, the term introduced by
Ludwig in his axiomatic treatment of quantum theory \cite{lud}.  It is
obvious that, for any effect $F$ and any state $\omega$, $0 \le
\omega(F) \le 1$.  Furthermore, given a collection $\set{F_\alpha}$ of
effects with $\sum_\alpha F_\alpha = \idty$, we will have $\sum_\alpha
\omega(F_\alpha) = 1$.  Thus, in the most general formulation, to each
outcome $o$ of an experiment performed on the system, classical or
quantum, we associate an effect $F_o$, such that $\omega(F_o)$ is the
probability of getting the outcome $o$ when the system is in the state
$\omega$.  Obviously, $\sum_{o \in \cO} \omega(F_o) = 1$, where $\cO$
is the set of all possible outcomes of the experiment.

Having said this, let us first treat states in the classical setting.
If $\omega$ is a state over the algebra $\cC(\sX)$, then it is clear
that $0 \le \omega(e_x) \le 1$ for all $x \in \sX$.  This follows from
the fact that $\omega(\idty) \equiv \sum_x \omega(e_x) = 1$ and from
the positivity of $\omega$.  Thus we see that any state $\omega$ over
the algebra $\cC(\sX)$ gives rise to a probability distribution
$\set{p_x}$ on $\sX$, where $p_x \defeq \omega(e_x)$.  Conversely,
given a probability distribution $\set{p_x}$ on $\sX$, we can define a
positive normalized linear functional on $\cC(\sX)$ in an obvious way.
Therefore there is a one-to-one correspondence between the states over
$\cC(\sX)$ and the probability distributions on $\sX$.  We have argued
this for the case of a finite $\sX$; in general, it is the content of
the Riesz-Markov theorem \cite[p. 107]{rs} that, given a compact
Hausdorff space $\sX$, there exists a one-to-one correspondence
between positive normalized linear functionals on $\cC(\sX)$ and
probability measures on $\sX$.

Similarly, given a state $\omega$ over the algebra $\cB(\sH)$ of a
quantum system, we can associate with it a matrix $\rho$ whose
elements in the basis $\set{e_i}$ will have the form $\rho_{ij} \defeq
\omega(e_{ji})$.  Thus, given any $X \in \cB(\sH)$, we will have
$\omega(X) = \sum_{i,j} X_{ij}\rho_{ji} \equiv \tr{(\rho X)}$.  The
matrix $\rho$ is easily seen to have unit trace because $\omega(\idty)
= \sum_i \omega(e_{ii}) = \sum_i \rho_{ii} \equiv \tr{\rho}$, and is
also positive semidefinite because, for any $\psi \in \sH$,
$\braket{\psi}{\rho\psi} = \omega(\ketbra{\psi}{\psi}) \ge
0$. Conversely, given a positive semidefinite matrix $\rho$ of unit
trace, we can define a state over $\cB(\sH)$ via $\omega(e_{ij})
\defeq \tr{(\rho e_{ij})} \equiv \rho_{ji}$.  Thus we see that, when
the Hilbert space $\sH$ of the system has finite dimension $n$, there
is a one-to-one correspondence between states over $\cB(\sH)$ and
positive semidefinite $n \times n$ matrices of unit trace (called {\em
  density matrices} or {\em density operators}).  This is not true in
the case when $\sH$ is infinite-dimensional:  not every state $\omega$
over $\cB(\sH)$ corresponds to a density operator.  Those states that
do have density operators associated with them are called {\em normal
  states}.

In light of the correspondence between states and probability measures
(in the classical case) or density operators (in the quantum case), we
will use the term \qt{state} interchangeably, referring either to the
functional on the corresponding algebra of observables, or to the
corresponding probability measure or the density operator.

The set $\cS(\cA)$ of states over a C*-algebra $\cA$ is a convex set
whose extreme points are referred to as {\em pure states}. The
adjective \qt{pure} reflects the fact that these are the states with
the least amount of \qt{randomness:}  being extreme points of the set
$\cS(\cA)$, they cannot be written as nontrivial convex combinations
of other states.  For this reason the pure states over an algebra of
observables play a crucial role.  In order to characterize the pure
states over the algebra $\cC(\sX)$, we invoke the fact that a state
$\omega$ over an abelian C*-algebra $\cA$ is pure if and only if
$\omega(AB)=\omega(A)\omega(B)$ for all $A,B \in \cA$, as well as the
fact that a state over $\cC(\sX)$ is determined by its action on the
basis functions $e_x$.  Because $e_x = (e_x)^2$, which means that
$e_x(x') = e_x(x')^2$ for any $x' \in \sX$, we have $\omega(e_x) =
\omega(e_x)^2$ for $\omega$ pure, which implies that $\omega(e_x) \in
\set{0,1}$ for each $x \in \sX$.  Since $\sum_x \omega(e_x) = 1$, we
conclude that, for each pure state $\omega$ over $\cC(\sX)$, there
exists a unique $y \in \sX$ such that
$$
\omega(e_x) = \left \{
	\begin{array}{ll}
		1 & {\rm if}\, x = y \\
		0 & {\rm if}\, x \neq y
	\end{array}
	\right..
$$
This pure state corresponds to the probability measure $\delta_y$
concentrated on the single point $y \in \sX$.  Such measures are
referred to as {\em point measures}.  Conversely, defining the state
$\omega_y$ corresponding to the point measure $\delta_y$, we can
easily convince ourselves that $\omega_y(fg) = \omega_y(f)\omega_y(g)$
for all pairs $f,g \in \cC(\sX)$.  Thus the pure states over the
algebra $\cC(\sX)$ are in a one-to-one correspondence with the point
measures over $\sX$.

As for quantum systems, we know that there is a one-to-one
correspondence between the set of states over $\cB(\sH)$ and the
convex set $\cS(\sH)$ of the density operators on $\sH$ (again, we
assume that the Hilbert space $\sH$ is finite-dimensional).
Furthermore this correspondence is {\em affine}, i.e., convex
combinations of states over $\cB(\sH)$ correspond to convex
combinations of density operators on $\sH$.  Hence there is a
one-to-one correspondence between the extreme points of the respective
sets.  The extreme points of $\cS(\sH)$ are the one-dimensional
projectors, i.e., those density matrices $\rho$ for which $\rho^2 =
\rho$.  Thus the pure states over $\cB(\sH)$ correspond precisely to
the one-dimensional projectors in $\cB(\sH)$ (or, equivalenty, to the
unit vectors in $\sH$).

States that are not pure are referred to as {\em mixed}; they
correspond to non-extreme points of the corresponding state spaces.
According to the Krein-Milman theorem \cite[p. 67]{sch}, any point of
a compact convex set $S$ in a locally convex topological space is a
limit of convex combinations of the extreme points of $S$.  In fact, a
stronger result due to Carath\'eodory \cite[p. 7]{phe} states that
any point in a compact convex subset $S$ of an $n$-dimensional space
is a convex combination of at most $n+1$ extreme points of $S$.
Therefore any mixed state over $\cC(\sX)$ can be represented as a
convex mixture of point measures on $\sX$, while any mixed state over
$\cB(\sH)$ is a convex mixture of one-dimensional projections on
$\sH$.  In either case the operation of forming a convex combination
of pure states can be thought of as introducing \qt{classical}
randomness.  In this respect an important role is played by the
so-called {\em maximally mixed} states, i.e., those states that are
\qt{most random.}  The maximally mixed state over the classical
algebra of observables $\cC(\sX)$ corresponds to the normalized
counting measure on $\sX$, i.e., to the measure that assigns the value
$1/\abs{\sX}$ to each $x \in \sX$.  The maximally mixed state over
$\cB(\sH)$ corresponds to the normalized identity matrix,
$\idty/{\dim{\sH}}$.  The reason for the name \qt{maximally mixed}
will become apparent when we discuss entropy in
Sec.~\ref{sec:entropy}.

States of composite systems are defined by means of the tensor product
construction.  In other words, a state of the system with the algebra
of observables $\tp{\cA}{\cB}$ is a positive normalized linear
functional over $\tp{\cA}{\cB}$.  Again, any state over
$\tp{\cA}{\cB}$ will be a convex combination of pure states.  In the
classical case, $\cA = \cC(\sX)$ and $\cB = \cC(\sY)$, pure states
correspond to the point measures $\delta_{(x,y)} \equiv \tp
{\delta_x}{\delta_y}$.  Thus any state over $\tp{\cC(\sX)}{\cC(\sY)}$
has the form
$$
\omega = \sum_{x,y} p_{xy} \tp{\delta_x}{\delta_y},\qquad 0 \le p_{xy}
\le 1, \sum_{x,y}p_{xy} = 1
$$
i.e., it can be written as a convex combination of product measures.
This is not so for the states over $\tp{\cB(\sH)}{\cB(\sK)}$, where
$\sH$ and $\sK$ are Hilbert spaces. Now the pure states correspond to
unit vectors in $\tp{\sH}{\sK}$, and it is a basic fact of the theory
of tensor products that not every vector in $\tp{\sH}{\sK}$ can be
written in the product form $\tp{\psi}{\phi}$ with $\psi \in \sH$ and
$\phi \in \sK$.  Consequently, not all states of a composite quantum
system are {\em separable} in the following sense.

\begin{definition}
\label{def:sepstate}
A state $\omega$ of a composite system with the algebra of observables
$\tp{\cA}{\cB}$ is called {\em separable} (or {\em classically
  correlated} in the terminology of Werner \cite{wer2}) if it can be
written as
\begin{equation}
\omega = \sum_i p_i \tp{\omega^\cA_i}{\omega^\cB_i},
\label{eq:sepstate}
\end{equation}
where $\omega^\cA_i \in \cS(\cA)$ and $\omega^\cB_i \in \cS(\cB)$,
with nontrivial weights $p_i$.  Otherwise $\omega$ is called {\em
  entangled}.
\end{definition}

On the contrary, {\em every} state of a composite classical system is
separable, as we have seen above.  This conclusion also follows for
very general cases from the observation that every such state is a
convex combination of point measures, but, because the point measures
on a Cartesian product of sets are precisely the product measures
\cite[p. 32]{sk}, the state is a convex combination of product
measures and hence separable.

There are many interesting examples of entangled states.  In the case
of $\tp{\sH}{\sH}$, where $\sH \simeq \bbc^2$, we can give an example
of a family of entangled states whose state vectors also form an
orthonormal basis of $\tp{\sH}{\sH}$.

\begin{example}
\label{ex:bellbasis}
({\bf the Bell basis})
Let $\ket{e_1}$ and $\ket{e_2}$ be an orthonormal basis of $\bbc^2$.
Then the pure states, whose vectors form the so-called {\em Bell
  basis},
\begin{eqnarray}
\ket{\Psi^+_1} & \defeq & \frac{1}{\sqrt{2}} (\ket{\tp{e_1}{e_1}} +
\ket{\tp{e_2}{e_2}}) \label{eq:bell1} \\
\ket{\Psi^+_2} & \defeq & \frac{1}{\sqrt{2}} (\ket{\tp{e_1}{e_2}} +
\ket{\tp{e_2}{e_1}}) \label{eq:bell2} \\
\ket{\Psi^-_1} & \defeq & \frac{1}{\sqrt{2}} (\ket{\tp{e_1}{e_1}} -
\ket{\tp{e_2}{e_2}}) \label{eq:bell3} \\
\ket{\Psi^-_2} & \defeq & \frac{1}{\sqrt{2}} (\ket{\tp{e_1}{e_2}} -
\ket{\tp{e_2}{e_1}}), \label{eq:bell4}
\end{eqnarray}
are entangled states.\endex

The theory of entanglement is a rich subfield of quantum information
theory, but, since we are not directly concerned with entanglement in
this work, we will limit ourselves to the very basic facts.  The
reader is encouraged to consult the survey article by M., P., and
R. Horodecki \cite{hhh} for further details.

Given an arbitrary state $\rho$, it is in general not an easy task to
decide whether it is entangled unless it is pure, in which case our
job reduces to the analysis of the so-called {\em Schmidt
  decomposition} of the corresponding state vector.  In order to
define the Schmidt decomposition, we first need to look at the
restriction of states to subsystems.

\begin{definition}
\label{def:restrstate}
Let $\omega$ be a state over the algebra of observables
$\tp{\cA}{\cB}$.  Then the {\em restriction of $\omega$ to $\cA$} is
the unique state $\omega_\cA$ determined by $\omega_\cA(A) \defeq
\omega(\tp{A}{\idty_\cB})$ for any $A \in \cA$.
\end{definition}

The number $\omega(\tp{A}{\idty_\cB})$ should be thought of as the
expected value of the observable $A$ which we measure on the subsystem
with the algebra $\cA$, completely ignoring the subsystem with the
algebra $\cB$.

In the classical case, where $\tp{\cA}{\cB} =
\tp{\cC(\sX)}{\cC(\sY)}$, observables of the form $\tp{A}{\idty}$ can
be written as
$$
\tp{A}{\idty} = \sum_{x,y} A(x) \tp{e_x}{e_y},
$$
so that the restriction to $\cA$ of the state corresponding to the
probability measure $p_{xy}$ on $\sX \times \sY$ is the state
corresponding to the probability measure $p_x = \sum_y p_{xy}$, i.e.,
$\omega_\cA(f) = \sum_{x,y}p_{xy}f(x)$. This is precisely the marginal
probability distribution obtained by integrating over the set
$\sY$. It is easy to see that any pure state of a bipartite classical
system restricts to a pure state on either subsystem.

In the case of a quantum system, the restriction of a state $\rho$
over $\tp{\cA}{\cB} = \cB(\tp{\sH}{\sK})$ to $\cA$ is determined by
$\tr{(\rho_\cA A)} = \tr{[\rho (\tp{A}{\idty_\sK})]}$, i.e., the
corresponding density operator $\rho_\cA$ is obtained by taking the
partial trace of $\rho$ over $\sK$, $\rho_\cA = \ptr{\sK}{\rho}$.
Contrary to the classical case, pure states over $\cB(\tp{\sH}{\sK})$
that are not elementary tensors (i.e., are not of the form
$\tp{\psi}{\phi}$) do not restrict to pure states over $\sH$ or over
$\sK$.  Indeed, the restriction to $\cB(\sH)$ of any the pure states
defined in Eqs.~(\ref{eq:bell1})-(\ref{eq:bell4}) is the maximally
mixed state $(1/2)\idty$.

Let $\cA$ and $\cB$ be algebras of observables. Given the restrictions
$\rho_\cA$ and $\rho_\cB$, it is generally impossible to reconstruct
the state $\rho$ over $\tp{\cA}{\cB}$ with these restrictions unless
it is known {\em a priori} that $\rho$ is pure.  In this case we have
the following theorem.

\begin{theorem}
\label{th:schmidt}
{\bf (Schmidt decomposition)} Let $\psi \in \tp{\sH}{\sK}$ be a unit
vector, and let $\rho_\sH$ be the restriction of the state
$\ketbra{\psi}{\psi}$ to the first system.  Let $\rho_\sH = \sum_i q_i
\ketbra{e_i}{e_i}$ be the spectral decomposition of $\rho_\sH$ with
$q_i > 0$.  Then there exists an orthonormal system $\set{f_i}$ in
$\sK$ such that
\begin{equation}
\psi = \sum_i \sqrt{q_i} \tp{e_i}{f_i}.
\label{eq:schmidt}
\end{equation}
Furthermore, the state $\ketbra{\psi}{\psi}$ is entangled if its
Schmidt decomposition (\ref{eq:schmidt}) has two or more terms.  The
number of terms is referred to as the {\em Schmidt number} of $\psi$.
\end{theorem}

\begin{proof}
By definition of the restricted state, we have
$$
\tr{\rho_\sH A} = \braket{\psi}{(\tp{A}{\idty_\sK})\psi},
$$
where $A \in \cB(\sH)$ is an arbitrary operator.  Writing $\psi =
\sum_i \tp{e_i}{v_i}$, where $v_i \in \sK$ are not normalized, we
obtain
$$
\tr{\rho_\sH A} = \sum_{i,j} \braket{e_i}{A e_j} \braket{v_i}{v_j}.
$$
We let $A = \ketbra{e_m}{e_n}$ to get $q_m \delta_{mn} =
\braket{v_m}{v_n}$.  Defining $f_i \defeq (1/\sqrt{q_i}) v_i$, we
obtain $\psi = \sum_i \sqrt{q_i} \tp{e_i}{f_i}$, which proves
Eq.~(\ref{eq:schmidt}).

Now suppose that $\psi$ is a product state.  Then the restriction of
$\ketbra{\psi}{\psi}$ to the first system is a one-dimensional
projection, and hence has only one nonzero eigenvalue, which means
that the Schmidt decomposition of $\ketbra{\psi}{\psi}$ has only one
term. \end{proof}

The Schmidt decomposition can also work \qt{in reverse,} as follows
from the following theorem.

\begin{theorem}
\label{th:purify}
{\bf (purification)}  Let $\sH$ be a Hilbert space.  For any state
$\rho \in \cS(\sH)$ there exist a Hilbert space $\sK$ and a pure state
$\psi \in \tp{\sH}{\sK}$, called the {\em purification} of $\rho$,
such that $\rho = \ptr{\sK}{\ketbra{\psi}{\psi}}$.  Furthermore, the
restriction $\ptr{\sH}{\ketbra{\psi}{\psi}}$ can be chosen to have no
zero eigenvalues, in which case the space $\sK$ and the vector $\psi$
are unique up to a unitary transformation.
\end{theorem}

\begin{proof}
Let $\rho = \sum^k_{i = 1}q_i \ketbra{e_i}{e_i}$ be the spectral
decomposition of $\rho$ with $q_i > 0$.  Choose $\sK$ isomorphic to
$\bbc^k$, and let $\set{f_i}^k_{i=1}$ be an orthonormal basis for
$\sK$.  Then the vector $\psi \defeq \sum^k_{i=1} \sqrt{q_i}
\tp{e_i}{f_i}$ is the desired purification.  Since the number $k$ and
the vectors $e_i$ are uniquely determined by $\rho$, the only freedom
in this construction is the orthonormal basis $\set{f_i}$, but any two
such bases are connected by a unitary transformation.
\end{proof}

With the aid of the Schmidt decomposition, we see that a pure state
over $\tp{\cA}{\cB}$ is separable if and only if it restricts to pure
states over both subsystems.  (Actually, Theorem ~\ref{th:schmidt}
implies the \qt{only if} part; the \qt{if} part is trivial.)  The
diametrical opposite of this situation is described in the following
definition.

\begin{definition}
\label{def:maxentstate}
A pure state of a bipartite system is called {\em maximally entangled}
if it restricts to maximally mixed states on either subsystem.
\end{definition}
For instance, the states forming the Bell basis are all maximally
entangled.  We will come back to the subject of maximally entangled
states in the next section.  Here we only mention that maximally
entangled states are a crucial resource in virtually every quantum
communication scheme and cryptographic protocol; see the survey by
Weinfurter and Zeilinger \cite{wz} for details.

\section{Channels}
\label{sec:channels}

After having introduced algebras of observables and states of
classical and quantum systems, we must provide the mathematical
description of any processing performed on these systems.  This is
done by means of the so-called {\em channels}.  From now on, we will
assume that all systems under consideration are quantum systems,
unless specified otherwise.

\subsection{Definitions}
\label{ssec:channeldef}

Let us consider the following situation.  Suppose that, after some
processing on the system with the algebra of observables $\cA$, the
result is a system with the corresponding algebra $\cB$.  On this
\qt{new} system, we measure an effect $F \in \cB$.  However, we can
also view this sequence of actions as the measurement of some effect
$\hat{F}\in \cA$ on the \qt{old} system.  Thus the processing step can
be thought of as a transformation $T$ that takes effects in $\cB$ to
effects in $\cA$, $\hat{F} = T(F)$ or, in general, as a mapping
$\map{T}{\cB}{\cA}$ that takes observables in $\cB$ to observables in
$\cA$.  Alternatively, we can view the processing step as a
transformation $T_*$ that takes states over $\cA$ to states over
$\cB$.  Obviously, these two interpretations of the processing step
must be equivalent in the statistical sense, so we require that, for
any state $\omega$ over $\cA$ and for any observable $X$ in $\cB$,
\begin{equation}
\left(T_*(\omega)\right)(X) = \omega \left(T(X)\right),
\label{eq:duality}
\end{equation}
which expresses the statement that the expectation values for the
outcome of any measurement must be the same for $T$ and for $T_*$.
Sometimes we will use the composition notation $\omega \circ T$ to
denote the state defined by $(\omega \circ T)(X) \defeq
(T_*(\omega))(X)$.

Already from this simple description we can glean the properties
required of the map $T$.  First of all, $T$ must map effects to
effects, which implies that $T$ must be a {\em positive map}, i.e., $X
\ge 0$ must imply $T(X) \ge 0$.  Secondly, the trivial measurement
corresponding to the effect $\idty_\cB$ must be mapped to the trivial
measurement $\idty_\cA$, $T(\idty_\cB) = \idty_\cA$.  These two
requirements can be summarized by saying that $T$ must be {\em
  positive} and {\em unital} (or {\em unit-preserving}).  Furthermore,
if $\omega$ is a state, then by hypothesis $T_*(\omega)$ is a state
also.  Hence the left-hand side of Eq.~(\ref{eq:duality}) is linear in
$X$, which means that the right-hand side must also be linear in $X$.
Thus $T$ must be a {\em linear} positive unital map $\cB \rightarrow
\cA$.  

The dual map $T_*$ on states can also be viewed as a map that takes
density operators in $\cA$ to density operators in $\cB$, which allows
us to rewrite Eq.~(\ref{eq:duality}) as
\begin{equation}
\tr{\left[T_*(\rho)X\right]} = \tr{\left[\rho T(X) \right]}.
\label{eq:duality_do}
\end{equation}
Since $T$ is unit-preserving, the linear map $T_*$ must be
trace-preserving, $\tr{T_*(\rho)} = \tr{\rho}$ [just substitute
  $\idty$ for $X$ in Eq.~(\ref{eq:duality_do})], and positive, so that
density operators are mapped to density operators.

Mere positivity of the maps $T$ and $T_*$, however, is not sufficient.
In many situations we need to consider parallel processing performed
on quantum systems, i.e., transformations of the form
$\map{\tp{S}{T}}{\tp{\cB_1}{\cB_2}}{\tp{\cA_1}{\cA_2}}$, where
$\cA_1,\cA_2,\cB_1,\cB_2$ are algebras of observables.  In order to
represent a physically meaningful processing step, the map $\tp{S}{T}$
must be a linear positive unital map.  However the tensor product of
positive maps may fail to be positive, as follows from the following
standard example \cite[p. 192]{con}.

\begin{example}
\label{ex:transp}
({\bf the transposition map}) Let the algebra $\cA$ be the space
$\cM_d$ of $d\times d$ complex matrices.  Matrices in $\cM_d$ act as
operators on the Hilbert space $\sH \simeq \bbc^d$.  Let
$\set{e_j}^d_{j=1}$ be an orthonormal basis of $\sH$. Consider the
{\em transposition map} $\map{\Theta}{\cA}{\cA}$, that is, the map
that sends $\ketbra{e_j}{e_k}$ to $\ketbra{e_k}{e_j}$.  Since $\Theta$
leaves each $\ketbra{e_j}{e_j}$ invariant, it is
trace-preserving and positive [given a positive operator $X$, write
  its spectral decomposition to see that $\Theta(X)$ is also
  positive].  Let us form the map $\tp{\Theta}{\id}$ on
$\tp{\cM_d}{\cM_d}$, where $\id$ is the identity map, in which case we
have
$$
\amap { \tp{\Theta}{\id} }
	{ \ketbra{\tp{e_j}{e_k}}{\tp{e_l}{e_m}} }
	{ \ketbra{\tp{e_l}{e_k}}{\tp{e_j}{e_m}} }.
$$
Now consider the operator
$$
A \defeq \sum^d_{j,k=1} { \ketbra{\tp{e_j}{e_j}}{\tp{e_k}{e_k}} }
$$
which is clearly positive.  Then
$$
F \defeq \tp{\Theta}{\id} (A) = \sum^d_{j,k=1} {
  \ketbra{\tp{e_k}{e_j}}{\tp{e_j}{e_k}} }
$$
is the so-called {\em flip operator} on $\tp{\bbc^d}{\bbc^d}$, that
is, for any pair $\psi,\phi \in \bbc^d$, $F(\tp{\psi}{\phi}) =
\tp{\phi}{\psi}$.  The flip operator is manifestly not positive
because, for the antisymmetric vector $\Psi \equiv \tp{\psi}{\phi} -
\tp{\phi}{\psi}$, we see that $F\Psi = - \Psi$.  Hence the operator
$F$ has a negative eigenvalue, and therefore cannot be positive.\endex

The above example shows that, even if a map $T$ is positive, the map
$\tp{T}{\id}$ may already fail to be positive, which in turn shows
that tensor products of positive maps do not have to be positive maps.
This is clearly unacceptable for the mathematical model of a channel.
A good way out of this difficulty is to restrict the class of
admissible maps to include only the so-called {\em completely
  positive} maps \cite[p. 25]{pau}.  

\begin{definition}
\label{def:cpmap}
Let $\map{T}{\cA}{\cB}$ be a map between operator algebras.  Define
the map $\map{T_n}{\tp {\cA}{\cM_n}}{\tp {\cB}{\cM_n}}$ via $T_n
\defeq \tp{T}{\id}$.  Then $T$ is called $n$-{\em positive} if $T_n$
is a positive map.  A map that is $n$-positive for all values of $n$
is termed {\em completely positive}.
\end{definition}

Now suppose that $\map{S}{\cB_1}{\cA_1}$ and $\map{T}{\cB_2}{\cA_2}$
are completely positive maps.  Let $m$ and $n$ be the dimensions of
the Hilbert spaces $\sH$ and $\sK$, where $\cB_2$ and $\cA_1$ are
subalgebras of $\cB(\sH)$ and $\cB(\sK)$ respectively.  Then the maps
$\map{\tp{S}{\id_m}}{\tp{\cB_1}{\cB_2}}{\tp{\cA_1}{\cB_2}}$ and
$\map{\tp{\id_n}{T}}{\tp{\cA_1}{\cB_2}}{\tp{\cA_1}{\cA_2}}$ are
positive.  Hence their composition, $\tp{S}{T}$, is well-defined and
positive.

\begin{diagram}
\tp{\cB_1}{\cB_2} && \rTo^{\scriptstyle{\tp{S}{T}}} && \tp{\cA_1}{\cA_2} \\
& \rdTo_{\scriptstyle{\tp{S}{\id_m}}} &&
\ruTo_{\scriptstyle{\tp{\id_n}{T}}} & \\
&& \tp{\cA_1}{\cB_2} &&& \\
\end{diagram}
This observation, pictured on the diagram above, motivates the
following definition.

\begin{definition}
\label{def:channel}
A {\em channel} converting systems with the algebra of observables
$\cA$ into systems with the algebra of observables $\cB$ is a
completely positive unital linear map $\map{T}{\cB}{\cA}$.  The dual
map $T_*$, related to $T$ via Eq.~(\ref{eq:duality}) is then a
completely positive trace-preserving linear map, and is referred to as
the {\em dual channel}.\end{definition}

\begin{rem}
We have been somewhat cavalier in our
  definition of the dual channel $T_*$ acting on states through the
  channel $T$ acting on observables, having ignored certain
  technicalities that arise when the Hilbert space $\sH$ is
  infinite-dimensional. These complications disappear in the
  finite-dimensional case, so we will not dwell on this point any
  further.
\end{rem}

We say that the channel $T$ corresponds to the {\em Heisenberg
  picture} of quantum dynamics, whereas the dual channel $T_*$
describes the {\em Schr\"odinger picture}.  This generalizes the
notions of the Heisenberg and the Schr\"odinger pictures, studied in
introductory courses on quantum mechanics.

\subsection{Examples}
\label{ssec:channelex}

It turns out that all physically meaningful examples of channels can
be constructed by putting together certain basic building blocks.  We
will get to this issue in a moment, but first we will provide several
examples of completely positive maps in general, and channels in
particular.  These examples can be found in Werner's survey
\cite{wer1}, but here we fill in the missing details.

\begin{example} \label{ex:starhomo} ({\bf $*$-homomorphisms}) Let
  $\cA$ and $\cB$ be C*-algebras, and consider a $*$-homomorphism
  $\map{\pi}{\cA}{\cB}$.  We know that $*$-homomorphisms map positive
  elements to positive elements, hence $\pi$ is a positive map.  Let
  us consider the map $\tp{\pi}{\id_n}$ that maps $\tp{\cA}{\cM_n}$ to
  $\tp{\cB}{\cM_n}$.  The tensor product $\tp{\cA}{\cM_n}$ is
  isomorphic to the algebra $\cM_n(\cA)$ of $n\times n$ matrices with
  $\cA$-valued entries; this follows from noting that any element of
  $\tp{\cA}{\cM_n}$ can be written in the form
  $\sum^n_{i,j=1}\tp{A_{ij}}{e_{ij}}$, where $A_{ij} \in \cA$ and
  $e_{ij}$ is the matrix unit with entries $\delta_{ij}$.  Thus it is
  natural to identify the element $A_{ij} \in \cA$ with the $(i,j)$th
  entry of an $n\times n$ $\cA$-valued matrix.  The product of
  elements in $\tp{\cA}{\cM_n}$ is given by
$$
\left(\sum_{i,j} \tp{A_{ij}}{e_{ij}}\right) \left(\sum_{k,l}
\tp{B_{kl}}{e_{kl}} \right) = \sum_{i,j} \sum_k
\tp{A_{ik}B_{kj}}{e_{ij}},
$$
where the $(i,j)$th entry is given by the usual laws of matrix
multiplication, but with elements of $\cA$ instead of complex numbers.
The other operations are defined similarly.  Furthermore, the action
of the map $\tp{\pi}{\id_n}$ on an element of $\cM_n(\cA)$ amounts to
the entrywise application of the $*$-homomorphism $\pi$.  It is an
easy task to show that the resulting map is also a $*$-homomorphism,
and hence positive.  This shows that $*$-homomorphisms between
C*-algebras are completely positive.\endex

\begin{example} \label{ex:conjug}({\bf conjugations}) Let $\sH$ and
  $\sK$ be Hilbert spaces, and let $\map{V}{\sH}{\sK}$ be a bounded
  operator.  Then the map $\map{T}{\cB(\sH)}{\cB(\sK)}$ defined by
  $T(X) = VXV^*$ is completely positive.  First of all, $T$ is
  obviously positive.  Indeed, given $X \ge 0$, there exists $Y$ such
  that $X = Y^*Y$, which implies that $T(X) = VY^*YV^* =
  (YV^*)^*(YV^*) \ge 0$.  Now, if $X = Y^*Y$ is a positive element of
  $\tp{\cB(\sH)}{\cM_n}$, then similarly $\tp{T}{\id_n}(X) =
  (\tp{V}{\idty})X(\tp{V^*}{\idty}) =
  \left(Y(\tp{V^*}{\idty})\right)^* \left(Y (\tp{V^*}{\idty})\right)
  \ge 0$.  This holds for all $n$, hence $T$ is completely positive.
  This example covers the special case of unitary conjugations, i.e.,
  the case when $V$ is a unitary operator.  Because $VV^* = V^*V =
  \idty$ for a unitary $V$, the corresponding conjugation is also a
  channel.\endex

\begin{example}
\label{ex:ptr}
({\bf restriction}) Let $\cA$ and $\cB$ be algebras, and consider the
map $\map{M_\cB}{\cA}{\tp{\cA}{\cB}}$ defined by $M_\cB(A) =
\tp{A}{\idty}$.  This map is clearly completely positive and unital.
Let us pass to the Schr\"odinger picture, where we expect that the
dual channel $M_{\cB *}$ is the operation of taking the partial trace
over the second system.  Indeed, consider a density operator
$$
\rho = \sum_{i,j,\mu,\nu} \rho_{ij,\mu\nu}
\ketbra{\tp{e_i}{e_\mu}}{\tp{e_j}{e_\nu}}.
$$
In the duality relation (\ref{eq:duality_do}), let $X$ be the matrix
unit $\ketbra{e_q}{e_p}$.  Then we obtain
$$
\braket{e_p}{M_{\cB *}(\rho)e_q} = \sum_\mu \rho_{pq,\mu\mu},
$$
which is precisely the $(p,q)$th matrix element of the partial trace
of $\rho$ over the second system.  Thus $M_{\cB *} \equiv
\ptr{\cB}{}$.\endex

\begin{example}
\label{ex:expand}
({\bf expansion}) A common operation in quantum information theory is,
given a system in some state $\rho$, to adjoin an auxiliary system in
some fixed state $\rho_0$.  In the Schr\"odinger picture, this
operation is a channel, and has the form $T_*(\rho) =
\tp{\rho}{\rho_0}$.  Let us determine the corresponding channel in the
Heisenberg picture. Let the two systems have $\cA$ and $\cB$
respectively as their algebras of observables.  The sought channel is a map
from $\tp{\cA}{\cB}$ to $\cA$.  Because any $X \in \tp{\cA}{\cB}$ can
be written in the form $X = \sum_i \tp{A_i}{B_i}$, where $A_i \in \cA$
and $B_i \in \cB$, the action of the channel $T$ is determined by its
effect on the elementary tensors $\tp{A}{B}$.  From the duality
relation (\ref{eq:duality_do}), we have
$\tr{\left[(\tp{\rho}{\rho_0})(\tp{A}{B})\right]} = \tr{\left[\rho
  T(\tp{A}{B})\right]}$, which can be rewritten as $\tr{(\rho A)}
\tr{(\rho_0 B)} = \tr{\left[\rho T(\tp{A}{B})\right]}$.  This must
hold for an arbitrary density operator $\rho$, which implies that
$T(\tp{A}{B}) = [\tr{(\rho_0 B)}]A$.  The action of $T$ can be extended
to the whole of $\tp{\cA}{\cB}$ by linearity.  Complete positivity
follows from the fact that $T_*$ is completely positive, and therefore
so is its dual map $T$.\endex

\begin{example}
\label{ex:msrmt}
({\bf measurement}) A measurement can be thought of as a channel that
converts quantum systems into classical systems.  Let $\sX$ be the set
of the measurement outcomes.  Then the act of measurement can be
represented by a mapping $\map{T}{\cC(\sX)}{\cA}$, where $\cA$ is the
algebra of observables of the quantum system.  The channel $T$ is
obviously determined by the operators $F_x \defeq T(e_x), x \in \sX$.
It is a basic result in the theory of completely positive maps that
any positive map $\map{T}{\cC(\sX)}{\cA}$, where $\sX$ is a compact
set and $\cA$ is an operator algebra, is automatically completely
positive \cite[p. 192]{con}.  Thus we must have $F_x \ge 0$.
Furthermore, because $T$ must be unital, the operators $F_x$ must form
a {\em resolution of identity} on $\cA$, i.e., $\sum_x F_x =
\idty$. The application of $T_*$ to a density operator $\rho$ yields a
function $f(x) = \tr{(\rho F_x)}$, i.e., the probability of obtaining
the outcome $x$ when the system is in the state $\rho$.  The
collection $\set{F_x}$ with $F_x \ge 0$ and $\sum_x F_x = \idty$ is an
example of a {\em positive operator-valued measure} (POVM).  We will
discuss POVM's in greater detail in Sec.~\ref{ssec:qdet}, when we talk
about quantum detection theory.  The \qt{old-school} projective (von
Neumann-L\"uders) measurement obtains when the effects $F_x$ have the
property $F_x F_y = \delta_{xy}F_x$.\endex

\begin{example}
({\bf irreversible quantum dynamics}) In Example \ref{ex:conjug}, we
have considered the case of unitarily implemented channels.  Such
channels arise whenever we talk about reversible quantum dynamics.  A
general theory of irreversible quantum dynaimcs proceeds as follows
\cite{dav}.  The system, initially in some state $\rho \in \cB(\sH)$,
is brought into contact with another system, the {\em reservoir},
initially in some fixed state $\rho_R \in \cB(\sK)$, where $\sK$ is
the Hilbert space of the reservoir.  The combined \qt{system +
  reservoir} entity is assumed to be closed.  Then the two are caused
to interact by means of a unitarily implemented channel, and the final
state of the system is obtained by tracing out the reservoir degrees
of freedom.  In the Schr\"odinger picture, this irreversible evolution
of the system is given by the channel $T_*(\rho) =
\ptr{\sK}{U(\tp{\rho}{\rho_R})U^*}$.  
\label{ex:irrev}\endex

Finally we give one more example, which has nothing to do with quantum
information theory {\em per se}, but rather serves to demonstrate the
all-encompassing nature of the definition of the channel.

\begin{example}
\label{ex:classch}
({\bf classical channel}) A classical channel is, roughly speaking,
a transformation that converts classical systems into classical
systems.  Hence a positive map $\map{T}{\cC(\sX)}{\cC(\sY)}$ is a
classical channel, which is uniquely determined by the functions
$\cC(\sY) \ni f_x \defeq T(e_x)$.  The dual map $T_*$ converts states
over $\cC(\sY)$ into states over $\cC(\sX)$ or, equivalently,
probability measures on $\sY$ into probability measures on $\sX$.
Specifically, we can expand $f_x = \sum_y f_{xy}e_y$, so that, for any
function $g \in \cC(\sX)$, we have $T(g) = \sum_{x,y}g(x)f_{xy}e_y$.
If $p = \set{p_y}$ a probability measure on $\sY$, the duality
relation (\ref{eq:duality}) says that
$$
\sum_y p_y (T(g))(y) = \sum_x (T_*(p))_xg(x),
$$
from which we get, upon expanding,
$$
\sum_x \sum_y p_y g(x) f_{xy} = \sum_x (T_*(p))_x g(x).
$$
Comparing coefficients, we obtain $(T_*(p))_x = \sum_y f_{xy}p_y$.
The positive numbers $f_{xy}$ form the {\em transition matrix} of the
channel, where $f_{xy}$ is the conditional probability $p(x |y)$ that
the symbol $x$ is received given that the symbol $y$ was transmitted.
Because $T$ is a channel, it is unital, i.e., $T(\idty) = \sum_x
T(e_x) = \idty \equiv \sum_y e_y$.  But $\sum_x f_x =
\sum_{x,y}f_{xy}e_y$, so we see, comparing coefficients, that $\sum_x
f_{xy} = 1$, i.e., the columns of the transition matrix add up to one.\endex

\subsection{The theorems of Stinespring and Kraus}
\label{ssec:stikr}

Up to this point, our treatment of channels has been largely
axiomatic.  However, we can adopt the pragmatic point of view and
demand that only those transformations that can be built up from
certain basic blocks can serve as channels.  We take our cue from
quantum theory of open systems \cite{dav} and say that any
\qt{physically acceptable} channel can be realized as a sequence of
the following steps: (a) adjunction of an auxiliary system (called the
{\em ancilla}\footnote{Latin for \qt{housemaid;} we choose not to
  dwell on the philosophical implications of this!\\} in the
terminology of Helstrom \cite{hel}) in some fixed initial state, (b)
unitarily implemented evolution of the enlarged system, and (c)
restriction to the original subsystem.  In other words, any channel
must be of the form described in Example \ref{ex:irrev}.  Luckily it
turns out that the two descriptions coincide; this is ultimately a
consequence of the Stinespring theorem \cite{sti} which we first
state, without proof, in the form given by Paulsen \cite[p. 43]{pau}.

\begin{theorem}
\label{th:stinespring}
{\bf (Stinespring)} Let $\cA$ be a C*-algebra with identity, and let
$\sH$ be a Hilbert space.  Then a linear map $\map{T}{\cA}{\cB(\sH)}$
is completely positive if and only if there exist a Hilbert space
$\sK$, a unital $*$-homomorphism $\map{\pi}{\cA}{\cB(\sK)}$, and a bounded
operator $\map{V}{\sH}{\sK}$ with $\norm{V}^2 = \norm{T(\idty)}$ such
that
\begin{equation}
T(A) = V^*\pi(A) V.
\label{eq:sti}
\end{equation}
for any $A \in \cA$.  We will refer either to Eq.~(\ref{eq:sti}) or to
the triple $(\sK,V,\pi)$ as the {\em Stinespring decomposition} of
$T$.
\end{theorem}
It immediately follows from the Stinespring theorem that if $T$ is
also a unital map, then $V$ is an isometry, i.e., $V^*V = \idty_\sH$.
The Stinespring theorem has a useful specialization \cite[p. 15]{egg},
\cite[p. 222]{tak} to the case when the algebra $\cA$ is an algebra of
operators in a Hilbert space.

\begin{theorem}
\label{th:sti_matr}
{\bf (Stinespring; the Hilbert-space version)} Let $\sH$ and $\sH_1$
be Hilbert spaces, and let $\map{T}{\cB(\sH_1)}{\cB(\sH)}$ be a
completely positive map with the Stinespring decomposition
$(\sK,V,\pi)$.  Then there exist a Hilbert space $\sH_2$ and a
unitary operator $\map{U}{\sK}{\tp{\sH_1}{\sH_2}}$ such that, for any
$A \in \cB(\sH_1)$,
\begin{equation}
T(A) = V^*  U^* (\tp{A}{\idty_{\sH_2}}) U V.
\label{eq:sti_matr}
\end{equation}
\end{theorem}
We can absorb the unitary $U$ and the mapping $V$ into a single
mapping to obtain the following corollary.

\begin{corollary}
Let $\sH$ and $\sK$ be Hilbert spaces, and let
$\map{T}{\cB(\sH)}{\cB(\sK)}$ be a completely positive map.  Then
there exist a Hilbert space $\sE$ and a bounded map
$\map{V}{\sK}{\tp{\sH}{\sE}}$ such that
\begin{equation}
T(A) =  V^*(\tp{A}{\idty_\sE})V
\label{eq:sti_final}
\end{equation}
for all $A \in \cB(\sH)$.  Furthermore, if $T$ is unital, then $V$ is
an isometry.
\end{corollary}

The following result \cite{kra}, which carries a great deal of
significance in quantum information theory, is a consequence of the
Stinespring theorem.  We provide the proof because it is instructive,
and because we will come to rely on some of the techniques used in it.

\begin{theorem}
\label{th:krausrep}
{\bf (the Kraus representation)} Let $\sH$ and $\sK$ be Hilbert
spaces, and let $\map{T}{\cB(\sH)}{\cB(\sK)}$ be a completely positive
map.  Then there exist bounded operators $\map{V_\alpha}{\sK}{\sH}$
such that
\begin{equation}
T(A) = \sum_\alpha V^*_\alpha A V_\alpha
\label{eq:krausrep}
\end{equation}
for all $A \in \cB(\sH)$, where the sum in Eq.~(\ref{eq:krausrep})
converges in the strong operator topology.  Furthermore, if the map
$T$ is unital, then $\sum_\alpha V^*_\alpha V_\alpha = \idty_\sK$.
The collection of operators $\set{V_\alpha}$ will be referred to as
the {\em Kraus decomposition} of $T$.
\end{theorem}

\begin{proof} Let $\sE$ and $V$ be given by Eq.~(\ref{eq:sti_final}).
  Now let $\set{\xi_\alpha}$ be an orthonormal basis of $\sE$.  Then,
  given any $\psi \in \sK$, we can expand
\begin{equation}
V\psi = \sum_\alpha \tp{V_\alpha\psi}{\xi_\alpha},
\label{eq:isokraus}
\end{equation}
where $\map{V_\alpha}{\sK}{\sH}$ are some operators. Let $\chi$ be an
arbitrary vector in $\sK$. Then the action of the adjoint $V^*$ on
elementary tensors $\tp{\psi}{\phi}\in \tp{\sH}{\sE}$ can be read off
from
$$
\braket{\chi}{V^*(\tp{\psi}{\phi})} = \braket{V\chi}{\tp{\psi}{\phi}}
= \sum_\alpha \braket{ \tp{V_\alpha \chi}{\xi_\alpha} }
{\tp{\psi}{\phi}} = \sum_\alpha \braket{\chi}{V^*_\alpha \psi}
\braket{\xi_\alpha}{\phi},
$$
which yields
\begin{equation}
V^*(\tp{\psi}{\phi}) = \sum_\alpha \braket{\xi_\alpha}{\phi}V^*_\alpha \psi.
\label{eq:kraus_step1}
\end{equation}

Now let $\psi$ be an arbitrary vector in $\sK$. For an arbitrary
operator $A \in \cB(\sH)$, we write
\begin{eqnarray}
T(A)\psi = V^*(\tp{A}{\idty})V\psi &=& V^*(\tp{A}{\idty})\sum_\alpha
\tp{V_\alpha \psi}{\xi_\alpha} = V^* \left(\sum_\alpha \tp{AV_\alpha
  \psi}{\xi_\alpha} \right) \nonumber \\
&=& \sum_{\alpha,\beta} \braket{\xi_\beta}{\xi_\alpha} V^*_\beta A
V_\alpha \psi = \sum_\alpha V^*_\alpha A V_\alpha \psi, \nonumber
\end{eqnarray}
which is Eq.~(\ref{eq:krausrep}). Now if $T$ is unital, then $V$ is an
isometry, which implies the normalization condition $\sum_\alpha
V^*_\alpha V_\alpha = \idty_\sK$.
\end{proof}

Given a completely positve map $T$, its Kraus decomposition is
obviously not unique.  As shown in the proof of Theorem~\ref{th:krausrep},
the operators $V_\alpha$ are determined by the map $V$ and by the
orthonormal basis $\set{\xi_\alpha}$ of $\sE$.  Thus we have the
freedom of choosing the basis of $\sE$; let $\set{\eta_\alpha}$ be
some other basis, and let $U$ be a unitary transformation such that
$U\xi_\alpha = \eta_\alpha$.  Then, for any $\psi \in \sK$, we can
expand $V\psi$ as
\begin{eqnarray}
V\psi &=& \sum_\alpha \tp{W_\alpha\psi}{\eta_\alpha} = \sum_\alpha
\tp{W_\alpha\psi}{U\xi_\alpha} = \sum_{\alpha,\beta,\gamma}
u_{\beta\gamma} \braket{\xi_\gamma}{\xi_\alpha} \tp{W_\alpha
  \psi}{\xi_\beta} \nonumber \\
&=& \sum_{\alpha,\beta} u_{\beta \alpha} \tp{W_\alpha \psi}{\xi_\beta}
= \sum_\alpha \tp{V_\alpha \psi}{\xi_\alpha}, \nonumber
\end{eqnarray}
where $V_\alpha \defeq \sum_\beta u_{\alpha \beta}W_\beta$, and it is
clear that both sets $\set{V_\alpha}$ and $\set{W_\alpha}$ form Kraus
decompositions of $T$. 

Now, if $\map{T}{\cB(\sH)}{\cB(\sK)}$ is a channel, then the dual map
$T_*$ transforms density operators on $\sK$ to density operators on
$\sH$.  Let $\set{V_\alpha}$ be the Kraus decomposition of $T$. Then
the duality relation (\ref{eq:duality_do}) implies that, for any
density operator $\rho$ on $\sK$, we have
\begin{equation}
T_*(\rho) = \sum_\alpha V_\alpha \rho V^*_\alpha.
\label{eq:kraus_dual}
\end{equation}
It follows from Eq.~(\ref{eq:kraus_dual}) that the dual channel $T_*$
can be extended to all trace-class operators on $\sK$, because any
trace-class operator can be written as a complex linear combination of
four density operators.

Finally, after all these tedious preparations, we are ready to state
and prove the result, due to Kraus \cite{kra}, that any channel can be
represented in the ancilla form.

\begin{theorem}
\label{th:ancilla}
{\bf (ancilla form)} Let $\map{T}{\cB(\sH)}{\cB(\sK)}$ be a channel.
Then there exist Hilbert spaces $\sF$ and $\sG$, a unit vector $\Omega \in
\sG$, and a unitary transformation
$\map{U}{\tp{\sK}{\sG}}{\tp{\sH}{\sF}}$ such that, for any density
operator $\rho$ on $\sK$,
\begin{equation}
T_*(\rho) = \ptr{\sF}U(\tp{\rho}{\ketbra{\Omega}{\Omega}})U^*.
\label{eq:ancilla}
\end{equation}
\end{theorem}

\begin{proof}  Let $V$ and $\sE$ be given by Eq.~(\ref{eq:sti_final}),
  and let $\sF$ and $\sG$ be Hilbert spaces such that $\tp{\sK}{\sG} \simeq
  \tp{\sH}{\sF}$ and $\dim{\sE} \le \dim{\sF}$.  Now pick a unit
  vector $\Omega \in \sG$ and consider the map
\begin{equation}
\hU(\tp{\psi}{\Omega}) \defeq V\psi
\label{eq:umap}
\end{equation}
for all $\psi \in \sK$. The vector on the right-hand side of
Eq.~(\ref{eq:umap}) is an element of $\tp{\sH}{\sE}$, hence an element
of $\tp{\sH}{\sF}$ because $\sE$ is, by construction, isomorphic to a
subspace of $\sF$.  Now if $\set{e_i}$ is
an orthonormal basis of $\sK$, then the vectors
$\hU(\tp{e_i}{\Omega})$ form an orthonormal system in $\tp{\sH}{\sF}$ because
$$
\braket{\hU(\tp{e_i}{\Omega})}{\hU(\tp{e_j}{\Omega})} =
\braket{Ve_i}{Ve_j} = \braket{e_i}{V^*Ve_j} = \braket{e_i}{e_j} =
\delta_{ij},
$$
where we have used the fact that $T$ is a channel, and therefore $V$
is an isometry.  Hence $\hU$ can be extended to a unitary map
$\map{U}{\tp{\sK}{\sG}}{\tp{\sH}{\sF}}$.  Furthermore, because $\sE$
is isomorphic to a subspace of $\sF$, we can express
the action of $U$ on the vectors of the form $\tp{\psi}{\Omega}$ using
a Kraus decomposition $\set{V_\alpha}$ of $T$ as
$$
U(\tp{\psi}{\Omega}) = \sum_\alpha \tp{V_\alpha \psi}{\xi_\alpha},
$$
where $\set{\xi_\alpha}$ is an orthonormal basis of $\sE$, determined
by $\set{V_\alpha}$ (cf. the proof of Theorem~\ref{th:krausrep}).
Then, for any $\psi \in \sK$, we have
\begin{eqnarray}
\ptr{\sF}{U(\tp{\ketbra{\psi}{\psi}}{\ketbra{\Omega}{\Omega}})U^*} &=&
\ptr{\sF}\sum_{\alpha,\beta} \tp{V_\alpha \ketbra{\psi}{\psi}V^*_\beta
}{\ketbra{\xi_\alpha}{\xi_\beta}}\nonumber \\
&=& \sum_\alpha V_\alpha
\ketbra{\psi}{\psi}V^*_\alpha \equiv
T_*(\ketbra{\psi}{\psi}), \nonumber
\end{eqnarray}
and the theorem is proved.
\end{proof}

\begin{rem}
When the Hilbert spaces $\sH$ and $\sK$ are isomorphic, the statement
of the theorem simplifies to the following.  There exist a Hilbert
space $\sE$, a unit vector $\Omega \in \sE$, and a unitary
$\map{U}{\tp{\sK}{\sE}}{\tp{\sH}{\sE}}$ such that $T_*(\rho) =
\ptr{\sE}{U(\tp{\rho}{\ketbra{\Omega}{\Omega}})U^*}$ for all $\rho \in
  \cS(\sK)$, where $\sE$ is determined by Eq.~(\ref{eq:sti_final}).
\end{rem}

\subsection{Duality between channels and bipartite states}
\label{ssec:duality}

There exists a correspondence between channels
$\map{T}{\cB(\sH)}{\cB(\sK)}$ and states over $\cB(\tp{\sH}{\sK})$ which, in
many situations, is more convenient than the Kraus representation or
the ancilla form.

First we make the following observation. Let $\sH$ and $\sK$ be
Hilbert spaces, and let $\map{A}{\sK}{\sH}$ be an operator which we
write as
\begin{equation}
A = \sum_{i,\mu}A_{i\mu} \braket{f_\mu}{\cdot}e_i,
\label{eq:opa}
\end{equation}
where $\set{e_i}$ and $\set{f_\mu}$ are orthonormal bases of $\sH$ and
$\sK$ respectively.  Using the Dirac notation, Eq.~(\ref{eq:opa}) can
be rewritten as $A = \sum_{i,\mu}A_{i\mu}\ketbra{e_i}{f_\mu}$.  We can
view the matrix elements $A_{i\mu}$ of $A$ as the coefficients, in the
basis $\set{\ket{\tp{e_i}{f_\mu}}}$, of a vector in $\tp{\sH}{\sK}$
which we denote by $\dket{A}$,
\begin{equation}
\dket{A} \defeq \sum_{i,\mu}A_{i\mu}\ket{\tp{e_i}{f_\mu}}.
\label{eq:opvect}
\end{equation}
The double-ket notation in Eq.~(\ref{eq:opvect}) is due to Royer
\cite{roy}.  We must caution the reader that, although the only object
appearing inside the double ket is the operator $A$, attention must be
paid to the choice of basis for the tensor product of the
corresponding Hilbert spaces.

The correspondence $\maps{A}{\dket{A}}$ yields a number of useful
formulas, which we summarize in the following lemma \cite{dps,roy}. We
omit the proof which consists in routine, but tedious, manipulations
with indices.

\begin{lemma}
\label{lm:dkets}
Let $\sH$ and $\sK$ be Hilbert spaces.  Then we have the following
relations for vectors in $\tp{\sH}{\sK}$:
\begin{eqnarray}
\dbraket{A}{B} &=& \tr{A^*B} \label{eq:dkets0} \\
(\tp{A}{B})\dket{C} &=& \dket{AC\trn{B}} \label{eq:dkets1} \\
\ptr{\sK}\dketbra{A}{B} &=& AB^* \label{eq:dkets2} \\
\ptr{\sH}\dketbra{A}{B} &=& \trn{A}\bar{B}, \label{eq:dkets3}
\end{eqnarray}
where $\trn{B}$ denotes the matrix transpose of $B$, and $\bar{B}$
denotes the operator whose matrix elements are obtained by taking the
complex conjugates of the matrix elements of $B$.
\end{lemma}

\begin{rem}
Once again, we point out that the relations stated in Lemma
\ref{lm:dkets} are valid as long as the matrix elements of the
operators $A$, $B$, and $C$ refer to the same choice of bases for
$\sH$ and $\sK$.
\end{rem}

Before proceeding to our main topic, we give a couple of examples, due
to D'Ariano, Lo Presti, and Sacchi \cite{dps}, that illustrate the
power of this approach.

\begin{example}
\label{ex:maxent}
({\bf maximally entangled states}) Let $\Psi$ be a pure state in
$\tp{\sH_A}{\sH_B}$, where $\sH_A$ and $\sH_B$ are Hilbert spaces of
the same (finite) dimension $N$. We claim that $\Psi$ is maximally
entangled if and only if it can be written in the form
$(1/\sqrt{N})\dket{U}$ for some unitary $\map{U}{\sH_B}{\sH_A}$.
Assume first that $\Psi = (1/\sqrt{N})\dket{U}$ for a unitary
$\map{U}{\sH_B}{\sH_A}$.  Then, using Eqs.~(\ref{eq:dkets2}) and
(\ref{eq:dkets3}), we see that the restrictions of
$\ketbra{\Psi}{\Psi}$ to $A$ and to $B$ are given by
\begin{eqnarray}
\ptr{B}{\ketbra{\Psi}{\Psi}} &=& (1/N)\ptr{B}\dketbra{U}{U} =
(1/N)UU^* =(1/N) \idty_A \nonumber \\
\ptr{A}{\ketbra{\Psi}{\Psi}} &=& (1/N)\ptr{A}\dketbra{U}{U} = (1/N)
\trn{U}\bar{U} = (1/N) \trn{(U^*U)} = (1/N) \idty_B,\nonumber
\end{eqnarray}
which shows that $\Psi$ is maximally entangled.  On the other hand,
suppose $\Psi$ is maximally entangled.  Let $\Psi = \dket{M}$, where
$\map{M}{\sH_B}{\sH_A}$ is some operator.  We have
\begin{eqnarray}
\ptr{B}{\ketbra{\Psi}{\Psi}} &=& \ptr{B}\dketbra{M}{M} = MM^* = (1/N)
\idty_A \nonumber \\
\ptr{A}{\ketbra{\Psi}{\Psi}} &=& \ptr{A}\dketbra{M}{M} =
\trn{M}\bar{M} = \trn{(M^*M)} = (1/N) \idty_B,\nonumber
\end{eqnarray}
which would hold if and only if $M = (1/\sqrt{N})U$ for some unitary
$U$.\endex

\begin{example}
\label{ex:schmidt}
({\bf the Schmidt decomposition})
Let $\dket{A} \in \tp{\sH_A}{\sH_B}$ be a pure state.  Write down the
polar decomposition of $A$, $A = V\sqrt{A^*A}$, where $V$ is unitary,
and choose a unitary operator $U$ such that $UA^*AU^*$ is diagonal.
Then
$$
\dket{A} = \dket{V\sqrt{A^*A}} =
(\tp{VU^*}{\trn{U}})\dket{U\sqrt{A^*A} U^*} = \sum_i \sqrt{\lambda_i}
\tp{e_i}{f_i},
$$
where $\set{\lambda_i,\psi_i}$ are the eigenvalues and the
eigenvectors of $\sqrt{A^*A}$, and we have defined the vectors $e_i
\defeq VU^*\psi_i, f_i \defeq \trn{U}\psi_i$.\endex

The matrix approach described above reveals its true strength in the
following characterization of channels due to D'Ariano and Lo Presti
\cite{dp1}.  Let $\map{T}{\cB(\sH)}{\cB(\sK)}$ be a completely
positive map, with the corresponding dual map
$\map{T_*}{\cS(\sK)}{\cS(\sH)}$.  Let $\set{e_i}$ be an orthonormal
basis of $\sK$, so that $\dket{\idty}\in \tp{\sK}{\sK}$ is the
unnormalized maximally entangled state $\sum_i \tp{e_i}{e_i}$.  Define
on $\tp{\sH}{\sK}$ the positive operator
\begin{equation}
R_T \defeq (\tp{T_*}{\id})(\dketbra{\idty}{\idty})
\label{eq:ropdef}
\end{equation}
(the positivity of $R_T$ follows from the complete positivity of
$T_*$).  Then the action of $T_*$ on an arbitrary $\rho \in \cS(\sK)$
can be given in terms of $R_T$ as
\begin{equation}
T_*(\rho) = \ptr{\sK}{\left[(\tp{\idty}{\trn{\rho}})R_T\right]},
\label{eq:ropaction}
\end{equation}
where the transpose operation is performed with respect to the basis
$\set{e_i}$.  Here is one way to prove Eq.~(\ref{eq:ropaction}). Pick
a Kraus decomposition $\set{V_\alpha}$ of $T$.  Then, using
Eq.~(\ref{eq:dkets1}), we get
$$
R_T = \sum_\alpha
(\tp{V_\alpha}{\idty})\dketbra{\idty}{\idty}(\tp{V^*_\alpha}{\idty}) =
\sum_\alpha \dketbra{V_\alpha}{V_\alpha}.
$$
Substituting this into the right-hand side of Eq.~(\ref{eq:ropaction})
and using Lemma \ref{lm:dkets} yields
$$
\sum_\alpha \ptr{\sK}{\left[ (\tp{\idty}{\trn{\rho}})
    \dketbra{V_\alpha}{V_\alpha} \right]} = \sum_\alpha \ptr{\sK}
\dketbra{V_\alpha \rho}{V_\alpha} = \sum_\alpha V_\alpha \rho
V^*_\alpha = T_*(\rho).
$$
In fact, the map $T_*$ defined in Eq.~(\ref{eq:ropaction}) can be
extended to a completely positive map on all operators $A \in
\cB(\sK)$.

The operator $R_T$ is the unique operator for which
Eq.~(\ref{eq:ropaction}) holds.  To see this, assume that, to the
contrary, Eq.~(\ref{eq:ropaction}) holds with some other operator $R$
in place of $R_T$.  Then, for any $\rho \in \cS(\sK)$,
$$
\ptr{\sE}{\left[(\tp{\idty}{\trn{\rho}}) (R_T - R) \right]} = 0 \in \cB(\sH).
$$
The fact that $R_T = R$ is now a consequence of the following lemma \cite{dp1}.

\begin{lemma}
Let $X$ be an operator on $\tp{\sH}{\sK}$.  Suppose that, for any
$\psi \in \sK$, the operator $\brakets{\psi}{\sK}{X\psi}{\sK} \in
\cB(\sH)$ is the zero operator.  Then $X$ is the zero operator on
$\tp{\sH}{\sK}$.
\end{lemma}

Thus we have shown that, for any completely positive
$\map{T}{\cB(\sH)}{\cB(\sK)}$, there exists a unique positive operator
$R_T \in \tp{\sH}{\sK}$ such that Eq.~(\ref{eq:ropaction}) holds.
However, this correspondence works in the reverse direction as well.
That is, given a positive operator $R \in \cB(\tp{\sH}{\sK})$, the map
\begin{equation}
T^R_*(A) \defeq \ptr{\sK}{\left[(\tp{\idty}{\trn{A}})R\right]}\qquad
\forall A \in \cB(\sK)
\label{eq:rmap}
\end{equation}
is completely positive.  In order to show this, we need the following
trivial lemma.

\begin{lemma}
\label{lm:posop}
Let $\sH$ be a Hilbert space.  An operator $X \in \cB(\sH)$ is
positive if and only if it can be written in the form $X = \sum_\alpha
\ketbra{\psi_\alpha}{\psi_\alpha}$ for some collection
$\set{\psi_\alpha}$ of vectors in $\sH$.
\end{lemma}

Now let a positive operator $R \in \cB(\tp{\sH}{\sK})$ be given.  Then
Lemma \ref{lm:posop} states that we can write $R = \sum_\alpha
\dketbra{V_\alpha}{V_\alpha}$, where $V_\alpha$ are some operators
from $\sK$ to $\sH$.  Substituting this form of $R$ into
Eq.~(\ref{eq:rmap}), we get $T^R_*(A) \defeq \sum_\alpha V_\alpha A
V^*_\alpha$, which is completely positive.  The map $T^R_*$ is then
the dual of the map $\map{T^R}{\cB(\sH)}{\cB(\sK)}$, which is also
completely positive.

We are interested in the specific case when
$\map{T}{\cB(\sH)}{\cB(\sK)}$ is a channel.  Then
$\map{T_*}{\cB(\sK)}{\cB(\sH)}$ is a trace-preserving map.  That is,
for any $A \in \cB(\sK)$, we must have
$$
\tr{T_*(A)} = \tr{(\trn{A}\ptr{\sH}{R_T})}=\tr{A}=\tr{\trn{A}},
$$
which implies that $\ptr{\sH}{R_T} = \idty_\sK$.  We can summarize
everything we have said up to now in the following theorem \cite{dp1}.

\begin{theorem}
\label{th:duality}
{\bf (duality between channels and bipartite states)}
Let $\sH$ and $\sK$ be finite-dimensional Hilbert spaces.  There
exists a one-to-one correspondence between channels
$\map{T}{\cB(\sH)}{\cB(\sK)}$ and density operators $\rho \in
\cS(\tp{\sH}{\sK})$ with $\ptr{\sH}{\rho} = (1/\dim{\sK}) \idty_\sK$,
given by
\begin{equation}
T_*(A) = \dim{\sK}\ptr{\sK}[(\tp{\idty}{\trn{A}})\rho] \qquad \forall
A \in \cB(\sK),
\label{eq:chstdual}
\end{equation}
where $\map{T_*}{\cB(\sK)}{\cB(\sH)}$ is the dual channel
corresponding to $T$.
\end{theorem}

This correspondence can be extended to tensor products of channels in
the following way \cite{cdkl,zan2}.  Let
$\map{S}{\cB(\sH_A)}{\cB(\sK_A)}$ and
$\map{T}{\cB(\sH_B)}{\cB(\sK_B)}$ be channels, and let $\set{e_i}$ and
$\set{f_\mu}$ be orthonormal bases of $\sK_A$ and $\sK_B$
respectively.  Define the vectors
\begin{eqnarray}
\dkets{\idty}{A} &\defeq & \sum_i \ket{\tp{e_i}{e_i}} \nonumber \\
\dkets{\idty}{B} &\defeq & \sum_\mu \ket{\tp{f_\mu}{f_\mu}}, \nonumber
\end{eqnarray}
and the operator
\begin{equation}
R_{\tp{S}{T}} \defeq
(\tp{S}{\id}\tp{}{}\tp{T}{\id})(\tp{\dketbras
{\idty}{A}{\idty}{A}}{\dketbras{\idty}{B}{\idty}{B}}).
\label{eq:ropdef_bipart}
\end{equation}
Then the action of $\tp{S_*}{T_*}$ on density operators $\rho \in
\cS(\tp{\sK_A}{\sK_B})$ is given by
$$
\tp{S_*}{T_*}(\tp{\rho_A}{\rho_B}) = 
\ptr{\sK_A}{}
\ptr
	{\sK_B}
	{
		\left[(
			\tp
				{\idty_{\tp{\sH_A}{\sH_B}}}
				{\trn{\rho_{\tp{\sK_A}{\sK_B}}}}
			)R_{\tp{S}{T}}
		\right]
	}.
$$
In the case of a product density operator $\tp{\rho_A}{\rho_B} \in
\cS(\tp{\sK_A}{\sK_B})$ we recover the correct relation
$\tp{S_*}{T_*}(\tp{\rho_A}{\rho_B}) = \tp{S_*(\rho_A)}{T_*(\rho_B)}$.

\section{Distinguishability measures for states}
\label{sec:statedist}

In quantum information theory, we frequently encounter the following
problem:  given two states $\omega_1$ and $\omega_2$, to what extent
does one of them approximate the other?  This problem is relevant,
e.g., for the circuit model of quantum computation \cite{bar},
whenever we need to determine how much the output states of an
\qt{ideal} quantum computer differ from the corresponding output
states of the quantum circuit that approximates it.  In this section
we concentrate on two such measures of closeness for states, the
trace-norm distance and the Jozsa-Uhlmann fidelity.  We will freely
use the concepts from the theory of trace ideals; the necessary
background information is given in Sec.~\ref{sec:trideals}.

\subsection{Trace-norm distance}
\label{ssec:trnorm}

Any state $\omega$ over an algebra of observables $\cA$ is given
essentially as a \qt{catalogue} of the expectation values $\omega(A)$
for all $A\in \cA$.  Therefore it makes intuitive sense to say that
two states $\omega_1$ and $\omega_2$ are close if the corresponding
expectations $\omega_1(A)$ and $\omega_2(A)$ are close for all $A \in
\cA$.  Without loss of generality, we can compare the expectations
$\omega_1(A)$ and $\omega_2(A)$ on the unit ball of $\cA$, i.e., the
set of all $A \in \cA$ with $\norm{A} \le 1$.  The corresponding
measure of closeness between $\omega_1$ and $\omega_2$ will thus be
given by the variational expression
$$
D(\omega_1,\omega_2) = \sup_{A \in \cA; \norm{A} \le
  1}\abs{\omega_1(A)-\omega_2(A)}.
$$
In fact, we can vary only over the group $\cU(\cA)$ of the unitary
elements of $\cA$ (i.e., those $U \in \cA$ for which $UU^* = U^*U =
\idty$).  This follows from the Russo-Dye theorem \cite[p. 25]{dav1}
which states that the unit ball in a C*-algebra $\cA$ with identity is
the closed convex hull of the unitary elements of $\cA$.  So we take
\begin{equation}
D(\omega_1,\omega_2) = \sup_{U \in \cU(\cA)}\abs{\omega_1(U)-\omega_2(U)}
\label{eq:vartrdist}
\end{equation}
as the putative measure of distance between $\omega_1$ and $\omega_2$.

Consider a concrete quantum system with the Hilbert space $\sH$, and
let $\rho_1$ and $\rho_2$ be a pair of density operators on $\sH$.
Then Eq.~(\ref{eq:vartrdist}) will take the form
\begin{equation}
D(\rho_1,\rho_2) = \sup_{U \in \cU(\sH)}\abs{\tr{(\rho_1U)}-\tr{(\rho_2U)}}.
\label{eq:trnorm_unitary}
\end{equation}
Now we can use the fact that, for any $A \in \cB(\sH)$, the maximum of
$\abs{\tr{(AU)}}$ over all unitaries $U$ is attained when $AU \ge 0$ and
equals the trace norm $\trnorm{A} \defeq \tr{(A^*A)^{1/2}} \equiv
\tr{\abs{A}}$ \cite[p. 43]{scha}.  In other words, $D(\rho_1,\rho_2)$
is precisely the trace-norm distance $\trnorm{\rho_1-\rho_2}$.\\

\begin{rem}Please note that the trace-norm distance $D(\rho_1,\rho_2)$
  defined here is twice the {\em trace distance} $D(\rho_1,\rho_2)$
  defined by Nielsen and Chuang \cite{nc}.  Therefore, in order to
  avoid confusion, we will no longer use the notation
  $D(\cdot,\cdot)$.
\end{rem}

The trace-norm distance is obviously a metric on the set $\cS(\sH)$ of
all density operators on $\sH$, and therefore possesses all the
properties that make \qt{geometrical sense} (e.g., the triangle
inequality).  In particular, because $\trnorm{\rho} = 1$ for any $\rho
\in \cS(\sH)$, we have $0 \le \trnorm{\rho_1-\rho_2} \le 2$.  It
follows from the standard properties of norms that the minimum value 0
is attained if and only if $\rho_1 = \rho_2$, and it can be shown
\cite{yue} that the maximum value 2 results if and only if $\rho_1
\rho_2 = 0$ (i.e., if and only if $\rho_1$ and $\rho_2$ have
orthogonal ranges).  When $\rho_1$ and $\rho_2$ are pure states, one
can readily derive the formula $\trnorm{\rho_1-\rho_2} = 2
\sqrt{1-\abs{\braket{\psi_1}{\psi_2}}^2}$, where $\psi_1$ and $\psi_2$
are the corresponding state vectors.  Furthermore, we have the
following key result.

\begin{theorem}
\label{th:trnorminv}
Let $\map{T}{\cB(\sH)}{\cB(\sK)}$ be a channel.  Then, for any
$\rho_1,\rho_2 \in \cS(\sK)$, we have the following.
\begin{enumerate} 
\item$\trnorm{T_*(\rho_1)-T_*(\rho_2)} \le \trnorm{\rho_1-\rho_2}$.
\item If $T$ is unitarily implemented, i.e., $T(A) = UAU^*$ for a
  unitary $\map{U}{\sH}{\sK}$, then $\trnorm{T_*(\rho_1) -
    T_*(\rho_2)} = \trnorm{\rho_1 - \rho_2}$.
\end{enumerate}
\end{theorem}

\begin{proof}
The proof of the first statement, due to Ruskai \cite{rus}, runs as
follows.  Write $\rho_1-\rho_2$ as a difference of two positive
operators $N_+,N_-$ with orthogonal ranges, so that
$\abs{\rho_1-\rho_2} = N_+ +N_-$.  Then
\begin{eqnarray}
\trnorm{T_*(\rho_1)-T_*(\rho_2)} &=& \trnorm{T_*(N_+) - T_*(N_-)} \nonumber \\
& \le & \trnorm{T_*(N_+)} + \trnorm{T_*(N_-)} \nonumber \\
& = & \tr{T_*(N_+)} + \tr{T_*(N_-)} \nonumber \\
& = & \tr{(N_+ + N_-)} \nonumber \\
& \equiv & \trnorm{\rho_1 - \rho_2}, \nonumber
\end{eqnarray}
which concludes the proof.

To prove the second statement, note that $\maps{V}{UVU^*}$ is a group
isomorphism between $\cU(\sH)$ and $\cU(\sK)$. Therefore substituting
$U^*\rho_1U$ and $U^*\rho_2U$ into Eq.~(\ref{eq:trnorm_unitary})
instead of $\rho_1$ and $\rho_2$ does not change the value of the
supremum.
\end{proof}

The trace-norm distance also has an operational characterization in
terms of generalized quantum measurements, and we will come back to it
in Sec.~\ref{ssec:qdet}.  

\subsection{Jozsa-Uhlmann fidelity}
\label{ssec:jufid}

Another useful distinguishability measure for quantum states, the {\em
  fidelity}, is given by the formidable-looking expression
\begin{equation}
F(\rho_1,\rho_2) \defeq
\left(\tr{\sqrt{\sqrt{\rho_1}\rho_2\sqrt{\rho_1}}}\right)^2,
\label{eq:statefid}
\end{equation}
where $\rho_1$ and $\rho_2$ are a pair of density operators.  The
fidelity (\ref{eq:statefid}) was introduced by Jozsa \cite{joz}, but
the original idea came from the work of Uhlmann \cite{uhl} who
generalized the notion of the \qt{transition probability}
$\braket{\psi}{\phi}$ for pure states to general states over
C*-algebras.  For this reason we will refer to the fidelity $F$ as the
{\em Jozsa-Uhlmann fidelity}.

The main appeal of the Jozsa-Uhlmann fidelity lies in the result known
as the Uhlmann theorem \cite{uhl}.  We state this theorem in the form
given by Jozsa \cite{joz}.

\begin{theorem}
\label{th:uhlmann}
{\bf (Uhlmann)} Let $\rho_1$ and $\rho_2$ be density operators on a
Hilbert space $\sH$.  Then
\begin{equation}
F(\rho_1,\rho_2) = \max_{\psi_1,\psi_2} \abs{\braket{\psi_1}{\psi_2}}^2,
\label{eq:uhlmann}
\end{equation}
where the maximum is taken over all purifications $\psi_1$ and
$\psi_2$ of $\rho_1$ and $\rho_2$ respectively in an extended Hilbert
space $\tp{\sH}{\sK}$. 
\end{theorem}

\begin{proof}
Without loss of generality, we may take $\sK \simeq \sH$ because only
the nonzero eigenvalues of a density operator are relevant for
constructing its purification.  Let $\set{e_i}$ be the eigenvectors of
$\rho_1$, and $\set{f_i}$ the eigenvectors of $\rho_2$.  We can write
all purifications of $\rho_1$ and $\rho_2$ in the form
$\dket{\sqrt{\rho_1}V}$ and $\dket{\sqrt{\rho_2}UW}$ with respect to
the basis $\set{e_i}$, where $f_i = Ue_i$, and $V$ and $W$ are the
unitaries corresponding to the choice of basis in the auxiliary
Hilbert space $\sK$ for each of the purifications.  Writing
$$
\dbraket{\sqrt{\rho_2}UW}{\sqrt{\rho_1}V} =
\tr{(W^*U^*\sqrt{\rho_2}\sqrt{\rho_1}V)} =
\tr{(VW^*U^*\sqrt{\rho_2}\sqrt{\rho_1})}
$$
and observing that $U$ is determined by $\rho_1$ and $\rho_2$, we see
that the maximum in Eq.~(\ref{eq:uhlmann}) can be written as
$\max_{V\in \cU(\sH)}\abs{\tr{(\sqrt{\rho_2}\sqrt{\rho_1}V})}^2$ and
hence equals $\left(\tr{\abs{\sqrt{\rho_2}\sqrt{\rho_1}}}\right)^2
\equiv \left(\tr{\sqrt{\sqrt{\rho_1}\rho_2\sqrt{\rho_1}}}\right)^2$.
This proves the theorem.
\end{proof}

Apart from its immediate physical significance, the Uhlmann theorem
allows us to derive the properties of the fidelity
(\ref{eq:statefid}).  We summarize these properties in the theorem
below, for the proof of which the reader is referred to the paper of
Jozsa \cite{joz}.

\begin{theorem}
{\bf (properties of the Jozsa-Uhlmann fidelity)}
\begin{enumerate}
\item $0 \le F(\rho_1,\rho_2) \le 1$ and $F(\rho_1,\rho_2) =1$ if and
  only if $\rho_1 = \rho_2$.
\item $F$ is a symmetric function: $F(\rho_1,\rho_2) = F(\rho_2,\rho_1)$.
\item If $\rho_1$ is pure, then $F(\rho_1,\rho_2) = \tr{(\rho_1\rho_2)}$
  for any $\rho_2$.  Otherwise, $F(\rho_1,\rho_2) \ge
  \tr{(\rho_1\rho_2)}$.
\item For a fixed $\rho$, $F(\rho,\cdot)$ is a concave function:
  $F(\rho,\lambda_1\rho_1 + \lambda_2\rho_2) \ge \lambda_1
  F(\rho,\rho_1)+\lambda_2 F(\rho,\rho_2)$ for any positive real numbers
  $\lambda_1,\lambda_2$ with $\lambda_1+\lambda_2 = 1$.
\item $F$ is multiplicative with respect to tensor products:
  $F(\tp{\rho_1}{\rho_2},\tp{\rho_3}{\rho_4}) =
  F(\rho_1,\rho_3)F(\rho_2,\rho_4)$.
\item If $T$ is a channel, then $F(T_*(\rho_1),T_*(\rho_2)) \ge
  F(\rho_1,\rho_2)$, where equality holds for all $\rho_1,\rho_2$ when
  $T$ is unitarily implemented.
\end{enumerate} 
\end{theorem}

It can be shown that the trace-norm distance and the Jozsa-Uhlmann
fidelity are equivalent distinguishability measures for quantum
states.  This follows from the following key theorem \cite{fvdg},
given here without proof.

\begin{theorem}
\label{th:fvdg}
{\bf (Fuchs-van de Graaf)}
For any two density operators $\rho_1,\rho_2$,
\begin{equation}
2-2\sqrt{F(\rho_1,\rho_2)} \le \trnorm{\rho_1-\rho_2} \le
2\sqrt{1-F(\rho_1,\rho_2)}.
\label{eq:fid_trnorm}
\end{equation}
\end{theorem}
Since the Jozsa-Uhlmann fidelity is easier to compute than the
trace-norm distance, the Fuchs-van de Graaf theorem provides a quick
and painless way to get tight estimates of the trace-norm distance.

\subsection{Quantum detection theory}
\label{ssec:qdet}

As we have shown in Example \ref{ex:msrmt}, any measurement performed
on a quantum system with the Hilbert space $\sH$ can be described by a
collection of effects $F_x$ that form a resolution of identity on
$\sH$, i.e., $\sum_x F_x = \idty$.  This is an example of a {\em
  positive operator-valued measure}, which is defined as follows
\cite{hol2}.

\begin{definition}
Let $(\sX,\Sigma)$ be a measurable space, i.e., $\sX$ is a set and
$\Sigma$ is a $\sigma$-algebra of subsets of $\sX$, and let $\sH$ be a
Hilbert space.  Then a {\em positive operator-valued measure (POVM)}
on $\sX$ with values in $\cB(\sH)$ is a map $F$ from $\sX$ to the
positive operators on $\sH$ which is
\begin{enumerate}
\item[(a)] normalized:  $F(\emptyset)=0,F(\sX)=\idty$.
\item[(b)] $\sigma$-additive:  for any countable collection of
  pairwise disjoint sets $\sS_i \in \Sigma$, $F(\cup_i \sS_i) = \sum_i
  F(\sS_i)$, where the sum converges in the strong operator topology.
\end{enumerate}
\end{definition}

The definition just given is the most general.  In this section we
will content ourselves with the case when the set $\sX$ is finite, so
we will not have to deal with $\sigma$-algebras and the like.  Then
any POVM is simply a collection of positive operators
$\setcond{F_x}{x\in \sX}$ with $\sum_x F_x = \idty$.  These operators
will be referred to as the {\em elements} of the POVM.

Consider the following problem.  We are presented with a quantum
system whose state is unknown, but we are told that it is drawn from
some known set $\set{\rho_m}^M_{m=1}$ according to the probability
distribution $\set{p_m}^M_{m=1}$.  Our task is to devise a measurement
that would maximize the probability of correctly identifying the
state.  This is known as the {\em $M$-ary quantum detection problem}
\cite{hel}.

Any measurement we would perform will be described by a POVM $F_m$ on
the $M$-element set $\set{1,\ldots,M}$.  Given the state $\rho$, the
probability of identifying $\rho$ as $\rho_m$ is equal to $\tr{(\rho
  F_m)}$.  Thus the average probability of correct decision using the
POVM $F \defeq \set{F_m}$ is given by
\begin{equation}
\bar{P}_c[F]\defeq\sum^M_{m=1}p_m \tr{(\rho_mF_m)}.
\label{eq:avpcm}
\end{equation}
The problem of designing the optimum $M$-ary quantum detector thus
amounts to finding the $M$-element POVM $F$ that would maximize
$\bar{P}_c[F]$.  

It is not possible to give a general closed-form expression for the POVM that
would maximize Eq.~(\ref{eq:avpcm}).  However, a theorem of Yuen,
Kennedy, and Lax \cite{ykl} gives necessary and sufficient
conditions for a given POVM $F$ to be a maximizer of $\bar{P}_c[F]$.
Usually the candidate POVM's are found by inspection or by taking
advantage of the problem's intrinsic symmetries, should they exist, in
which case the Yuen-Kennedy-Lax theorem gives a quick way to verify
the optimality. For many interesting examples, the reader is invited
to consult their article \cite{ykl}, as well as the book by Helstrom
\cite{hel}.

We give a complete solution of the binary quantum detection problem
($M=2$).  In this case we are considering two-element POVM's
$\set{F,\idty-F}$, so there is only one independent operator $F$ that
must satisfy the condition $0 \le F \le \idty$.  The corresponding
variational expression is
$$
\bar{P}_c = \max_{0 \le F \le \idty} \set{p_1\tr{(\rho_1 F)}+p_2 \tr{[\rho(\idty-F)]}},
$$
which simplifies to
\begin{equation}
\bar{P}_c = p_2+ \max_{0 \le F \le \idty} \tr{[(p_1\rho_1-p_2\rho_2)F]}.
\label{eq:binaryopt}
\end{equation}
We have the following theorem \cite{hel}.

\begin{theorem}
\label{th:optbinary}
{\bf (optimum binary quantum detection)} Consider the binary quantum
detection problem for the density operators $\rho_1$ and $\rho_2$ and
the probabilities $p_1$ and $p_2$.  Then the optimum average
probability of correct decision is given by
\begin{equation}
\bar{P}_c = \frac{1}{2}+\frac{1}{2}\trnorm{p_1\rho_1-p_2\rho_2},
\label{eq:optbinary}
\end{equation}
and the elements of the optimum POVM can be chosen to be projection operators.
\end{theorem}

\begin{proof}
Write down the orthogonal decomposition $p_1\rho_1-p_2\rho_2 = R_+ -
R_-$, where $R_\pm \ge 0$ and $R_+R_- = 0$.  Because $R_- \ge 0$, we
have $\tr{R_-F}\ge 0$ for any $F\ge 0$, so
$$
\max_{0 \le F \le \idty} \tr{[(R_+-R_-)F]} \le \max_{0 \le F \le \idty}
\tr{(R_+F)} \le \tr{R_+},
$$
where the maximum is achieved by the projection operator $P$ with
$PR_+ = R_+$ and $PR_-=0$.  Thus
$$
\max_{0\le F \le \idty}\tr{[(R_+-R_-)F]} = \tr{R_+}.
$$
Because $\tr{(p_1\rho_1-p_2\rho_2)} = p_1-p_2$, we have $\tr{R_-} =
\tr{R_+} + p_2-p_1$.  Also
$$
\trnorm{p_1\rho_1 - p_2\rho_2} = \tr{\abs{p_1\rho_1-p_2\rho_2}} =
\tr{R_+}+\tr{R_-} = 2\tr{R_+} + p_2 - p_1,
$$
whence it follows that
$$
\max_{0 \le F \le \idty}\bar{P}_c[F] = p_2 + \max_{0\le F \le \idty}
\tr{[(p_1\rho_1-p_2\rho_2)F]} =
\frac{1}{2}+\frac{1}{2}\trnorm{p_1\rho_1-p_2\rho_2}.
$$
The optimizing POVM is then given by $\set{P,\idty-P}$.
\end{proof}

Theorem \ref{th:optbinary} clearly exhibits the prominent role played
by the trace-norm distance in the quantitative characterization of the
performance of generalized quantum measurements.  In particular, the
probability of correct discrimination between two equiprobable states
$\rho_1$ and $\rho_2$ equals $1/2+(1/4)\trnorm{\rho_1-\rho_2}$.
Furthermore, the maximum average probability (\ref{eq:optbinary}) of
correct decision equals unity if and only if $\rho_1$ and $\rho_2$ are
such that the trace norm $\trnorm{p_1\rho_1-p_2\rho_2}$ attains its
maximum value of $p_1+p_2 \equiv 1$, which happens if and only if
$\rho_1\rho_2=0$.  In the case when $\rho_1$ and $\rho_2$ are pure
states, this reduces to the requirement that the corresponding state
vectors be orthogonal.

The trace-norm distance between states $\rho_1$ and $\rho_2$ can also
be expressed as \cite[p. 405]{nc}
\begin{equation}
\trnorm{\rho_1-\rho_2} = 2\sup_{\set{F_m}} \sum_m
\abs{\tr{(\rho_1F_m)}-\tr{(\rho_2F_m)}},
\label{eq:trnormpovm}
\end{equation}
where the supremum is taken with respect to all POVM's whose elements
belong to $\cB(\sH)$, where $\sH$ is the Hilbert space on which the density
operators $\rho_1$ and $\rho_2$ act.  A similar expression can be derived for
$\trnorm{p_1\rho_1-p_2\rho_2}$, which shows that it is sufficient to
consider only two-element POVM's for the solution of the binary
quantum detection problem.  The sum on the right-hand side of
Eq.~(\ref{eq:trnormpovm}) is the so-called Kolmogorov distance between
the probability distributions $\set{\tr{(\rho_1F_m)}}^M_{m=1}$ and
$\set{\tr{(\rho_2F_m)}}^M_{m=1}$.

Incidentally, the Jozsa-Uhlmann fidelity (\ref{eq:statefid})
can likewise be given an intuitive operational meaning in terms of
generalized quantum measurements by means of the formula
\cite[p. 412]{nc}
\begin{equation}
\sqrt{F(\rho_1,\rho_2)} = \inf_{\set{F_m}} \sum_m \sqrt{ 
	\tr{(\rho_1 F_m)} \tr{(\rho_2 F_m)}}.
\label{eq:fidpovm}
\end{equation}
The sum in the right-hand side of (\ref{eq:fidpovm}) is the so-called
Fisher metric \cite[p. 29]{an}, a Riemannian metric on the manifold of
probability distributions on an $m$-element sample space.  The
physical meaning of Eqs.~(\ref{eq:trnormpovm}) and (\ref{eq:fidpovm})
is apparent:  whenever we engage in the business of distinguishing
quantum states, we are essentially distinguishing probability
distributions describing the outcomes of generalized quantum
measurements.

\section{Distinguishability measures for channels}
\label{sec:chdist}

In the preceding section we have discussed ways in which we can
compare quantum states.  It is also important to have at our disposal
some tools for the comparison of channels.  In this section we
describe two distinguishability measures for channels, the cb-norm
distance and the channel fidelity.  The latter distinguishability
measure was defined and studied by the present author \cite{rag1}.

\subsection{Norm of complete boundedness}
\label{ssec:cbnorm}

Just as we have defined a distinguishability measure for states as a
metric induced by the trace norm, it should likewise be possible to
construct a distinguishability measure for channels from $\cB(\sH)$ to
$\cB(\sK)$ in a natural way from a suitable metric on the set of all
completely positive maps from $\cB(\sH)$ to $\cB(\sK)$.  One possible
candidate is the metric induced by the operator norm,
\begin{equation}
\norm{T} \defeq \sup_{X \in \cB(\sH); \norm{X}=1} \norm{T(X)}.
\label{eq:opnorm}
\end{equation}
Unfortunately, the operator norm is rather ill-behaved: it is not
stable with respect to tensor products.  In particular, there are some
positive maps $T$, for which the norm  $\norm{\tp{T}{\id_n}}$ will
increase with $n$, as the following example \cite{pau} shows.

\begin{example}
({\bf transposition map revisited}) Consider the transposition map
  $\Theta$ (cf. Example \ref{ex:transp}) on the algebra $\cM_2$, and
  define the map $\Theta_2 \defeq \tp{\Theta}{\id_2}$ on
  $\tp{\cM_2}{\cM_2}$.  Let $F$ be the flip operator
$$
F = \sum^2_{i,j=1} \ketbra{\tp{e_j}{e_i}}{\tp{e_i}{e_j}} \equiv
	\left(
	\begin{array}{cc}
	e_{11} & e_{21} \\
	e_{12} & e_{22} \end{array} \right),
$$
for which we have $\norm{F} = 1$.  Now
$$
\Theta_2(F) = \left(
\begin{array}{cc}
\Theta(e_{11}) & \Theta(e_{21}) \\
\Theta(e_{12}) & \Theta(e_{22}) \end{array}
\right) = \left(
\begin{array}{cc}
e_{11} & e_{12} \\
e_{21} & e_{22} \end{array} \right),
$$
which has norm 2.  Thus $\norm{\Theta_2} \ge 2$.\endex

A good choice then is the metric induced by the stabilized version of
the operator norm (\ref{eq:opnorm}), namely the {\em norm of complete
  boundedness} (or cb-norm for short), defined by \cite{pau}
\begin{equation}
\cbnorm{T} \defeq \sup_n \norm{\tp{T}{\id_n}}.
\label{eq:cbnorm}
\end{equation}
For any operator $X \in \cB(\sH)$ and any two maps $S,T$ on
$\cB(\sH)$ with finite cb-norm (in the case of finite-dimensional
$\sH$, this is always true \cite{pau}), we have the relations
\begin{eqnarray}
\norm{T(X)} &\le & \cbnorm{T}\norm{X}, \label{eq:cbnorm1} \\
\cbnorm{ST} &\le & \cbnorm{S}\cbnorm{T}, \label{eq:cbnorm2} \\
\cbnorm{\tp{S}{T}} &=& \cbnorm{S}\cbnorm{T}.
\label{eq:cbnorm3}
\end{eqnarray}
Furthermore, for any completely positive map $T$, we have \cite{pau}
$\cbnorm{T} = \norm{T(\idty)}$.  This implies, in particular, that
$\cbnorm{T}=1$ for any channel.

We have defined the cb-norm (\ref{eq:cbnorm}) for channels that act on
observables, but we also need a similar norm for the corresponding
dual channels that act on states (and, by linear extension, on
trace-class operators).  Thus let $\map{T}{\cB(\sH)}{\cB(\sK)}$ be a
channel, and let $\map{T_*}{\cT_1(\sK)}{\cT_1(\sH)}$ be its dual.
Then we define the norm of $T_*$ as
$$
\norm{T_*} \defeq \sup_{X \in \cT_1(\sK); \trnorm{X}=1} \trnorm{T_*(X)},
$$ 
and the corresponding cb-norm as
\begin{equation}
\cbnorm{T_*} \defeq \sup_n \norm{\tp{T_*}{\id_n}}.
\label{eq:cbnormdual}
\end{equation}
Luckily, the cb-norms of $T$ and $T_*$ agree, as follows from the
following argument. We have
$$
\sup_{X \in \cT_1(\sK); \atop \trnorm{X}=1} \sup_{Y \in \cB(\sH);
  \atop \norm{Y} =
  1} \abs{\tr{[T_*(X)Y]}} = \sup_{X \in \cT_1(\sK); \atop \trnorm{X}=1}
  \trnorm{T_*(X)} \equiv \norm{T_*}
$$
and
\begin{eqnarray}
\sup_{X \in \cT_1(\sK); \atop \trnorm{X}=1} \sup_{Y \in \cB(\sH);
  \atop \norm{Y} =
  1} \abs{\tr{[T_*(X)Y]}} &=& \sup_{Y \in \cB(\sH); \atop \norm{Y}=1}
  \sup_{X \in \cT_1(\sK); \atop \trnorm{X} = 1} \abs{\tr{[XT(Y)]}}
  \nonumber \\
&=& \sup_{Y \in \cB(\sH); \atop \norm{Y}=1} \norm{T(Y)} \equiv
  \norm{T}, \nonumber
\end{eqnarray}
which shows that $\norm{T}=\norm{T_*}$ for any two maps
$\map{T}{\cB(\sH)}{\cB(\sK)}$ and $\map{T_*}{\cT_1(\sK)}{\cT_1(\sH)}$
that are connected via the duality relation
$$
\tr{[XT(Y)]} = \tr{[T_*(X)Y]},\qquad \forall X \in \cT_1(\sK),\forall
Y \in \cB(\sH)
$$
and such that at least one of them has finite norm. Therefore we
have, for any $n$, $\norm{\tp{T}{\id_n}} = \norm{\tp{T_*}{\id_n}}$;
taking the supremum of both sides with respect to $n$, we get
$\cbnorm{T}=\cbnorm{T_*}$. This equality holds for completely positive
maps in particular, and for completely bounded maps in general (e.g.,
for sums and differences of completely positive maps). Thus the
properties similar to (\ref{eq:cbnorm1})-(\ref{eq:cbnorm3}) also hold
for the cb-norm (\ref{eq:cbnormdual}), but with obvious modifications
(e.g., with the operator norm replaced by the trace norm).  In
particular, we have $\cbnorm{T_*}=1$ for any channel $T$.  The
\qt{dual} cb-norm (\ref{eq:cbnormdual}) has appeared, under different
guises, in the work of Aharonov, Kitaev, and Nisan \cite{akn}, Giedke \etal
\cite{gbcz}, and Kitaev \cite{kit}. 

If two channels $T,S$ are close in cb-norm, then, for any density
operator $\rho$, the corresponding states $T_*(\rho),S_*(\rho)$ are close
in trace norm since, from Eq.~(\ref{eq:cbnorm1}), it follows that
$$
\trnorm{T_*(\rho)-S_*(\rho)} = \trnorm{(T_*-S_*)(\rho)} \le
\cbnorm{T_*-S_*} = \cbnorm{T-S}.
$$
In fact, the above estimate cannot be loosened by adjoining a
second system with the Hilbert space $\sK$ in some state $\rho_\sK$,
entangling the two systems through some channel $K$ on
$\cB(\tp{\sH}{\sK})$, and then comparing the channels
$\tp{T}{R}$ and $\tp{S}{R}$, where $\map{R}{\cB(\sK)}{\cB(\sK)}$ is some
suitably chosen channel. This is evident from the
estimate
$$
\trnorm{ (\tp{T_*}{R_*}) K_* (\tp{\rho}{\rho_\sK})-
(\tp{S_*}{R_*})K_*(\tp{\rho}{\rho_\sK})} \le \cbnorm{T_*-S_*},
$$
which can be easily obtained by repeated application of
Eqs.~(\ref{eq:cbnorm1})-(\ref{eq:cbnorm3}).  In other words, as far as
the cb-norm distinguishability criterion is concerned, entangling the
system with an auxiliary system will not improve distinguishability of
the channels $T$ and $S$.  The cb-norm, however, is an extremely
strong distinguishability measure: its definition already accounts for
optimization with respect to entanglement and input states over
Hilbert spaces of very large (but finite) dimension. There exist
weaker measures of channel distinguishability (such as the channel
fidelity presented below) that describe how channels may be
distinguished with bounded resources.  Using these weaker criteria,
one may show that the use of entanglement does entail an improvement
in the practical distinguishability of both states and channels
\cite{dpp}.

\subsection{Channel fidelity}
\label{ssec:chfid}

Recall that the Jozsa-Uhlmann fidelity (\ref{eq:statefid}) can be
given an intuitive operational meaning in terms of the Fisher metric
on the manifold of probability distributions.  This suggests that any
experiment designed to distinguish between two given quantum states
amounts to distinguishing a pair of suitable probability
distributions.

\begin{figure}
\includegraphics{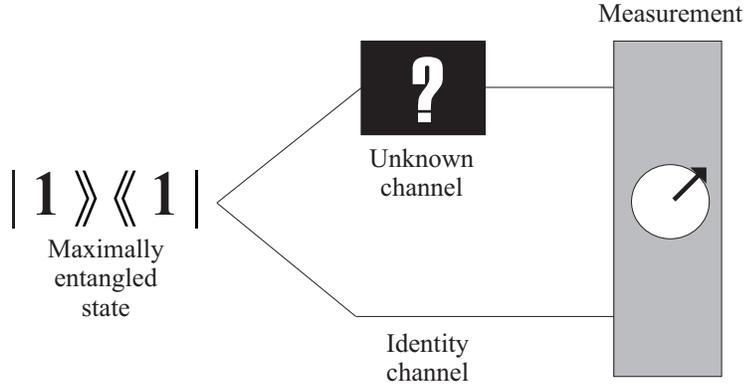}

\caption{Using entanglement to distinguish between quantum channels.}
\label{fig:chfid}
\end{figure}

It is tempting to apply the same idea to distinguishability of
channels.  Consider the situation portrayed in Fig.~\ref{fig:chfid}.
Namely, suppose that we are given a \qt{black box} that effects one of
two channels $S,T$.  In order to tell what
the \qt{black box} does, we can exploit the correspondence
(\ref{eq:ropdef}) between channels and bipartite states.  That is, we
prepare two systems, $A$ and $B$, in the maximally entangled state
$\dketbra{\idty}{\idty}$, and then let the \qt{black box} act on $A$,
while leaving $B$ untouched.  The resulting state is, up to
normalization, the $R$-operator (\ref{eq:ropdef}) of the unknown
channel.  In order to distinguish between $S$ and $T$, we simply
perform a measurement that would distinguish between the states
$(\tp{S_*}{\id})(\dketbra{\idty}{\idty})$ and
$(\tp{T_*}{\id})(\dketbra{\idty}{\idty})$.  In fact, D'Ariano, Lo
Presti, and Paris have recently shown \cite{dpp} that strategies of
this kind generally result in improved distinguishability.

The correspondence (\ref{eq:ropdef}) between channels
$\map{T}{\cB(\sH)}{\cB(\sK)}$ and positive
operators $R_T \in \cB(\tp{\sH}{\sK})$ is, as we already stated,
bijective:  namely, $R_S = R_T$ if and only if $S=T$.  Furthermore, it
is easy to see that $\rho_T \defeq (1/d)R_T$, where $d = \dim \sK$, is
a density operator.  Hence it seems natural to define the {\em
  fidelity $\cF(S,T)$ between two channels $S$ and $T$} as the
Jozsa-Uhlmann fidelity between the density operators $\rho_S$ and
$\rho_T$:
\begin{equation}
\cF(S,T) \defeq F(\rho_S,\rho_T).
\label{eq:chfid}
\end{equation}
Being expressed in terms of the Jozsa-Uhlmann fidelity $F$, the
channel fidelity $\cF$ inherits many of its natural properties. We now
summarize these properties with brief proofs \cite{rag1}.

\begin{theorem}
\label{th:chfidprop}
{\bf (properties of the channel fidelity)}
Let $S,T$ be channels $\cB(\sH) \rightarrow \cB(\sK)$, and let $d =
\dim \sK$.  Then the channel fidelity $\cF$ has the following
properties.
\begin{enumerate}
\item $0 \le \cF(S,T) \le 1$, and $\cF(S,T) = 1$ if and only if $S=T$.

\item $\cF(S,T) = \cF(T,S)$ (symmetry).

\item For any two unitarily implemented channels $\hat{U}$ and
  $\hat{V}$ [i.e., $\hat{U}_*(\rho)=U\rho U^*$ and $\hat{V}_*(\rho) =
  V\rho V^*$ with unitary $U$ and $V$], $\cF(\hat{U},\hat{V}) =
  (1/d^2)\abs{\tr{(U^* V)}}^2$.

\item For any real $\lambda$ with $0 \le \lambda \le 1$,
  $\cF(S,\lambda T_1 + (1-\lambda)T_2) \ge \lambda \cF(S,T_1) +
  (1-\lambda)\cF(S,T_2)$ (concavity).

\item $\cF(\tp{S_1}{S_2},\tp{T_1}{T_2}) = \cF(S_1,T_1)\cF(S_2,T_2)$
  (multiplicativity with respect to tensoring).

\item $\cF$ is invariant under composition with unitarily implemented
  channels, i.e., for any unitarily implemented channel $\hat{U}$,
  $\cF(S\hat{U},T\hat{U}) = \cF(S,T)$ and $\cF(\hU S,\hU T) =
  \cF(S,T)$.

\item $\cF$ does not decrease under composition with arbitrary
  channels, i.e., for any channel $R$, $\cF(SR, TR) \ge \cF(S,T)$.

\end{enumerate}\end{theorem}

\begin{proof}
\begin{enumerate}
\item[1, 2.] These hold because $\rho_S$ and $\rho_T$ are
  density operators, and because $\maps{T}{\rho_T}$ is a
  bijection.
\item[3.] $R_{\hat{U}} = \dketbra{U}{U}$, and similarly for
  $R_{\hat{V}}$. Thus both $\rho_{\hat{U}}$ and $\rho_{\hat{V}}$ are
  pure states. Since for pure states $\psi,\phi$ we have $F =
  \abs{\braket{\psi}{\phi}}^2$, it follows that $\cF(\hat{U},\hat{V})
  = (1/d^2) \abs{\dbraket{U}{V}}^2 =(1/d^2) \abs{\tr{(U^* V)}}^2$.
\item[4.] Note that the map $\maps{T}{\rho_T}$ is linear.  Thus, for $T =
  \lambda T_1 + (1-\lambda)T_2$, we have $\rho_T = \lambda \rho_{T_1}
  + (1-\lambda) \rho_{T_2}$, and the concavity of $\cF$ follows from
  the concavity of the mixed-state fidelity (\ref{eq:statefid}).
\item[5.] It follows from Eq.~(\ref{eq:ropdef_bipart}) that
  $\rho_{\tp{S}{T}} = \tp{\rho_S}{\rho_T}$, and the multiplicativity
  property of $\cF$ follows from the corresponding property of $F$.
\item[6.] Write $\tp{T\hU}{\id} = (\tp{T}{\id})(\tp{\hU}{\id})$ to obtain
  $\rho_{T\hU} = (\tp{U}{\idty})\rho_T(\tp{U^*}{\idty})$, and do the
  same for $S\hU$.  Since the mixed-state fidelity $F$ is invariant
  under unitary transformations, the same property holds for the
  channel fidelity $\cF$.  For $\hU T$, we have $\rho_{\hU T} =
  (\tp{T_*}{\id}) \dketbra{U}{U} =
  (\tp{\idty}{\trn{U}})\rho_T(\tp{\idty}{(\trn{U})^*})$, and the same
  holds for $\hU S$.  Because $U$ is unitary, $\trn{U}$ is unitary
  also. Thus $\rho_T$ and $\rho_{\hU T}$ are unitarily equivalent, as
  are $\rho_S$ and $\rho_{\hU S}$, and the desired conclusion again
  follows from the unitary invariance of the Jozsa-Uhlmann fidelity.
\item[7.] The same reasoning as before, except now we have to use the
  property that $F((\tp{R_*}{\id})\rho_S,(\tp{R_*}{\id})\rho_T) \ge
  F(\rho_S,\rho_T)$.\end{enumerate}\end{proof}

\begin{rem}
Property 6 (invariance of $\cF$ under unitary transformations) implies
that our definition of the channel fidelity is good in the sense that
we could have used {\em any} maximally entangled pure state to define
the density operators $\rho_S$ and $\rho_T$ for the channels $S$ and
$T$ to obtain the same numerical value for the fidelity $\cF(S,T)
\defeq F(\rho_S,\rho_T)$.
\end{rem}

Our next step is to obtain a meaningful analogue of Uhlmann's theorem
for the channel fidelity $\cF$.  In order to do that, we must draw the
connection between the channel $T$ and purifications of the density
operator $\rho_T$. As a warm-up, let us first prove the following
lemma \cite{rag1}.

\begin{lemma}
Given a channel $\map{T}{\cB(\sH)}{\cB(\sK)}$, where $\sH$ and $\sK$
are isomorphic Hilbert spaces with $d=\dim{\sH}=\dim{\sK}$, the
density operator $\rho_T$ is pure if and only if the channel $T$ is unitarily
implemented. 
\end{lemma}

\begin{proof}
Proving the forward implication is easy:  $\rho_{\hat{U}} =
(1/d)\dketbra{U}{U}$, which is a pure state.  Let us now prove the
reverse implication. Suppose that,
given the channel $T$, the state $\rho_T$ is pure.  It follows from
Theorem \ref{th:duality} that the reduced density operator
$\ptr{\sH}{\rho_T}$ is a multiple of the identity, i.e., a maximally
mixed state.  Since $\rho_T$ is pure, the reduced density operators
$\ptr{\sK}{\rho_T}$ and $\ptr{\sH}{\rho_T}$ have the same nonzero
eigenvalues \cite{lin}.  All eigenvalues of $\ptr{\sH}{\rho_T} \equiv
(1/d)\idty$ are equal and positive.  Since $\sH \simeq \sK$ by
assumption, $\ptr{\sK}{\rho_T}$ and $\ptr{\sH}{\rho_T}$ are
isospectral.  Hence $\rho_T$ is a maximally entangled state and
therefore has the form $(1/d)\dketbra{U}{U}$ for some unitary $U$
(cf. Example \ref{ex:maxent}).  Using Eq.~(\ref{eq:dkets1}), we can
write
$$
\dketbra{U}{U} = (\tp{U}{\idty})\dketbra{\idty}{\idty}(\tp{U^*}{\idty}),
$$
which implies that $T$ is a unitarily implemented channel.
\end{proof}

Therefore, for any two unitarily implemented channels $\hat{U}$ and
$\hat{V}$, the states $\rho_{\hat{U}},\rho_{\hat{V}}$ are already pure
and, as stated in Theorem \ref{th:chfidprop},
\begin{equation}
\cF(\hat{U},\hat{V}) = \frac{1}{d^2} \abs{\dbraket{U}{V}}^2 \equiv
\frac{1}{d^2} \abs{\tr{(U^*V)}}^2,
\label{eq:fidunitary}
\end{equation}
which is nothing but the squared normalized Hilbert-Schmidt inner product of
the operators $U$ and $V$.  As we shall now show, the fidelity
$\cF(S,T)$ for {\em arbitrary channels}
$\map{S,T}{\cB(\sH)}{\cB(\sK)}$ can be expressed as a maximum of
expressions similar to the right-hand side of
Eq.~(\ref{eq:fidunitary}), but with the difference that, in place of
the unitaries $U$ and $V$, there will appear certain {\em isometries}
from $\sK$ to $\tp{\sH}{\sE}$, where $\sE$ is a suitably defined
auxiliary Hilbert space.  The following theorem \cite{rag1} states
this in precise terms.

\begin{theorem}
\label{th:uhl_chfid}
Let $\map{S,T}{\cB(\sH)}{\cB(\sK)}$ be channels, where $\sH$ and $\sK$
are finite-dimensional Hilbert spaces.  Then we can choose a Hilbert
space $\sE$ and two isometries $\map{V,W}{\sK}{\tp{\sH}{\sE}}$ such
that, for any $A \in \cB(\sH)$,
\begin{eqnarray}
S(A) &=& V^*(\tp{A}{\idty_\sE})V \label{eq:s_sti} \\
T(A) &=& W^*(\tp{A}{\idty_\sE})W, \label{eq:t_sti}
\end{eqnarray}
and the isometries $V,W$ are unique up to a unitary transformation of
$\sE$.  Furthermore,
\begin{equation}
\cF(S,T) = \left(\frac{1}{\dim \sK}\right)^2 \max_{V,W} \abs{\tr{(V^*W)}}^2,
\label{eq:uhl_chfid}
\end{equation}
where the maximum is taken over all such isometries $V$ and $W$.
\end{theorem}

\begin{proof}
Consider the channel $S$.  Given any Kraus decomposition
$\set{V_\alpha}$ of $S$, we can define the isometry $V$ through
Eq.~(\ref{eq:isokraus}).  It can be shown that we can always choose a
Kraus decomposition of $S$ in such a way that the operators forming it
are linearly independent in the sense of Hilbert-Schmidt; such a
decomposition, referred to as the {\em minimal Kraus decomposition}
\cite{ncsb}, will consist of at most $\dim \sH \cdot \dim \sK$
operators.  The same holds for the channel $T$.  Then we can choose
the Hilbert space $\sE \simeq \tp{\sH}{\sK}$ and add as many zero
operators to the given minimal Kraus decompositions $\set{V_\alpha}$
and $\set{W_\alpha}$ of $S$ and $T$ as necessary.

Assuming that this has been done, we can write
$$
\rho_S = \frac{1}{\dim\sK} \sum_\alpha \dketbra{V_\alpha}{V_\alpha},
$$
and construct the purification of $\rho_S$ in $\tp{\sH}{\sK}\tp{}{\sE}$,
$$
\psi_S \defeq \frac{1}{\sqrt{\dim\sK}} \sum_\alpha
\tp{\dket{V_\alpha}}{\ket{e_\alpha}}.
$$
Let us define the isometry
$\map{V'}{\tp{\sK}{\sK}}{\tp{\sH}{\sK}\tp{}{\sE}}$ through
$$
V'(\tp{\psi}{\phi}) \defeq \sum_\alpha
\tp{(V_\alpha\psi)}{\phi}\tp{}{e_\alpha},
$$
in which case we have $\psi_S = (1/\sqrt{\dim\sK}) V'\dket{\idty}$.
Do the same thing for the channel $T$ to arrive at the purification
$\psi_T = (1/\sqrt{\dim\sK}) W'\dket{\idty}$ of $\rho_T$.  From
Uhlmann's theorem (Theorem \ref{th:uhlmann}) we have
$$
F(\rho_S,\rho_T) = \max_{\psi_S,\psi_T} \abs{\braket{\psi_S}{\psi_T}}^2.
$$
It is easily shown that $\braket{\psi_S}{\psi_T} = (1/\dim\sK)
\dbraket{\idty}{(V'^*W')\idty} \equiv (1/\dim\sK) \tr{V^*W}$.  Hence
the maximization over the purifications $\psi_S$ and $\psi_T$ of
$\rho_S$ and $\rho_T$ is equivalent to the maximization over the
isometries $V$ and $W$, and the theorem is proved.
\end{proof}

Finally we consider the relation of the channel fidelity $\cF$ to the
cb-norm.  Using the properties of the latter, as well as the Fuchs-van de
Graaf theorem (Theorem \ref{th:fvdg}), we easily obtain the inequality
\begin{equation}
2-2 \sqrt{\cF(S,T)} \le \cbnorm{S - T}.
\label{eq:lobound}
\end{equation}
It is certainly an interesting and important problem to derive an
upper bound on $\cbnorm{S-T}$ in terms of $\cF(S,T)$.  We can expect
that this upper bound will not be nearly as tight as the lower bound
because, as we have indicated above, the cb-norm distance is a much
stronger distinguishability criterion than the channel fidelity.

However, in the case when one of the channels is the identity channel,
and the other one is an arbitrary channel
$\map{T}{\cB(\sH)}{\cB(\sH)}$, we actually {\em can} bound the channel
fidelity in terms of the cb-norm both above and below.   For this
purpose we need the {\em off-diagonal fidelity} of the channel
$\map{T}{\cB(\sH)}{\cB(\sH)}$, defined by \cite{wer1}
$$
\cF_\%(T) \defeq \sup_{\psi,\phi \in \sH}
	\rp {\braket {\phi} {T_*(\ketbra{\phi}{\psi})\psi} },
$$
for which we have the inequality
$$
\cbnorm{T-\id} \le 4 \sqrt{1-\cF_\%(T)}.
$$
Then $\cF(T,\id) \le \cF_\%(\tp{T}{\id})$, so that, using the fact
that the cb-norm is multiplicative with respect to tensor products, we
get
\begin{equation}
\cbnorm{T-\id} \le 4\sqrt{1-\cF(T,\id)}.
\label{eq:upbound}
\end{equation}
Combining inequalities (\ref{eq:lobound}) and (\ref{eq:upbound}) yields
\begin{equation}
\left(1-\frac{1}{2}\cbnorm{T-\id}\right)^2 \le \cF(T,\id) \le 1 -
\frac{1}{16}\cbnorm{T-\id}^2.
\label{eq:fcb}
\end{equation}
The upper bound in this inequality is not nearly as tight as the lower
bound.  Indeed, when $\cbnorm{T-\id}$ equals its maximum value of 2,
the fidelity $\cF(T,\id)$ can take any value between 0 and 3/4.  This
serves as yet another indication that the cb-norm is a much more
stringent distinguishability criterion than the channel fidelity
$\cF$.  

%% file: chap3.tex
\chapter{Strictly contractive channels}
\label{ch:scc}

%\vskip 1in

Among the basic postulates laid down by the founding fathers of
statistical physics, there is the so-called {\em zeroth law of
  thermodynamics} \cite[p. 3]{dor}, \cite[p. 18]{sim} which expresses
formally the empirical fact that any large system will normally be
observed in an {\em equilibrium state} characterized by a few
macroscopic parameters, and that any system not in equilibrium will
rapidly approach it.  This process of return to equilibrium is
normally referred to as {\em relaxation} \cite{kry}.  One of the main
parameters of a relaxation process is the {\em relaxation time}
$\tau_{\rm relax}$ --- if we disturb a large system at $t=0$ and then
observe it again at $t \gg \tau_{\rm relax}$, we will find, with high
probability, that the system is in a state arbitrarily close to
equilibrium.

The classic example of a relaxation process is the phenomenon of {\em
  thermalization}, i.e., when a physical system reaches thermal
equilibrium with its surroundings, the latter being maintained at an
absolute temperature $T$.  It is a basic result in statistical
mechanics \cite[p. 153]{str} that the corresponding equilibrium state
is precisely the {\em canonical} (or {\em Gibbs}) state
\begin{equation}
\rho_\beta \defeq \frac{e^{-\beta H}}{Z_\beta},
\label{eq:gibbs}
\end{equation}
where $H$ is the Hamiltonian of the system, $\beta \defeq 1/\kb T$ is
the {\em inverse temperature}, and the normalizing factor $Z_\beta
\defeq \tr{e^{-\beta H}}$ is known as the {\em canonical partition
  function}.  We will discuss the Gibbs state in greater detail in
Ch.~\ref{ch:enterg}, when we talk about the Gibbs variational
principle.  

The ultimate goal of the experimental research into quantum
information processing is the construction of a reliable large-scale
quantum computer.  Such a computer will necessarily be a macroscopic
system subject to the laws of thermodynamics; therefore it makes sense
to deal with such things as approach to equilibrium, and relaxation
processes in general, in the context of quantum information theory.

\section{Relaxation processes and channels}
\label{sec:chrelax}

How can we model a relaxation process using the tools of quantum
information theory?  A good way to do this is by means of a
discrete-time version of a quantum dynamical semigroup
\cite{dav}. Consider a quantum system with the algebra of observables
$\cA$, and let $\map{T}{\cA}{\cA}$ be a channel\footnote{We have made
  the unfortunate choice of denoting a typical channel by the letter
  $T$, the same letter also being used for temperature.  We hope that
  it will always be clear from the context what the letter $T$ stands
  for.\\}. Then the dynamics is given by the semigroup $\set{T^n}_{n
  \in \bbn}$, so that $T$ represents a single step of the dynamics.
Given any initial state $\omega_0 \in \cS(\cA)$, we can track the
evolution of the system under the semigroup dynamics by following the
orbit $\set{\omega_n}$, where $\omega_n \defeq \omega_0 \circ T^n$.
The dynamics defined in this way is {\em stationary} (i.e., the
evolution law is independent of time) and {\em Markovian} (i.e., the
state $\omega_n$ depends only on the initial state $\omega_0$).

A good model of relaxation should satisfy the following natural
requirements. (1) There should exist a unique state $\omega_T \in
\cS(\cA)$ such that $\omega_T \circ T = \omega_T$. (2) For any choice
of the initial state $\omega_0$, the sequence $\set{\omega_n}$ should
converge weakly* to $\omega_T$, in the sense of the following
definition.
\begin{definition}
Let $\sX$ be a Banach space with the dual Banach space $\sX^*$.  Then
the sequence $\set{\omega_n}$ in $\sX^*$ is said to converge to some
$\omega \in \sX^*$ {\em in the weak* sense}, written as $\wslim_{n
  \rightarrow \infty}\omega_n = \omega$, if for any $X \in \sX$ we
have $\lim_{n \rightarrow \infty}\omega_n(X) = \omega(X)$.
\label{def:wstar}
\end{definition}
It can be shown \cite[p. 68]{br} that weak* convergence of a
sequence $\set{\omega_n}$ of normal states implies trace-norm
convergence of the sequence $\set{\rho_n}$ of the corresponding
density operators, and {\em vice versa}.  (3) The convergence of the
orbit $\set{\omega_n}$ to the equilibrium state $\omega_T$ should be
exponential, i.e., there should exist a constant $k$ with $0 < k < 1$
such that, for any $A \in \cA$, $\abs{\omega_n(A)-\omega_T(A)} \le C_A
k^n$, where $C_A$ is a constant depending on $A$.  Our reason for
insisting on this is dictated essentially by the zeroth law.

The first requirement states that the channel $T$ must be {\em
  ergodic}, according to the following definition.

\begin{definition}
Let $\cA$ be an algebra of observables.  A channel $\map{T}{\cA}{\cA}$
is called {\em ergodic} if there exists a unique state $\omega_T \in
\cS(\cA)$ such that $\omega_T \circ T = \omega_T$.
\label{def:ergch}
\end{definition}
Ergodicity of the channel $T$ implies that, for any state $\omega \in
\cS(\cA)$, the {\em ergodic mean} 
\begin{equation}
\bar{\omega}_N \defeq \frac{1}{N+1} \sum^N_{n=0} \omega \circ T^n
\label{eq:ergmean}
\end{equation}
converges weakly* to the unique $T$-invariant state $\omega_T$.  In
terms of the corresponding density operators, we would then have
$$
\lim_{N \rightarrow \infty} \trnorm{ \frac{1}{N+1} \sum^N_{n=0}
  T^n_*(\rho) - \rho_T } = 0,
$$
where $T_*$ is the dual channel corresponding to $T$, and $\rho_T$ is
the density operator corresponding to the unique $T$-invariant state,
i.e., $T_*(\rho_T)=\rho_T$.

To prove the weak* convergence of the ergodic mean (\ref{eq:ergmean})
to $\omega_T$, we first note that, because the state space of a
C*-algebra with identity is weakly* compact \cite[p. 53]{br}, the
sequence $\set{\bar{\omega}_N}$ has a weakly* convergent subsequence.
Furthermore, for any $A \in \cA$,
$$
\abs{(\bar{\omega}_N \circ T)(A) - \bar{\omega}_N(A)} =
\frac{1}{N+1}\abs{(\omega \circ T^{N+1})(A) - \omega(A)} \le \frac{2
  \norm{A}}{N+1},
$$
i.e., any weakly* convergent subsequence of $\set{\bar{\omega}_N}$ has
a $T$-invariant limit.  Because the $T$-invariant state $\omega_T$ is
unique, we see that {\em every} weakly* convergent subsequence of
$\set{\bar{\omega}_N}$ converges to $\omega_T$.  Thus it follows that
$$
\wslim_{N \rightarrow \infty}\frac{1}{N+1}\sum^N_{n=0} \omega \circ
T^n = \omega_T \qquad \forall \omega \in \cS(\cA)
$$
by weak* compactness of $\cS(\cA)$. 

However, mere ergodicity of $T$ is not sufficient; we also demand
weak* convergence of the orbit $\set{\omega \circ T^n}$ for any
$\omega \in \cS(\cA)$.  Thus, for any observable $A \in \cA$, the
sequence of the expectation values $\set{(\omega \circ T^n)(A)}$
should converge to the expectation value $\omega_T(A)$, i.e., the
dynamics should be {\em mixing}.  In terms of the corresponding
density operators, we must have
$$
\lim_{n \rightarrow \infty} \trnorm{T^n_*(\rho) - \rho_T} = 0
%\label{eq:trnorm_do}
$$
for all $\rho$. While mixing implies ergodicity, the converse is not
necessarily true. It turns out, however, that, at least in the case
when the underlying Hilbert space $\sH$ is finite-dimensional, there
is a condition under which ergodicity and mixing are equivalent. This
is the content of the following result of Werner, stated in the
article by Terhal and DiVincenzo \cite{tdv}.

\begin{theorem}
{\bf (Werner)}
Let $\sH$ be a Hilbert space of finite dimension $d$.  Let
$\map{T}{\cB(\sH)}{\cB(\sH)}$ be a channel with the dual channel
$T_*$.  Suppose that the map $T_*$, extended linearly to all of
$\cB(\sH)$, has a unique fixed point $\rho_T \in \cS(\sH)$.  Then
there exist a polynomial $P$ and a constant $k \in (0,1)$ such that,
for any $\rho \in \cS(\sH)$,
\begin{equation}
\trnorm{T^n_*(\rho) - \rho_T} \le C_d P(n)k^n,
\label{eq:tdv}
\end{equation}
where the constant $C_d$ depends on the dimension $d$.  Furthermore,
if we view $T_*$ as a linear operator on $\cB(\sH)$ with the
eigenvalues $\mu_m$, then $k = \max_{m; \mu_m \neq 1} \abs{\mu_m}$.
If $T_*$ is diagonalizable, then the estimate (\ref{eq:tdv}) holds
with $P \equiv 1$.
\label{th:mixerg}
\end{theorem}

The main requirement of Theorem \ref{th:mixerg} is the uniqueness of
the fixed point of $T_*$ in the {\em entire} algebra $\cB(\sH)$, not
just in the state space $\cS(\sH)$.  However, if the only information
we have is that there is a unique $T$-invariant {\em state}, the above
criterion may not apply.  Fortunately, there exist other methods of
proving the mixing property, such as the following theorem
\cite[p. 52]{str}.

\begin{theorem}
{\bf (Liapunov's direct method)}
Let $\sX$ be a separable compact space, and let $\map{\tau}{\sX}{\sX}$
be a continuous map.  Suppose that there exists a {\em strict Liapunov
  function for $\tau$}, i.e., a continuous functional $f$ on $\sX$
such that, for any $x \in \sX$, $(f \circ \tau)(x) > f(x)$ unless
$\tau(x) = x$.  Suppose also that $\tau$ has a unique fixed point
$x_\tau \in \sX$.  Then, for any $x \in \sX$, the sequence
$\set{\tau^n(x)}$ converges to $x_\tau$.
\label{th:liapunov}
\end{theorem}
\begin{rem}
In order for Theorem \ref{th:liapunov} to be applicable, the topology
of $\sX$ must be such that (a) $\sX$ is separable and compact, (b)
$\tau$ is continuous, and (c) $f$ is continuous.  Then the sequence
$\set{\tau^n(x)}$ converges in this topology.
\end{rem}

Finally we come to the last desideratum on our list, namely the
exponential convergence of the sequence $\set{T^n_*(\rho)}$.  When the
algebra of observables is finite-dimensional, Theorem \ref{th:mixerg}
says that this holds whenever $T_*$ has a unique fixed point (which
would necessarily be a density operator \cite{tdv}), and is a
diagonalizable linear operator on $\cB(\sH)$. However, if the only
piece of information we have to go on is the uniqueness of the
$T$-invariant {\em state}, then Theorem \ref{th:mixerg} will not
apply.  We can only say that, in general, the exact convergence rate
of the orbit $\set{T^n_*(\rho)}$ will depend on the spectrum of $T_*$.

Explicit models of relaxation processes were constructed using the
tools of quantum information theory in the articles of Scarani \etal
\cite{szsgb} and Ziman \etal \cite{zsbhsg}. Also, an interesting paper
by Terhal and DiVincenzo \cite{tdv} investigates the possibility of
using quantum computers to {\em simulate} relaxation processes. In
this chapter we describe another approach to this problem, via the
so-called {\em strictly contractive channels}. Our exposition closely
follows Ref.~\cite{rag2}.

Before we go on, we make one important comment concerning notation.  For
the most part of our discussion in this chapter, as well as in the
next chapter, we will deal with transformations of {\em states}
(i.e., the Schr\"odinger picture).  Therefore, if we are given a
system with the Hilbert space $\sH$, we will use the term \qt{channel}
to refer to any completely positive trace-preserving linear map
$\map{T}{\cS(\sH)}{\cS(\sH)}$, and we will also omit the asterisk
subscript in order to avoid cluttered equations.  On those rare
occasions when we do talk about the Heisenberg picture, the
corresponding map will be denoted by $\hT$.

\section{Strictly contractive channels}
\label{sec:scc}
\subsection{Definition}
\label{ssec:sccdef}

Recall Theorem \ref{th:trnorminv}, which states that, for any channel
$T$ and any two density operators $\rho,\sigma$ in its domain,
$$
\trnorm{T(\rho) - T(\sigma)} \le \trnorm{\rho-\sigma}.
$$
In other words, any channel is a {\em contraction} in the trace norm
on the set of density operators.  Now consider the following
definition.

\begin{figure}
\includegraphics{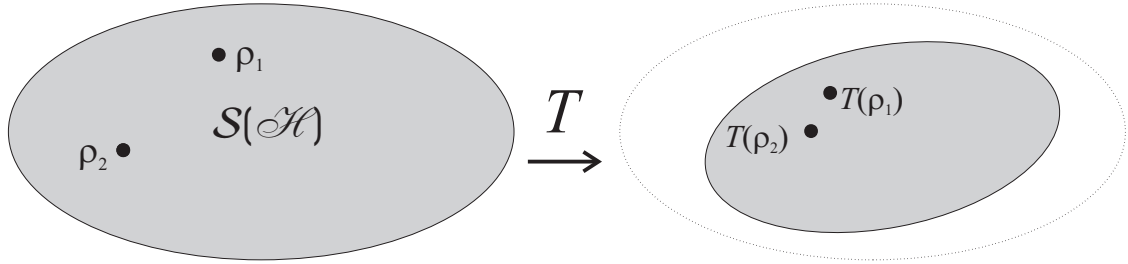}

\caption{The effect of a strictly contractive channel $T$ on the state
  space $\cS(\sH)$ of the quantum system.}
\label{fig:scc}
\end{figure}

\begin{definition}
A channel $\map{T}{\cS(\sH)}{\cS(\sH)}$ is called {\em strictly
  contractive} if there exists a constant $k \in [0,1)$, called the
  {\em contractivity modulus}, such that
\begin{equation}
\trnorm{T(\rho) - T(\sigma)} \le k \trnorm{\rho - \sigma}
\label{eq:scontract}
\end{equation}
for any pair of density operators $\rho,\sigma \in \cS(\sH)$.
\end{definition}
As shown in Fig.~\ref{fig:scc}, the action of a strictly contractive
channel on the set $\cS(\sH)$ can be visualized as a uniform shrinking
of the trace-norm distance between any two density operators $\rho,\sigma$.

It is easily seen that any strictly contractive channel satisfies our
requirements for a relaxation dynamics. First of all, the set
$\cS(\sH)$ is a closed subset of the Banach space $\cT_1(\sH)$ of the
trace-class operators on $\sH$.  Then the contraction mapping
principle (cf. Section \ref{sec:fpt}) tells us that there exists a
{\em unique} density operator $\rho_T \in \cS(\sH)$ such that
$T(\rho_T) = \rho_T$.  Furthermore, given any pair $\rho_0,\sigma_0
\in \cS(\sH)$, consider the orbits $\set{T^n(\rho_0)}$ and
$\set{T^n(\sigma_0)}$.  Strict contractivity shows that these orbits
get exponentially close to each other with $n$ because
$$
\trnorm{T^n(\rho_0) - T^n(\sigma_0)} \le k^n\trnorm{\rho_0 - \sigma_0}.
$$
Furthermore, each orbit converges to $\rho_T$ as $n \rightarrow
\infty$.  Thus the image of $\cS(\sH)$ under the iterates $T^n$
shrinks to a point (namely, $\rho_T$) exponentially fast.  These
features naturally lead us toward mixing, and hence ergodicity,
because the trace-norm convergence of the orbit $\set{T^n(\rho)}$
implies the convergence of the expectation values $\tr{[A T^n(\rho)]}$
to $\tr{(A \rho_T)}$ for any $A \in \cB(\sH)$.  

Another feature of strictly contractive channels is that they render
the states of the system less distinguishable in the sense of quantum
detection theory (see Section \ref{ssec:qdet}).  To see this, we
observe that no two density operators in the image of $\cS(\sH)$ under
a strictly contractive channel can be farther than $2k$ from one
another in terms of the trace-norm distance.  This puts an upper bound
on the optimum probability of correct discrimination between any two
equiprobable density operators in the image of a strictly contractive
channel in a binary quantum detection scheme, namely
$$
\bar{P}_c \le \frac{1+k}{2}.
$$
Thus there is always a nonzero probability of making an error, which
satisfies the bound
$$
\bar{P}_e \ge \frac{1-k}{2}.
$$
In any realistic setting, hardly any event occurs with probability exactly
equal to unity.  For instance, we can never prepare a pure state
$\ketbra{\psi}{\psi}$, but rather a mixture
$(1-\epsilon)\ketbra{\psi}{\psi} + \epsilon \rho$, where both
$\epsilon$ and $\rho$ depend on the particulars of the preparation
procedure.  Similarly, the measuring device that would ideally
identify $\ketbra{\psi}{\psi}$ perfectly will instead be realized
by $(1-\delta)\ketbra{\psi}{\psi} + \delta F$, where $\delta$ and
the operator $F, 0 \le F \le \idty$, are again determined by
practice.  If we assume that, in any physically realizable
quantum computer, all state preparation, manipulation, and registration
procedures can be carried out with finite precision, then it is
reasonable to expect that there exist strict bounds on all
probabilities that figure in the description of the computer's
operation.

As we argued in Chapter \ref{ch:intro}, any imprecision in a nonideal
quantum computer can be traced back to our inability to distinguish
between quantum states beyond a certain resolution threshold.  In
other words, in any experimental situation there will always be some
small $\epsilon_0$ such that any two states with $\trnorm{\rho -
  \sigma} < \epsilon_0$ must be considered {\em practically}
indistinguishable.  It follows from the discussion above that strictly
contractive channels capture this intuition mathematically.
Furthermore, as we will see later, {\em any} channel can be
approximated arbitrarily closely in cb-norm by a strictly contractive
channel.  This means that, for any channel $T$ and any $\epsilon > 0$,
there exists some strictly contractive channel $T'$ such that
$\cbnorm{T-T'} < \epsilon$.  Then, using Eq.~(\ref{eq:lobound}), we
see that $\cF(T,T') > (1-\epsilon/2)^2 \simeq 1-\epsilon$ for
$\epsilon$ sufficiently small.  This, of course, means that the
channels $T$ and $T'$ cannot be distinguished by any experimental
procedure whenever $\epsilon$ is less than the threshold resolution
$\epsilon_0$.

We note that strictly contractive channels have been mentioned in the
text of Nielsen and Chuang \cite{nc}, but none of their properties,
apart from the uniqueness of the fixed point, were described.

\subsection{Examples}
\label{ssec:sccex}

In this section we give a few examples of strictly contractive
channels.  All of these channels have been extensively studied by the
researchers in the field of quantum information theory.\\

\begin{example}
({\bf degenerate channel}) Let $\rho$ be a density operator on $\sH$,
  and consider the map $\amap{K_\rho}{X}{(\tr{X})\rho}$.  This is a
  completely positive trace-preserving map of $\cT_1(\sH)$ into itself,
  and its restriction to $\cS(\sH)$ is the channel that maps any
  density operator $\sigma \in \cS(\sH)$ to $\rho$. It is easy to see
  that $\rho$ is the only fixed point of $K_\rho$, and that $K_\rho$
  is strictly contractive with $k=0$. Channels of this form are called
  {\em degenerate} (in the terminology of Davies \cite{dav}).
\label{ex:degench}\endex

\begin{example}
({\bf depolarizing channel}) Let $\sH$ be a Hilbert space of finite
  dimension $d$.  For any $p \in (0,1]$, define the map
$$
D_p \defeq (1-p) \id + p K_{\idty/d}.
$$
The restriction of $D_p$ to $\cS(\sH)$ is the so-called {\em
  depolarizing channel}
$$
D_p(\rho) = (1-p) \rho + p \frac{\idty}{d}.
$$
For any pair $\rho,\sigma \in \cS(\sH)$, we have
$$
\trnorm{D_p(\rho) - D_p(\sigma)} = (1-p) \trnorm{\rho - \sigma},
$$
which shows that $D_p$ is strictly contractive with $k = 1-p$.  The
unique $T$-invariant density operator is the maximally mixed state
$\idty/d$.  Channels that preserve the maximally mixed state are
called {\em bistochastic}.
\label{ex:depolarch}\endex

\begin{example}
({\bf two-Pauli channel})  Consider the channel on $\cS(\bbc^2)$
  with the Kraus operators
$$
V_1 = \sqrt{p}\idty,\quad V_2 = \sqrt{(1-p)/2}\paulix,\quad V_3 = -i
\sqrt{(1-p)/2}\pauliy.
$$
This channel is bistochastic and strictly contractive with $k =
\max{\set{p,2p-1}}$.
\label{ex:bfs}\endex

\begin{example}
({\bf amplitude damping})
The channel on $\cS(\bbc^2)$ with the Kraus operators
$$
V_1 = \left(
	\begin{array}{cc}
	1 & 0 \\
	0 & \sqrt{1-\gamma}
	\end{array} \right),\qquad
V_2 = \left(
	\begin{array}{cc}
	0 & \sqrt{\gamma} \\
	0 & 0
	\end{array} \right)
$$
is strictly contractive with $k = \sqrt{1-\gamma}$.  Its unique fixed
point is the pure state $\ketbra{\psi_+}{\psi_+}$ with $\pauliz \psi_+
= \psi_+$.
\label{ex:ad}\endex

\begin{example}
({\bf thermalization of a qubit}) Consider the Hamiltonian $H =
  E\pauliz$ with $E > 0$, and the corresponding Gibbs state
$$
\rho_\beta = \frac{1}{2 \cosh{\beta E}} \left(
	\begin{array}{cc}
	e^{-\beta E} & 0 \\
	0 & e^{\beta E}
	\end{array} \right).
$$
Let $p \defeq \exp{(-\beta E)}/(2 \cosh{\beta E})$. Consider the map
$\map{T}{\cS(\bbc^2)}{\cS(\bbc^2)}$ given in the Kraus form $T(\rho) =
\sum^4_{n=1}V_n \rho V^*_n$ with
\begin{eqnarray}
V_1 &=& \sqrt{p} \left(
	\begin{array}{cc}
	1 & 0 \\
	0 & \sqrt{1-\gamma} 
	\end{array}
	\right), \qquad
V_2 = \sqrt{p} \left(
	\begin{array}{cc}
	0 & \sqrt{\gamma} \\
	0 & 0
	\end{array}
	\right) \nonumber \\
V_3 &=& \sqrt{1-p} \left(
	\begin{array}{cc}
	\sqrt{1-\gamma} & 0 \\
	0 & 1 
	\end{array}
	\right), \qquad
V_4 = \sqrt{1-p} \left(
	\begin{array}{cc}
	0 & 0 \\
	\sqrt{\gamma} & 0
	\end{array}
	\right), \nonumber
\end{eqnarray}
where $\gamma$ is a constant between 0 and 1.  Then we have
$T(\rho_\beta) = \rho_\beta$, and a straightforward calculation shows
that $T$ is strictly contractive with $k = \sqrt{1-\gamma}$.  The
constant $\gamma$ can be given a direct physical interpretation.  Let
us write $\gamma = 1 - e^{-2/\Delta}$.  Then for $n \ge \Delta$ we
will have $\trnorm{T^n(\rho)-\rho_\beta} \le e^{-1} \trnorm{\rho -
  \rho_\beta}$.
\label{ex:gad}\endex

\subsection{Strictly contractive channels on $\cS(\bbc^2)$}
\label{ssec:sccqubit}

Consider a channel $\map{T}{\cS(\bbc^n)}{\cS(\bbc^n)}$.  Because the
trace class $\cT_1(\bbc^n)$ is the linear span of $\cS(\bbc^n)$, and
because we also have $\cT_1(\bbc^n) = \cB(\bbc^n) \equiv \cM_n$,
the map $T$ can be uniquely extended to all of $\cM_n$.  We can
naturally identify the space $\cM_n$ of $n \times n$ complex matrices
with $\bbc^{n^2}$.  Thus any linear map of $\cM_n$ to itself can be
naturally regarded as an $n^2 \times n^2$ complex matrix.

Under this identification, it is possible to parametrize all
completely positive maps on $\cM_n$ (see, e.g., the article of
Fujiwara and Algoet \cite{fa}), but the analysis turns out to be quite
involved already in the case of $\cM_2$ (cf. King and Ruskai \cite{kr}
or Ruskai, Szarek, and Werner \cite{rsw}).  In this section we show
how the contraction properties of a channel
$\map{T}{\cS(\bbc^2)}{\cS(\bbc^2)}$ can be read off directly from its
matrix representation.

First of all, recall that any $2 \times 2$ complex matrix $M$ can be
written as a linear combination of the identity matrix and the Pauli
matrices
$$
\paulix = \left( \begin{array}{cc}
	0 & 1 \\
	1 & 0 \end{array} \right), \quad
\pauliy = \left( \begin{array}{cc}
	0 & -i \\
	i & 0 \end{array} \right), \quad
\pauliz = \left( \begin{array}{cc}
	1 & 0 \\
	0 & - 1 \end{array} \right).
$$
In particular, if $M$ is Hermitian, then the coefficients in this
expansion will be real.  The set $\set{\idty/\sqrt{2}, \paulix/\sqrt{2}, \pauliy/\sqrt{2},
  \pauliz/\sqrt{2}}$ forms an orthonormal basis of $\cM_2$, when the latter is
viewed as a Hilbert space with the Hilbert-Schmidt inner product
$\inp{A}{B} \defeq \tr{(A^*B)}$. We will refer to this basis as the
{\em Pauli basis}. The upshot is that we can represent any matrix
$$
M = m_0 \idty + m_1 \paulix + m_2 \pauliy + m_3 \pauliz
$$
as a vector in $\bbc^4$ with the components $m_i, i = 0,1,2,3$.
Furthermore, we have $m_0 = \tr{M}/2$ and $m_i = \tr{(M\sigma_i/2)}$.

Now it is easy to show that any density matrix $\rho \in \cS(\bbc^2)$
can be written as
$$
\rho = \frac{1}{2}(\idty + r_1 \paulix + r_2 \pauliy + r_3 \pauliz)
\equiv \frac{1}{2}(\idty + \mathbf{r} \cdot \bbsigma),
$$
where $r_i \in \bbr$ and $r^2_1 + r^2_2 + r^2_3 \le 1$ with equality
if and only if $\rho$ is a pure state.  Thus there is a one-to-one
correspondence between the density matrices in $\cM_2$ and the points
in the closed unit ball in $\bbr^3$.  Under this identification, this
ball is known as the {\em Bloch-Poincar\'e ball}.  Given $\rho \in
\cS(\bbc^2)$, we will refer to the corresponding vector $\mathbf{r}
\in \bbr^3$ as the {\em Bloch-Poincar\'e vector} of $\rho$.

Our characterization of strictly contractive channels on $\cS(\bbc^2)$
hinges on the following important theorem \cite{kr}.
\begin{theorem}
{\bf (King-Ruskai)} Let $T$ be a channel on $\cS(\bbc^2)$.  Then there
exist unitaries $U,V$ and vectors $\mathbf{t},\mathbf{v} \in \bbr^3$
such that
\begin{equation}
T(\rho) = U T_{\mathbf{t},\mathbf{v}}(V\rho V^*)U^*,
\label{eq:kr}
\end{equation}
where the channel $T_{\mathbf{t},\mathbf{v}}$ is defined by
\begin{equation}
T_{\mathbf{t},\mathbf{v}} \left(w_0 \idty + \sum^3_{i=1}w_i
\sigma_i\right) = w_0 \idty + \sum^3_{i=1} (w_0 t_i + v_i w_i)
\sigma_i.
\label{eq:tvt}
\end{equation}
\label{th:kr}
\end{theorem}

\begin{proof}
First observe that, with respect to the Pauli basis, any
trace-preserving map $\map{T}{\cM_2}{\cM_2}$ can be written in block
form as
\begin{equation}
T = \left(
\begin{array}{c|c}
1 & \mathbf{0} \\
\hline
\mathbf{u} &  \tT  \\
\end{array} \right),
\label{eq:channelblock}
\end{equation}
where $\mathbf{u} \in \bbc^3$, and $\tT$ is a $3 \times 3$
complex matrix. Furthermore, if $T$ is a positive map, then
$\mathbf{u} \in \bbr^3$ and $\tT$ is a real matrix; this follows from
the fact that any positive map on a C*-algebra $\cA$ leaves invariant
the set of self-adjoint elements of $\cA$ \cite{sto}.

Any $n\times n$ matrix $A$ can be written in the form $A = VDW^*$,
where $V$ and $W$ are unitary, and $D$ is a positive semidefinite
diagonal matrix.  This is referred to as the {\em singular value
  decomposition} of $A$ \cite{bha}.  If $A$ is real, the matrices $V$
and $W$ can be chosen real orthogonal.  Write down the singular value
decomposition $\tT = V D \trn{W}$, where $V, W \in O(2)$.  Any matrix
in $O(2)$ is a rotation (modulo sign), so we can write
$$
\tT = R_1\tD\trn{R_2},
$$
where $R_1$ and $R_2$ are rotations, and the sign has been absorbed
into the matrix $\tD$. We can then decompose $T$ as follows:
\begin{equation}
T = \left(
\begin{array}{c|c}
1 & \mathbf{0} \\
\hline
\mathbf{0} &  R_1  \\
\end{array} \right) \left(
\begin{array}{c|c}
1 & \mathbf{0} \\
\hline
\trn{R_1}\mathbf{u} &  \tD  \\
\end{array} \right)
\left(
\begin{array}{c|c}
1 & \mathbf{0} \\
\hline
\mathbf{0} &  \trn{R_2}  \\
\end{array} \right).
\label{eq:weirddecomp}
\end{equation}
Now put $\mathbf{t} \defeq \trn{R_1}\mathbf{u}$, and let $\mathbf{v}$
be the vector whose components are the diagonal entries of $\tD$.  The
middle matrix on the right-hand side of Eq.~(\ref{eq:weirddecomp}) is
precisely the matrix representation, with respect to the Pauli basis,
of the map $T_{\mathbf{v},\mathbf{t}}$ defined in Eq.~(\ref{eq:tvt}),
while the first and last matrices correspond to unitary conjugations
in $\cM_2$.  This proves the theorem.
\end{proof}

Using Theorem \ref{th:kr}, we can read off the contraction properties
of $T$ from the channel $T_{\mathbf{v},\mathbf{t}}$ (the unitary
conjugations $U \cdot U^*$ and $V \cdot V^*$ are irrelevant for this
purpose because of unitary invariance of the trace norm).  Thus
consider two density operators $\rho,\rho'$ with the Bloch-Poincar\'e
vectors $\mathbf{r},\mathbf{r}'$.  Letting $\mathbf{\Delta} \defeq
\mathbf{r} - \mathbf{r}'$, we have
\begin{eqnarray}
\trnorm{T(\rho)-T(\rho')} &=& \trnorm{T_{\mathbf{v},\mathbf{t}}(\rho -
  \rho')}\nonumber \\
& =& \frac{1}{2}\trnorm{\sum^3_{i=1} v_i \Delta_i \sigma_i } \nonumber \\
&\le &  \frac{1}{2} \left(\max_i \abs{v_i} \right)
  \trnorm{\sum^3_{i=1} \Delta_i \sigma_i} \nonumber \\
&\equiv & \left(\max_i \abs{v_i} \right) \trnorm{\rho - \rho'}. 
\label{eq:qubitcontract}
\end{eqnarray}
Clearly, the upper bound in Eq.~(\ref{eq:qubitcontract}) is achieved
whenever the only nonzero component of $\mathbf{\Delta}$ corresponds
to the direction in which $\abs{v_i}$ is largest.  Hence, if $T$ is
strictly contractive, we have $k = \max_i \abs{v_i}$.  Writing $T$ in
the form (\ref{eq:channelblock}), we see that $k$ is also the largest
singular value of the matrix $\tT$, i.e., $k = \norm{\tT}$.

Given two density operators $\rho,\rho' \in \cM_2$, let $\mathbf{r}$
and $\mathbf{r}'$ be their Bloch-Poincar\'e vectors. Then we have the
following useful observation. The trace-norm distance $\trnorm{\rho -
  \rho'}$ can be expressed geometrically in terms of the Euclidean
distance between $\mathbf{r}$ and $\mathbf{r}'$ as
$$
\trnorm{\rho-\rho'} = \sqrt{ \Delta^2_1 + \Delta^2_2 + \Delta^2_3},
$$
where $\mathbf{\Delta}$ is defined as before.  The proof is easy once
we note that the matrix $\rho - \rho'$ has eigenvalues $\pm (1/2)
\sqrt{\Delta^2_1 + \Delta^2_2 + \Delta^2_3}$.  Thus the action of any
channel $T$ on $\cS(\bbc^2)$ can be visualized, modulo a rotation and
a translation, as the shrinking of the Bloch-Poincar\'e ball into an
ellipsoid, and the contractivity modulus of $T$ is precisely the
half-length of the longest symmetry axis of this ellipsoid.  In this
sense, the term \qt{strictly contractive} becomes especially apt.  For
instance, the action of the depolarizing channel $D_p$ on
$\cS(\bbc^2)$ is tantamount to the rescaling of the Bloch-Poincar\'e
ball by the factor of $1-p$.

Tensor products of strictly contractive channels do not lend
themselves as easily to an intuitive geometric interpretation, apart
from some special cases.  In particular, when $T$ and $T'$ are
bistochastic strictly contractive channels on $\cS(\bbc^2)$, with the
respective contractivity moduli $k$ and $k'$, it can be shown that the
product channel $\tp{T}{T'}$ is also strictly contractive with the
contractivity modulus $\max{\set{k,k'}}$.  To see this, we first note
that any density matrix in $\cM_4$ can be written as \cite[Ch. 2]{mw}
$$
\rho = \frac{1}{4} \left( \idty + \sum^3_{i=1} r_i \tp{\sigma_i}{\idty
  + \sum^3_{i=1} s_i \tp{\idty}{\sigma_i} + \sum^3_{i,j=1} \theta_{ij}
  \tp{\sigma_i}{\sigma_j}} \right),
$$
where the vectors $\mathbf{r},\mathbf{s} \in \bbr^3$ are referred to
as the {\em coherence vectors} of the first and second qubit
respectively, and the $3 \times 3$ real matrix $\Theta$ is called the
{\em correlation tensor} of $\rho$. Hence each density operator $\rho
\in \cM_4$ can be uniquely described by the ordered triple
$(\mathbf{r},\mathbf{s},\Theta)$, so we will write $\rho \sim
(\mathbf{r},\mathbf{s},\Theta)$.  The contraction properties of the
channel $\tp{T}{T'}$ can be read off from the corresponding channel
$\tp{T_{\mathbf{t},\mathbf{v}}}{T_{\mathbf{t}',\mathbf{v}'}}$.
Consider two density operators $\rho \sim
(\mathbf{r},\mathbf{s},\Theta)$ and $\rho \sim
(\mathbf{r}',\mathbf{s}',\Theta')$ and define $\mathbf{\Gamma} \defeq
\mathbf{r}-\mathbf{r}'$, $\mathbf{\Delta} \defeq
\mathbf{s}-\mathbf{s}'$, and $\Xi \defeq \Theta - \Theta'$. We then
have
\begin{eqnarray}
\| (T &\otimes & T') (\rho - \rho') \|_1 \nonumber \\
&=& \| (T_{\mathbf{t},\mathbf{v}} \otimes T_{\mathbf{t}',\mathbf{v}'}
)(\rho - \rho') \|_1 \nonumber \\
&=& \frac{1}{4}\trnorm{ \sum^3_{i=1}v_i \Gamma_i \tp{\sigma_i}{\idty}
  + \sum^3_{i=1} v'_i \Delta_i \tp{\idty}{\sigma_i} + \sum^3_{i,j=1}
  v_i v'_j \Xi_{ij} \tp{\sigma_i}{\sigma_j}} \nonumber \\
&\le & \frac{1}{4} \max {\left\{ \abs{v_1},\ldots,\abs{v_3},\atop
  \abs{v'_1},\ldots,\abs{v'_3}\right\} } \trnorm{\sum^3_{i=1}\Gamma_i
  \tp{\sigma_i}{\idty} + \sum^3_{i=1} \Delta_i \tp{\idty}{\sigma_i} +
  \sum^3_{i,j=1}\Xi_{ij} \tp{\sigma_i}{\sigma_j}} \nonumber \\
&\equiv & \max{\set{k,k'}} \trnorm{\rho - \rho'}, 
\label{eq:2qubitcontract}
\end{eqnarray}
where we have used the fact that $\abs{v_i},\abs{v'_i} < 1$ for all
$i$ because $T$ and $T'$ are strictly contractive. Again, the bound
(\ref{eq:2qubitcontract}) can be achieved by choosing $\rho$ and
$\rho'$ suitably, so we conclude that $\tp{T}{T'}$ is strictly
contractive, and that its contractivity modulus equals the greater of
$k$ and $k'$.

\subsection{The density theorem for strictly contractive channels}
\label{ssec:sccdense}

Suppose we are presented with some quantum system in an unknown state
$\rho$, and we are trying to determine this state.  Any physically
realizable apparatus will have finite resolution $\epsilon$, so that
all states $\rho'$ with $\trnorm{\rho-\rho'} < \epsilon$ are
considered indistinguishable from $\rho$.  Now, if $\sH$ is the
Hilbert space associated with the system, and if $\Sigma$ is a dense
subset of $\cS(\sH)$, then, by definition of a dense subset, for
any $\epsilon > 0$ and any $\rho \in \cS(\sH)$, there will always
be some $\sigma \in \Sigma$ such that $\trnorm{\rho-\sigma} <
\epsilon$.

The same reasoning also applies to distinguishability of quantum
channels, except now the appropriate measure of closeness is
furnished by the cb-norm. Thus, if an experiment utilizes some
apparatus with resolution $\epsilon$, then any two channels $T,S$
with $\cbnorm{T-S} < \epsilon$ are considered indistinguishable
from each other.  There is, however, no fundamental difference
between distinguishability of states and distinguishability of
channels because any experiment purporting to distinguish between two
given channels $T$ and $S$ consists in preparing the apparatus in some
state $\rho$ and then making some measurements that would tell the
states $T(\rho)$ and $S(\rho)$ apart from each other.  Then, since for
any state $\rho$, $\trnorm{T(\rho)-S(\rho)} \le \cbnorm{T-S}$, the
resolving power of the apparatus that will distinguish between $T$ and
$S$ is limited by the resolving power of the apparatus that will
distinguish between $T(\rho)$ and $S(\rho)$.

The main result of this section is summarized in the following theorem
\cite{rag2}.
\begin{theorem} Let $C(\sH)$ be the set of all
channels on $\cS(\sH)$.  Then the set $C_{\rm
sc}(\sH)$ of all strictly contractive channels on $\cS(\sH)$ is a
$\cbnorm{\cdot}$-dense convex subset of $C(\sH)$.
\label{th:densityscc}
\end{theorem}

\begin{proof}  We show convexity first.  Suppose $T_1,T_2
\in C_{\rm sc}(\sH)$.  Define the channel $S \defeq \lambda T_1 +
(1-\lambda)T_2$, $0 < \lambda < 1$.  Then, for any $\rho,\sigma
\in \cS(\sH)$, we have the estimate
\begin{eqnarray}
\trnorm{S(\rho)-S(\sigma)} & \le &  \lambda \trnorm{T_1(\rho)-T_1(\sigma)}
+
(1-\lambda) \trnorm{T_2(\rho)-T_2(\sigma)} \nonumber \\
& \le & \left[ \lambda k(T_1) + (1-\lambda)k(T_2)\right]
\trnorm{\rho-\sigma}, \nonumber
\end{eqnarray}
where $k(T_i)$ is the contractivity modulus of $T_i$, $i \in
\set{1,2}$. Defining $k \defeq \max{\set{k(T_1),k(T_2)}}$, we get
$$
\trnorm{S(\rho)-S(\sigma)} \le k\trnorm{\rho-\sigma}.
$$
Since $T_1,T_2$ are strictly contractive, $k < 1$, and
therefore $S \in C_{\rm sc}(\sH)$. To prove density, let us fix
some $\sigma \in \cS(\sH)$.  Given $\epsilon > 0$, pick some
positive $n$ such that $1/n < \epsilon$.  For any $T \in C(\sH)$,
define
$$
T_n \defeq \frac{1}{2n}K_\sigma + \left(1-\frac{1}{2n}\right)T.
$$
Clearly, $T_n \in C_{\rm sc}(\sH)$, and the estimate
$$
\cbnorm{T-T_n} = \frac{1}{2n}\cbnorm{T-K_\sigma} \le \frac{1}{n} <
\epsilon
$$
finishes the proof.
\end{proof}

This theorem indicates that, as far as physically realizable
(finite-precision) measurements go, there is no way to distinguish
a given channel $T$ from some strictly contractive $T'$ with
$\cbnorm{T-T'} < \epsilon$, where $\epsilon$ is the resolution of
the measuring apparatus.  We can also consider the channel fidelity as
the measure of distinguishability, in which case we have the following
corollary.

\begin{corollary}
For any channel $\map{T}{\cS(\sH)}{\cS(\sH)}$ and any $\epsilon > 0$,
there exists a strictly contractive channel
$\map{T'}{\cS(\sH)}{\cS(\sH)}$ such that $\cF(T,T') > 1 - \epsilon$.
\end{corollary}

\begin{proof}
Given $\epsilon$, Theorem \ref{th:densityscc} says that there exists a
strictly contractive channel $T'$ with $\cbnorm{T - T'} <
2(1-\sqrt{1-\epsilon})$.  Then Eq.~(\ref{eq:lobound}) implies that
$\cF(T,T') > 1 - \epsilon$.
\end{proof}

We also mention that any channel $T$ with $\cbnorm{T-\id} < \epsilon$
(for some sufficiently small $\epsilon > 0$) cannot be distinguished
from a depolarizing channel. Indeed, it suffices to pick some
$$
n > \frac{ \cbnorm{K_{\idty/d}-\id} } {\epsilon - \cbnorm{T-\id} },
$$
where $d = \dim \sH$, so that
$$
\cbnorm{T - D_{1/n}} \le \cbnorm{T-\id} + (1/n)\cbnorm{K_{\idty/d}-\id} <
\epsilon.
$$

We note that the channel formed by taking a convex combination of any
channel with a strictly contractive channel is a strictly contractive
channel.  Let $T \in C$ be an arbitrary channel, and suppose that $T'
\in C_{\rm sc}$ [from now on, we will not mention the Hilbert space $\sH$ when
talking about channels on $\cS(\sH)$, unless this omission might
cause ambiguity]. Define, for some $0 < \lambda <1$, the channel
$S \defeq \lambda T + (1-\lambda)T'$.  Then
\begin{eqnarray}
\trnorm{S(\rho) - S(\sigma)} & \le & \lambda \trnorm{T(\rho)-T(\sigma)} +
(1-\lambda)
\trnorm{T'(\rho) - T'(\sigma)} \nonumber \\
& \le & \left[\lambda + (1-\lambda)k(T')\right]
\trnorm{\rho-\sigma}. \nonumber
\end{eqnarray}
Since $\lambda + (1-\lambda)k(T') < 1$, we conclude that $S
\in C_{\rm sc}$.

Finally, we mention that a method similar to that in the proof of
Theorem~\ref{th:densityscc} can be used to show that the
set $C_{\rm sc}^\idty$ of all bistochastic strictly contractive
channels is a dense convex subset of the set $C^\idty$ of all bistochastic
channels.

\section{Strictly contractive dynamics of quantum registers and computers}
\label{sec:noisyqmem}

So far we have established two important properties of strictly
contractive channels.  Firstly, any channel $T$ can be
approximated arbitrarily closely by a strictly contractive channel
$T'$, i.e., for any $\epsilon > 0$, we can find a strictly contractive
channel $T'$ such that $\cF(T,T') > 1 - \epsilon$. Secondly, any
quantum decision strategy that would, in principle, distinguish some
pair $\rho,\rho'$ of density operators with certainty, will fail with
probability at least $[1-k(T)]/2$ in the presence of a
strictly contractive error channel $T$. The latter statement can
also be phrased as follows:  no two density operators in the image
$T\cS(\sH)$ of $\cS(\sH)$ under some $T \in C_{\rm sc}$ have
orthogonal supports; furthermore, the trace-norm distance between
any two density operators in $T\cS(\sH)$ is bounded from above by
$2k$.

In this section we obtain dimension-independent estimates on
decoherence rates of quantum memories and computers under the
influence of strictly contractive noise and without any error
correction (the possibility of error correction will be addressed
in the next two sections).

We treat quantum memories (registers) first.  Suppose that we want
to store a state $\rho_0 \in \cS(\sH)$ for time $t$ in the
presence of errors modeled by some strictly contractive channel
$T$.  Let $\tau$ be the decoherence timescale, with $\tau \ll t$,
and let $n = \lceil t/\tau \rceil$. The final state of the
register is then $\rho_n = T^n(\rho_0)$. If $\rho_T$ is the unique
fixed point of $T$, then
$$
\trnorm{\rho_n - \rho_T} = \trnorm{T^n(\rho_0) - T^n(\rho_T)} \le
k(T)^n \trnorm{\rho_0 - \rho_T}.
$$
In other words, the state $\rho_0$, stored in a quantum register
in the presence of strictly contractive noise $T$, evolves to the
unique $T$-invariant state $\rho_T$, and the convergence is incredibly
rapid.  For the sake of concreteness let us consider a numerical
example.  Suppose that $k(T) = 0.9$, and that initially the states
$\rho_0$ and $\rho_T$ have orthogonal supports, so $\trnorm{\rho_0 -
  \rho_T} = 2$.  Then, after $n=10$ iterations (i.e., $t=10\tau$), we
have $\trnorm{\rho_n - \rho_T} \le 0.697$, and the probability of
correct discrimination between $\rho_n$ and $\rho_T$ is only 0.674.
Note that the decoherence rate estimate $$
r(n;\rho_0,T) \defeq \frac{\trnorm{\rho_n -
\rho_T}}{\trnorm{\rho_0-\rho_T}} \le k(T)^n \label{eq:decmem}
$$
does not depend on the dimension of $\sH$, but only on the
contractivity modulus $k(T)$ and on the relative storage
duration $n$.  In other words, quantum registers of {\em any} size
are equally sensitive to strictly contractive errors with the same
contractivity modulus.

Obtaining estimates on decoherence rates of computers is not so
simple because, in general, the sequence $\set{\rho_n}$, where
$\rho_n$ is the overall state of the computer after $n$
computational steps, does not have to be convergent.  Let us first
fix the model of a quantum computer.  We define an
{\em ideal quantum circuit of size $n$} to be an ordered $n$-tuple
of unitaries $U_i$, where each $U_i$ is a tensor product of
elements of some set $\cG$ of universal gates \cite{bar}, which must
be a dense subgroup of the group $\cU(\sH)$ of all unitary operators
on $\sH$.  For some error channel $T$, a {\em $T$-noisy quantum
  circuit of size $n$ with $r$ error locations} is an ordered
$(n+r)$-tuple containing $n$ channels $\hat{U}_i \defeq U_i \cdot
U^*_i$, where the unitaries $U_i$ are of the form described above, as
well as $r$ instances of $T$.  We will assume, for simplicity, that
each $T$ is preceded and followed by some $\hU_i$ and
$\hU_{i+1}$. Based on this definition, the \qt{noisiest} computer for
fixed $T$ and $n$ is represented by a $T$-noisy quantum circuit of
size $n$ with $n$ error locations, i.e., by a $2n$-tuple of the form
$(\hat{U}_1,T,\hat{U}_2,T,\ldots,\hat{U}_n,T)$.  If the initial state
of the computer is $\rho_0$, then we will use the notation
\begin{equation}
\rho_n = \left(\prod^n_{i=1}T\hat{U}_i\right)(\rho_0)
\label{eq:ncomp}
\end{equation}
to signify the state of the computer after $n$ computational
steps. In the above expression, the product sign should be
understood in the sense of composition $T \circ \hat{U}_n \circ
\ldots \circ T \circ \hat{U}_1$.

Given an arbitrary sequence of computational steps, the sequence
$\set{\rho_n}$, defined by Eq.~(\ref{eq:ncomp}) (assuming that $n$
is suficiently large, i.e., the computation is sufficiently long)
need not be convergent.  However, if the channel $T$ is strictly
contractive, then for any $\epsilon > 0$ there will exist some $N_0$
such that, for any pair of initial states $\rho_0,\rho'_0 \in
\cS(\sH)$, the states $\rho_n,\rho'_n$, $n \ge N_0$, will be
indistinguishable from each other.  In other words, any two
sufficiently lengthy computations will yield nearly the same final state.

Using Eq.~(\ref{eq:ncomp}), as well as unitary invariance of the
trace norm, we obtain
\begin{eqnarray}
\trnorm{\rho_n - \rho'_n} &=&
\trnorm{\left(\prod^n_{i=1}T\hat{U}_i\right) (\rho_0 - \rho'_0)}
\nonumber \\
& \le & k(T)^n \trnorm{\rho_0 - \rho'_0}. \nonumber
%\label{eq:deccomp}
\end{eqnarray}
Now suppose that at the end of the computation we perform a
measurement with precision $\epsilon$, i.e., any two states
$\rho,\rho'$ with $\trnorm{\rho-\rho'} < \epsilon$ are
considered indistinguishable.  Then, if the computation takes at
least $N_0 = \lceil \log(\epsilon/2)/\log k(T) \rceil$ steps,
we will have $\trnorm{\rho_n - \rho'_n} < \epsilon$ for all $n
\ge N_0$.  For a numerical illustration, we take $k(T) = 0.9$
and $\epsilon = 0.01$, which yields $N_0 = 50$.  In other words,
the result of any computation that takes more than 50 steps in the
presence of a strictly contractive channel $T$ with
$k(T)=0.9$ is untrustworthy since we will not be able to
distinguish between any two states $\rho$ and $\rho'$ with
$\trnorm{\rho-\rho'} < 0.01$. Again, $N_0$ depends only on the
contractivity modulus of $T$ and on the measurement precision
$\epsilon$, not on the dimension of $\sH$, at least not
explicitly.  We note that, if the state of the computer is a
density operator over a $2^s$-dimensional Hilbert space, then any
efficient quantum computation will take $O({\rm Poly}(s))$ steps,
and therefore the sensitivity of the computer's algorithm to errors
grows exponentially with $s$.

There are, however, some cases when the sequence $\set{\rho_n}$ does
converge. Suppose first that the channel $T \in C_{\rm sc}$ is
bistochastic.  Then, since each channel $\hat{U}_i$ is bistochastic as
well, the sequence $\set{\rho_n}$ converges exponentially fast to the
maximally mixed state $\idty/d$, where $d = \dim \sH$. Also, if the
computation employs a static algorithm, i.e., $\hat{U}_i = \hat{U}$
for all $i$ (this is true, e.g., in the case of Grover's search
algorithm \cite{gro}), then the channel $S \defeq T\hat{U}$ is also
strictly contractive, and $k(S) = k(T)$ by unitary invariance of the
trace norm.  Denoting the fixed point of $S$ by $\rho_S$, we then have
$$
\trnorm{\rho_n - \rho_S} = \trnorm{S^n(\rho_0) - S^n(\rho_S)} \le
k(T)^n \trnorm{\rho_0 - \rho_S},
$$
i.e., the output state of any sufficiently lengthy computation
with a static algorithm will be indistinguishable from the fixed
point $\rho_S$ of $S = T\hat{U}$.

\section{Error correction and strictly contractive channels}
\label{sec:ecscc}

After we have seen that quantum
memories and computers are ultrasensitive to errors modeled by
strictly contractive channels, we must address the issue of error
correction (stabilization of quantum information). Since we have
not made any specific assumptions (beyond strict contractivity)
about the errors affecting the computer, it is especially
important to investigate the possibility of error correction, if
only to determine the limitations on the robustness of physically
realizable quantum computers from the foundational standpoint.

\subsection{The basics of quantum error correction}
\label{ssec:qecbasics}

The simplest scheme for protecting quantum information is a
 straightforward adaptation of a classical error-correcting code
 \cite{ms}.  The basic object in the construction of a quantum
 error-correcting code is defined as follows.

\begin{definition} An {\em $(n,k)$ quantum code} is an isometry
 $\map{V}{\sH_0}{\sK}$ from the $2^k$-dimensional Hilbert space
 $\sH_0$ to the $2^k$-dimensional subspace $\sK$ (the {\em code}) of a
 $2^n$-dimensional {\em coding space} $\sH$.
\end{definition}
In other words, an input state $\rho \in \cS(\sH)$ gives rise to the
encoded state $V\rho V^* \in \cS(\sK)$.  The encoded state is acted
upon by some known channel $T$, which models the errors.  Then we say
that $V$ is a {\em $T$-correcting code} if and only if there exists
the {\em recovery channel} $\tR$ such that, for any $\rho \in
\cS(\sH_0)$, we have
\begin{equation}
(\tR \circ T)(V\rho V^*) = \rho.
\label{eq:qecc_recovery}
\end{equation}
The channel $\tR \circ T \circ (V \cdot V^*)$ can be viewed as the
composition of the encoding step, the noisy channel, and the decoding
step.  The isometry $V$ can be eliminated from
Eq.~(\ref{eq:qecc_recovery}) by writing it as
\begin{equation}
(R \circ T)(\rho) = \rho\qquad \forall \rho \in \cS(\sK).
\label{eq:recovery2}
\end{equation}
We then say that $\sK$ is a $T$-correcting code if and only if
Eq.~(\ref{eq:recovery2}) holds for some channel $R$.  

The following theorem, due to Knill and Laflamme \cite{kl}, is a tool
for determining whether a given subspace $\sK$ of the coding space can
serve as a $T$-correcting code. 

\begin{theorem}
{\bf (Knill-Laflamme)} Let $\sH$ be a Hilbert space, and consider a
channel $\map{T}{\cS(\sH)}{\cS(\sH)}$.  Let $\set{V_\alpha}$ be a
Kraus decomposition of $T$. Then a subspace $\sK$ of $\sH$ is a
$T$-correcting code if and only if, for all $\psi,\phi \in \sK$,
$$
\braket{\psi}{(V^*_\alpha V_\beta)\phi} = c_{\alpha \beta} \braket
{\psi}{\phi},
$$
where $c_{\alpha \beta}$ is some constant that depends only on
$V_\alpha$ and $V_\beta$.
\label{th:kl}
\end{theorem}

We do not give the proof of Theorem \ref{th:kl} because it is not
important for our purposes; the interested reader can consult either
the original proof of Knill and Laflamme \cite{kl}, or an alternative
argument due to Nielsen \etal \cite{ncsb}.  Knill and Laflamme have
also given another criterion \cite{kl} in terms of maximally entangled
pure states on $\tp{\sK}{\sK}$.  Please note that our proof differs
from the original argument in that it relies on the concept of the
channel fidelity (cf. Section \ref{ssec:chfid}).

\begin{theorem}
{\bf (Knill-Laflamme)} Let $\sH$ be a Hilbert space, and consider a
channel $\map{T}{\cS(\sH)}{\cS(\sH)}$.  Then a subspace $\sK$ of $\sH$
is a $T$-correcting code if and only if there exists a channel
$\map{R}{\cS(\sH)}{\cS(\sH)}$ such that, for any orthonormal basis
$\set{e_i}$ of $\sK$,
\begin{equation}
(\tp{(R \circ T)}{\id})\left(\sum_{i,j}
  \ketbra{\tp{e_i}{e_i}}{\tp{e_j}{e_j}}\right) = \sum_{i,j}
  \ketbra{\tp{e_i}{e_i}}{\tp{e_j}{e_j}}.
\label{eq:kl2}
\end{equation}
\label{th:kl2}
\end{theorem}

\begin{proof}
We note that Eq.~(\ref{eq:kl2}) is equivalent to the statement that
$\cF(\left. R \circ T \right |_\sK,\id_{\cB(\sK)}) = 1$, where $\left. R
\circ T \right |_\sK$ denotes the restriction of the channel $R \circ
T$ to $\cS(\sK)$.  However, this will hold if and only if $\left. R
\circ T \right |_\sK \equiv \id_{\cB(\sK)}$, which proves the theorem. 
\end{proof}

The Knill-Laflamme theory provides also for approximately correctable
channels.  That is, let $\set{V_\alpha}$ be a Kraus decomposition of
some channel $T$ on $\cS(\sH)$.  For any subset $\Lambda$ of
$\set{V_\alpha}$, we can define the completely positive map
$T_\Lambda$ via
$$
T_\Lambda(X) \defeq \sum_{V_\alpha \in \Lambda}V_\alpha X V^*_\alpha,\quad
\forall X \in \cB(\sH).
$$
Then a subspace $\sK$ of $\sH$ can serve as a
$T_\Lambda$-correcting code if there exists some channel $R$ on
$\cS(\sH_c)$ such that, for all $\rho \in \sK$,
$$
(R \circ T)(\rho) \propto \rho.
$$
If $\cbnorm{T-T_\Lambda}$ is sufficiently small, then it makes sense
to say that the noisy channel $T$ is {\em approximately correctable}.

The method of quantum error-correcting codes is ill-suited for dealing
with correlated errors.  A more general approach to the stabilization
of quantum information is described in the work of Knill, Laflamme,
and Viola \cite{klv} and Zanardi \cite{zan}, the essence of which we
now summarize.  Given some quantum system with the associated
finite-dimensional Hilbert space $\sH$, we consider the error channel
$T$ with the Kraus operators $V_\alpha$.  We define the {\em
  interaction algebra $\cV$ of $T$} as the $*$-algebra generated by
the $V_\alpha$'s (i.e., as the norm closure of the set of all
polynomials in $V_\alpha$ and their adjoints).  It is obvious that
$\cV$ is an algebra with identity because of the condition
$\sum_\alpha V^*_\alpha V_\alpha = \idty$.  However, since the Kraus
representation of a channel $T$ is not unique, we must make sure that,
for any two choices $\set{V_\alpha}$ and $\set{W_\alpha}$ of Kraus
decompositions of $T$, the corresponding interaction algebras are
equal.  Using the fact that any two Kraus decompositions of a channel
are connected via
$$
V_\alpha = \sum_\beta v_{\alpha\beta}W_\beta,
$$
where $v_{\alpha \beta}$ are the entries of a matrix $V$ with $V^*V =
\idty$, we see that it is indeed the case that the interaction algebra
of a channel $T$ does not depend on the particular choice of the Kraus
operators.

The existence of noiseless subsystems with respect to
$T$ hinges on the reducibility of the interaction algebra
$\cV$. Since $\cV$ is, by definition, a uniformly closed
$*$-subalgebra of $\cB(\sH)$, it is a finite-dimensional C*-algebra. A
basic result from representation theory \cite{af} tells us that $\cV$
is isomorphic to a direct sum of $r$ full matrix algebras, each of
which appears with multiplicity $m_i$ and has dimension $n^2_i$ (i.e.,
it is an algebra of $n_i \times n_i$ complex matrices).  Thus $\dim {\cV} =
\sum^r_{i=1}n^2_i$. The {\em commutant} $\cV'$ of $\cV$
is defined as the set of all operators $X \in \cB(\sH)$ that
commute with all $V \in \cV$. From the Wedderburn theorem
\cite[p. 61]{zhe} it follows that each $V \in \cV$ has the form
\begin{equation}
V = \bigoplus^r_{i=1} \tp{\idty_{m_i}}{V_i},\qquad V_i \in
\cM_{n_i}, \label{eq:intalg1}
\end{equation}
and that each $V' \in {\cV}'$ has the form
\begin{equation}
V' = \bigoplus^r_{i=1} \tp{V^\prime_i}{\idty_{n_i}},\qquad
V^\prime_i \in \cM_{m_i}. \label{eq:intalg2}
\end{equation}
Thus $\dim {\cV}' = \sum^r_{i=1}m^2_i$.  We have the
corresponding isomorphism
\begin{equation}
\sH \simeq \bigoplus^r_{i=1} \tp{{\mathbb C}^{m_i}}{{\mathbb
C}^{n_i}}, \label{eq:hsisomorph}
\end{equation}
and each factor ${\mathbb C}^{m_i}$ is referred to as a {\em
noiseless subsystem} because it is effectively decoupled from the
error channel $T$.  It is rather obvious that, in order to be of
any use, a noiseless subsystem must be nontrivial, i.e., at least
two-dimensional.  Now, if the interaction algebra ${\cV}$ is
irreducible, then $\dim {\cV}' = 1$, and no noiseless
subsystems exist.  There is a simple necessary and sufficient
condition for irreducibility of an algebra, Schur's lemma
\cite[p. 47]{br}, which states that a $*$-algebra $\cA$ is irreducible
if and only if its commutant $\cA'$ consists of complex multiples of
the identity.

\subsection{Impossibility of perfect error correction}
\label{ssec:nonns}

We now draw our attention to the correctability of errors modeled by
strictly contractive channels.  Consider a strictly contractive
channel $\map{T}{\cS(\sH)}{\cS(\sH)}$.  It is easy to see that there
does not exist a subspace of $\sH$ that could serve as a
$T$-correcting code.  Indeed, suppose to the contrary that $\sK$ is
such a subspace.  Then there exists a channel
$\map{R}{\cS(\sH)}{\cS(\sH)}$ such that Eq.~(\ref{eq:recovery2})
holds.  However, because $T$ is strictly contractive, the channel $R
\circ T$ is also strictly contractive with $k(R \circ T) \le k(T)$.
Therefore, for any $\rho,\rho' \in \cS(\sK)$ we have
$$
\trnorm{(R \circ T)(\rho - \rho')} \le k(T) \trnorm{\rho-\rho'}.
$$
On the other hand, Eq.~(\ref{eq:recovery2}) implies that
$$
\trnorm{(R \circ T)(\rho - \rho')} = \trnorm{\rho - \rho'},
$$
which would be true only for $k(T) \ge 1$.  This is a contradiction,
so we see that no strictly contractive channel $T$ admits a {\em
  perfect} quantum error-correcting code. On the other hand, keeping
in mind the fact that the Knill-Laflamme theory allows for approximate
correctability of errors, we can conclude that the nonexistence of
perfect error-correcting codes is not likely to be a serious problem.  

It turns out, however, that the property of strict contractivity is so
strong that no strictly contractive channel admits noiseless
subsystems.  As a warm-up, let us prove the special case when the
channel in question is also bistochastic \cite{rag2}.

\begin{theorem}
Consider a bistochastic strictly contractive channel
$\map{T}{\cS(\sH)}{\cS(\sH)}$, where $\sH$ is a finite-dimensional
Hilbert space.  Then the interaction algebra $\cV$ of $T$ is
irreducible, i.e., $T$ admits no noiseless subsystems.
\label{th:nns_bist}
\end{theorem}

\begin{proof}
We first observe that any operator $X$ that belongs to the commutant
of $\cV$ must necessarily be a fixed point of $T$ in $\cB(\sH)$.
Indeed, if $X \in {\cV}'$, then
$$
T(X) = \sum_\alpha V_\alpha X V^*_\alpha = X \sum_\alpha V_\alpha
V^*_\alpha = X,
$$
where we used the fact that $\sum_\alpha V_\alpha V^*_\alpha = \idty$
for a bistochastic channel.  Now if $X \in {\cV}'$, then we also have
$X^* \in {\cV}'$. This implies that, for any $X \in \cB(\sH)$, $X_1
\defeq (X+X^*)/2$ and $X_2 \defeq (X-X^*)/2i$ are in ${\cV}'$ whenever
$X$ is.  We can therefore restrict ourselves to self-adjoint operators
in the commutant of $\cV$.

For any self-adjoint operator $X$, the operator $\abs{X} \equiv
(X^*X)^{1/2}$ belongs to the algebra generated by $X^2$
\cite[p. 34]{br}, whence
$$
X = X^* \in {\cV}' \Longrightarrow X_\pm \defeq \frac{\abs{X}\pm X}{2}
\in {\cV}'.
$$
Since $X = X_+ - X_-$ and $X_\pm \ge 0$, we reduce our task to showing
that any positive $X$ in the commutant of $\cV$ is a multiple of the
identity.  Without loss of generality we may assume that $\trnorm{X} =
1$, which, together with the positivity of $X$, implies that $X$ is a
density operator.  But the only density operator left invariant by $T$
is the maximally mixed state $\idty/\dim{\sH}$, so we conclude that
$\cV' = \bbc \idty$, i.e., $\cV$ is irreducible by Schur's lemma.
\end{proof}

\begin{rem}
Incidentally, one can use Theorem \ref{th:nns_bist} to show that the
multiples of identity are the only operators in $\cB(\sH)$ that are
left invariant by the Heisenberg-picture channel $\hT$.  This is a
consequence of a theorem of Fannes, Nachtergaele, and Werner
\cite{bjkw, fnw}, which in the finite-dimensional case states that if
there exists an invertible density operator left invariant by $T$,
then the fixed-point set of $\hT$ is precisely the commutant of the
interaction algebra of $T$.  
\end{rem}

We now prove the general case \cite{rag2}.  Whereas the proof of
Theorem \ref{th:nns_bist} relied only on the existence and uniqueness
of the $T$-invariant state, the proof below directly exploits the
property of strict contractivity.

\begin{theorem}
Consider a strictly contractive channel $\map{T}{\cS(\sH)}{\cS(\sH)}$,
where $\sH$ is a finite-dimensional Hilbert space.  Then the
interaction algebra $\cV$ of $T$ is irreducible, i.e., $T$ admits no
noiseless subsystems.
\label{th:nns}
\end{theorem}

\begin{proof}
Let $\cV$ be the interaction algebra
of the channel $T$.  Let us suppose, contrary to the statement of
the theorem, that $T$ admits at least one noiseless subsystem
(i.e., $\cV$ is reducible).  That is, there exists at least
one $j \in \set{1,\ldots,r}$ such that $m_j,n_j \ge 2$ in
Eqs.~(\ref{eq:intalg1})-(\ref{eq:hsisomorph}).  Let $\sK$ be some
closed subspace of $\sH$.  Restricting the channel $T$ to the set
$$
\cS(\sK) \defeq \setcond{\rho \in \cS(\sH)}{{\rm supp}\,\rho
\subseteq \sK}
$$
(where ${\rm supp}\,\rho$ is the orthogonal complement of $\ker
\rho$), we note that, by definition, the contractivity modulus of the
restricted channel cannot exceed the contractivity modulus of $T$. Let
$\sH_j$ be the $j$th direct summand $\tp{{\mathbb C}^{m_j}}{{\mathbb
    C}^{n_j}}$ in Eq.~(\ref{eq:hsisomorph}). Define the channel $T_j$
as the restriction of $T$ to $\cS(\sH_j)$. Then any Kraus operator of
$T_j$ has the form $\tp{\idty_{m_j}}{V_\mu}$ where $V_\mu \in \cM_{n_i}$ and
$$
\sum_\mu V^*_\mu V_\mu = \idty_{n_i}.
$$
Furthermore $k(T_j) \le k(T) < 1$.  Now $T_j$ is the
channel of the form $\tp{\id}{S_j}$, where $S_j$ is the
channel on $\cS({\mathbb C}^{n_j})$ with Kraus operators $V_\mu$.
As can be easily seen, channels of this
form are not strictly contractive (they have infinitely many fixed points).
Thus $k(T_j) = 1$, and the theorem is proved, {\em
reductio ad absurdum}.
\end{proof}

The statement of Theorem \ref{th:nns} is quite shocking as it
unequivocally rules out the existence of noiseless subsystems for any
strictly contractive channel.  From the standpoint of foundations of
quantum theory, the importance of Theorem \ref{th:nns} lies in the
fact that it establishes nonexistence of noiseless subsystems for a
wide class of physically realizable quantum computers on the basis
of a minimal set of assumptions.  Furthermore, from the
mathematical point of view, it is rather remarkable that strict
contractivity of a channel already implies irreducibility of its
interaction algebra.  We must, however, hasten to emphasize that,
despite its sweeping generality, Theorem \ref{th:nns} should not be
considered as a proof of impossibility of building a reliable quantum
computer.  It merely rules out the possibility of building quantum
computers with {\em perfect} protection against errors modeled by strictly
contractive channels.

\subsection{Approximate error correction}
\label{ssec:approxec}

At this point we must realize that the results of the previous
section are not as unexpected as they may seem.  After all,
nothing is perfect in the real world!  Therefore, our
error correction schemes must, at best, come as close as possible
to the perfect scenario.  Of course, the precise criteria for
determining how close a given error correction scheme is to the
\qt{perfect case} will vary depending on the particular situation,
but we can state perhaps the most obvious criterion in terms of
distinguishability of channels.

Let us first phrase everything in abstract terms. Let the error
mechanism affecting the computer be modeled by some channel $T$.
We assume that there exists some positive $\delta < 1$ which, in
some way, characterizes the channel $T$ (it could be given, e.g.,
by the minimum of the operator norms of the Kraus operators of
$T$, and thus quantify the ``smallest'' probability of an error
occurring).  Let $\sH$ be the Hilbert space associated with the
computer.  Then, for each $\epsilon > 0$, we define an {\em
$(\epsilon,\delta)$-approximate error-correcting scheme for $T$}
to consist of the following objects:
\begin{enumerate}
\item[(1)] an integer $n > 1$,
\item[(2)] a Hilbert space $\sH_{\rm ext}$ with $\dim \sH_{\rm ext} \ge \dim
\sH$,
\item[(3)] a channel $\map{E}{\cS(\sH)}{\cS(\sH_{\rm ext})}$,
\item[(4)] a channel $\map{\tilde{T}}{\cS(\sH_{\rm ext})}{\cS(\sH_{\rm ext})}$,
and
\item[(5)] a completely positive (CP) map $\map{T_{\rm corr}}{\cS(\sH_{\rm
ext})}{\cS(\sH_{\rm ext})}$,
\end{enumerate}
such that the channel $\tilde{T}$ depends uniquely on $n$,
$\sH_{\rm ext}$, $T$, and $E$; the CP map $T_{\rm corr}$ is correctable
(say, in the Knill-Laflamme sense, or through other means, depending
on the particular situation); and we have the estimate
\begin{equation}
\cbnorm{\tilde{T}-T_{\rm corr}} < \delta^n < \epsilon.
\label{eq:approxec}
\end{equation}

Let us give a concrete example in order to illustrate the above
definition. Suppose that the channel $T$ is of the form $\id + S$
with $\cbnorm{S} < \delta$.  Then, for any $n$, we can write
\begin{equation}
T^{\otimes n} = \id + \sum_{A \subset \set{1,\ldots,n} \atop 0<
\abs{A} < n} \bigotimes^n_{k=1} S^{\iota_A(k)} + S^{\otimes n},
\label{eq:excorr1}
\end{equation}
where $\abs{A}$ denotes the cardinality of the set $A$, and
$\map{\iota_A}{\set{1,\ldots,n}}{\set{0,1}}$ is the indicator
function of $A$. We use the convention that, for any map $M$, $M^0
= \id$.  In other words, the summation on the right-hand side of
Eq.~(\ref{eq:excorr1}) consists of tensor product terms with one
or more identity factors.  For the last term, we have
$\cbnorm{S^{\otimes n}} < \delta^n$.

In this case, given some $\epsilon > 0$, we pick such $n$ that
$\delta^n < \epsilon$ and let $\sH_{\rm ext} \defeq \sH^{\otimes
n}$.  If the CP map given by the sum of the first two terms on the
right-hand side of Eq.~(\ref{eq:excorr1}) is correctable on some
subspace $\sK$ of $\sH_{\rm ext}$, then the channel $E$ is defined
in a natural way through the composition of the following two
operations:  (a) adjoining additional $n-1$ copies of $\sH$
in some suitable state $\rho_0$, and (b) restricting to the
subspace $\sK$. This way, we obviously have $\tilde{T} \defeq
T^{\otimes n}$ and
$$
T_{\rm corr} \defeq  \id + \sum_{A \subset \set{1,\ldots,n} \atop
0< \abs{A} < n} \bigotimes^n_{k=1} S^{\iota_A(k)}.
$$
The estimate (\ref{eq:approxec}) holds because $\tilde{T}-T_{\rm
corr} = S^{\otimes n}$.  We note that this construction results in
a quantum error-correcting code that corrects any $n-1$ errors. We
can use similar reasoning to describe quantum codes that correct
$k < n$ errors.

Constructing $\sH_{\rm ext}$ as a tensor product of a number of
copies of $\sH$, the Hilbert space of the computer, evidently
leads to the usual schemes for fault-tolerant quantum computation
\cite{pre}.  Other solutions, such as embedding the
finite-dimensional Hilbert space $\sH$ in a suitable
infinite-dimensional Hilbert space (e.g., encoding a qubit in a
harmonic oscillator \cite{gkp}), can also be formulated in a
manner consistent with our definition above.

Let us now address approximate correctability of strictly
contractive errors.  We have previously
demonstrated that, in the absence of error correction, the
sensitivity of quantum memories and computers to such errors grows
exponentially with storage and computation time respectively.  Let
$T$ be a strictly contractive error channel.  It is obvious that
the appropriate approximate error correction scheme must be such
that the contraction rate of the \qt{encoded} computer, where the
errors are now modeled by the channel $\tilde{T}$, is effectively
slowed down.  In some cases, straightforward tensor-product
realization may prove useful (e.g., when the product channel
$\tp{T}{T}$ is not strictly contractive). We must recall
that, for any channel $S$, a necessary condition for
correctability is $k(S) = 1$. Thus, if we can find a suitable
approximate error-correcting scheme where $\tilde{T}$ would be
well approximated by some channel $T_{\rm corr}$ with
$k(T_{\rm corr})=1$, we may effectively slow down the
contraction rate by protecting the encoded computer against errors
modeled by $T_{\rm corr}$.  A more ingenious approach may call
for replacing circuit-based quantum computation with that in
massively parallel arrays of interacting particles; several such
implementations have already been proposed (see, e.g., \cite{br2}).
It is quite likely that the possible \qt{encodings} of quantum
computation in these massively parallel systems may offer a
more efficient implementation of approximate error correction.

Finally, we should mention that the idea of \qt{approximate} noiseless
subsystems has already been explored by Bacon, Lidar, and Whaley
\cite{blw}.  In their work, it is argued that the symmetry, which is
required of a channel in order for noiseless subsystems to exist, is
generally broken by perturbing the channel.  They show that, if the
perturbations of the channel are \qt{reasonable,} then the noiseless
subsytem is stable to second order in time.  We must reiterate that
the negative results we have stated in the previous section refer only
to nonexistence of \qt{perfectly} noiseless subsystems; in the real
world, we have no choice but to settle for \qt{almost perfect}
anyway.

\section{Implications for quantum information processing}
\label{sec:imp}

\subsection{General considerations}
\label{ssec:gencons}

As we have seen in Section \ref{sec:noisyqmem}, the maximum number
$n_{\rm max}$ of operations that can be carried out on a physically
realizable quantum computer in the presence of strictly contractive
noise is limited by the contraction rate $k$ and the measurement
precision $\epsilon$, and is equal to $\log{(\epsilon/2)}/\log{k}$.
The measurement precision $\epsilon$ depends on the measuring
apparatus, while the contraction rate $k$ is determined by the
decoherence mechanism.  In the next section we will present an
elementary analysis of noisy bulk spin-resonance quantum computation
\cite{cfh,gc} in terms of the strictly contractive decoherence model;
here we focus on the quantitative conclusions that can be drawn
regardless of the type of \qt{hardware} used for building the quantum
computer.

First of all, let us make an obvious observation that the number of
operations that can be carried out within the \qt{coherence time} of
the computer is related not to the {\em size} of the corresponding
quantum circuit (i.e., the total number of gates used to construct
it), but rather to the {\em depth} of the circuit (i.e., the maximum
number of gates acting on any qubit throughout the computation).  It is
quite clear that the {\em complexity-theoretic} circuit depth is irrelevant
here; what matters is the {\em physical} circuit depth, which is, of course,
determined by the particular realization of the computer.  With that
in mind, let $D_A(n)$ denote the physical circuit depth for some
quantum algorithm $A$ with the input state of $n$ qubits.  Then, if
the contraction rate $k$ is fixed, the required measurement precision
is easily seen to be given by
\begin{equation}
\epsilon = 2 k^{D_A(n)}.
\label{eq:prec}
\end{equation}
Because the contraction rate can be written as $1/2^\alpha$, where
$\alpha$ is some large positive number, we can rewrite
Eq.~(\ref{eq:prec}) as
$$
\epsilon = \frac{1}{2^{c D_A(n)}},
$$
where $c$ is a constant that depends on the decoherence mechanism and
increases as the noise gets stronger.  Thus we see that the required
measurement precision grows exponentially with the physical circuit
depth.  

Alternatively we can consider the case when we are given $\epsilon$
and $D_A(n)$, and need to determine the maximum tolerable error
rate. Then, whenever $k \ge (\epsilon/2)^{1/D_A(n)}$, we will have
$n_{\rm max} \ge D_A(n)$.  When the noise is sufficiently weak, we can
approximate it with a depolarizing channel (cf. Section
\ref{ssec:sccdense}), in which case $k = 1 - \eta$, and the
depolarization constant $\eta$ can be thought of as the error rate.
If the computation is to be concluded within the coherence time, then
the maximum allowable error rate is given by
$1-(\epsilon/2)^{1/D_A(n)}$, whence we see that, in order to build
fault-tolerant circuit-based quantum computers, we need high-precision
measurements and shallow circuits.  To make an (admittedly academic)
illustration of this, we provide in the table below the values of the
threshold error rate for algorithms with various physical circuit
depths for the case when the measurement precision is on the order of
$\sqrt{\hbar}$, comparable to the so-called {\em standard quantum
  limit} (SQL) \cite{bk2}.

\begin{center}
\begin{tabular}{|l|c|c|c|c|c|}
\hline
 & \multicolumn{5}{c|}{$n$ (number of qubits)} \\
\cline{2-6}
$D(n)$ & 20 & 40 & 60 & 80 & 100 \\
\hline \hline
$\log{n}$ & $\sim 1$ & 0.999 & 0.999 & 0.998 & 0.998\\
\hline
$n$ & 0.863 & 0.630 & 0.485 & 0.392 & 0.328\\
\hline
$n^3$ & $4.96 \times 10^{-3}$ & $6.22 \times 10^{-4}$ & $1.84 \times 10
^{-4}$ & $7.78 \times 10^{-5}$ & $3.98 \times 10^{-5}$ \\
\hline
$\sqrt{2^n}$ & 0.038 & $3.80 \times 10^{-5}$ & $3.71 \times 10^{-8}$ &
$3.62 \times 10^{-11}$ & $3.53 \times 10^{-14}$ \\
\hline
\end{tabular}
\end{center}
Despite the fact that the measurement precision we have assumed is
ridiculously high ($\epsilon \sim 10^{-17}$), the maximum tolerable
error rate is still prohibitively low for circuits of polynomial and
superpolynomial depth, even when the number of qubits is quite modest.
It is worth noticing, however, that the threshold error rate starts
off very close to unity and rolls off fairly slowly when the quantum
circuit has logarithmic or linear depth.  We will come back to this
point in Section \ref{ssec:where}, when we talk about parallelization
as a means of protecting the computer against noise.

\subsection{Case study:  ensemble quantum computation using nuclear
  magnetic resonance}
\label{ssec:nmr}

Looking back to the formula $n_{\rm max} =
\log{(\epsilon/2)}/\log{k}$, we can pose the following question.
Given a particular experimental scheme for realizing a quantum
computer, what can we say about the measurement precision and about
the noise strength (contraction rate)?  In this section we carry out a
simple analysis in order to answer this question for ensemble
quantum computation using nuclear magnetic resonance, proposed in 1997
independently by Cory, Fahmy, and Havel \cite{cfh}, and by Gershenfeld
and Chuang \cite{gc}. From now on we will use the term \qt{NMR quantum
  computation} to refer to this scheme; a more descriptive, and also
more cumbersome, term would be \qt{high-temperature liquid-state NMR
  quantum computation.}

The basic idea behind NMR quantum computation is the following.  An
$N$-spin NMR quantum computer operates on a sample solution containing
a huge number of molecules (on the order of $10^{23}$), each of which
accommodates $N$ two-level nuclear spins.  The sample, which is placed
in a strong unidirectional magnetic field, is subjected to a temporal
sequence of radio-frequency pulses, and each molecule functions as an
autonomous computational unit.  The result of the computation, which
is read off by means of the usual techniques of NMR spectroscopy \cite{ebw}, is
the ensemble average of the computer outputs taken with respect to the
state of all the molecules in the sample.  Because NMR experiments are
conducted at room temperature ($\sim 300$ K), the initial state of the
sample is the thermal equilibrium state $\exp{(-\beta H)}/Z_\beta$,
where $H$ is the Hamiltonian of a single molecule.

On a more formal level, the inner workings of an NMR quantum computer
can be described using the concept of an {\em effective pure state},
which is defined as follows \cite{cgkl}.

\begin{definition}
\label{def:eps}
Let $\sH$ be a Hilbert space.  Consider a unit vector $\psi \in \sH$,
a channel $\map{T}{\cS(\sH)}{\cS(\sH)}$, and a set $\set{X_i}$ of
observables in $\cB(\sH)$.  Then the state $\rho \in \cS(\sH)$ is called
an {\em effective pure state} for $\psi$ with respect to $T$ and
$\set{X_i}$ if there exists another channel $T'$ and a constant
$\alpha$ such that, for each $i$,
\begin{equation}
\tr{[T'(\rho)X_i]} = \alpha \braket{\psi}{\hT(X_i)\psi}.
\label{eq:eps}
\end{equation}
\end{definition}

Here is a concrete example \cite{cgkl} to illustrate this abstract
definition.  Let $T$ be a bistochastic channel, i.e., $T(\idty) =
\idty$.  Then, for any $\alpha \in (0,1)$, the state
\begin{equation}
\rho_\alpha \defeq (1-\alpha) \idty/\dim{\sH} + \alpha \ketbra{\psi}{\psi}
\label{eq:ralpha}
\end{equation}
is an effective pure state, with $T' = T$, for the pure state
$\ketbra{\psi}{\psi}$ with respect to $T$ and any set of traceless
observables.  Indeed, for any $X$ with $\tr{X}=0$, we have
$$
\tr{\left[T(\rho_\alpha)X\right]} = \frac{1-\alpha}{\dim{\sH}}
\tr{[T(\idty) X]} + \alpha \braket{\psi}{\hT(X)\psi} = \alpha
\braket{\psi}{\hT(X)\psi},
$$
so that the condition (\ref{eq:eps}) is satisfied.

The significance of the above formalism for NMR quantum computation
comes from the fact that the Gibbs state of the liquid sample is well
approximated by the state of the form (\ref{eq:ralpha}) with
$\dim{\sH} = 2^N$ and $\alpha = N \hbar \Omega \beta/2^{N+1}$, where
$\hbar \Omega$ is the average difference between the excited-state and
the ground-state energies of the nuclear spins in a strong magnetic
field \cite{gc,sc2}. The quantity $\hbar \Omega \beta/2$ is referred
to as the {\em Boltzmann factor} \cite{cgkl}. To give a feel for the
orders of magnitude involved, the average resonant frequency of a
nuclear spin in a typical NMR experiment is on the order of 200 MHz
\cite{gc}, which at room temperature corresponds to $\alpha \sim 1.6
\times 10^{-5} N/2^N$.  One crucial feature of an effective pure state
of the form (\ref{eq:ralpha}) is that, for any bistochastic channel
$T$, we have
$$
T(\rho_\alpha) = (1-\alpha)2^{-N}\idty + \alpha T(\ketbra{\psi}{\psi}),
$$
i.e., the polarized spins that participate in the actual computation
evolve independently of the unpolarized spins forming the \qt{thermal
  background} in the sample.  Then, provided that a suitable set of
traceless observables is measured at the end of the computation, the
only detectable signal comes from the pure portion of $\rho_\alpha$,
the obvious disadvantage being that the corresponding signal strength
is $O(N/2^N)$, which decreases rapidly as the number of spins per
molecule increases.

An NMR quantum computer is usually run many times, and each time a
single-spin observable is measured; the measurement results are then
processed on a classical computer \cite{cgkl}.  Let
$\paulix^n,\pauliy^n,\pauliz^n$ denote the Pauli spin matrices acting
on the Hilbert space of the $n$th spin.  A typical observable measured
on the $n$th spin after a single run of the computer is equal, up to a
multiplicative constant, to $M_n \defeq \paulix^n + i \pauliy^n$, so that the
experimentally detected output is proportional to the transverse
magnetization of the sample, $N_S \tr{(\rho M_n)}$, where $N_S$ is the
number of molecules in the sample and $\rho$ is the state of the
sample after the computation \cite{gc}.  Using this information, we
can give a concrete interpretation of the measurement precision
$\epsilon$.  Consider the measurement of an arbitrary observable $A$.
For any two density operators $\rho$ and $\rho'$, we have the bound
\begin{equation}
\abs{\tr{(\rho A)} - \tr{(\rho' A)}} \le \norm{A} \trnorm{\rho - \rho'}.
\label{eq:maxdiff}
\end{equation}
Now let $\Delta_A$ a typical (e.g., r.m.s.) fluctuation of $A$. If we
stipulate that the resolution of the measurement of $A$ is limited by
$\Delta_A$, then Eq.~(\ref{eq:maxdiff}) suggests that any two density
operators $\rho,\rho'$ with $\trnorm{\rho - \rho'} <
\Delta_A/\norm{A}$ can be considered indistinguishable.  When we are
talking about the number $N_S$ of molecules in a macroscopic sample,
the corresponding fluctuation is given by $\sqrt{N_S}$ \cite[Ch. 1]{ll}.
Therefore, because $\norm{M_n} = 2$, we have $\epsilon =
1/2\sqrt{N_S}$, which yields, for $N_S \sim 10^{23}$, the value
$\epsilon \sim 10^{-12}$.  This gives us a rough (order-of-magnitude)
estimate of the measurement precision in NMR quantum computers.

As far as the decoherence mechanism is concerned, we need only
consider single-spin dynamics because, after each run of the computer,
only the single-spin observables are measured.  There are two main
sources of decoherence \cite{gc,vsb}, namely the thermal relaxation
and the phase damping; they are described explicitly as follows
\cite{vsb}. The channel that models thermal relaxation is precisely
the channel shown in Example \ref{ex:gad}; this channel is strictly
contractive with $k = e^{-\tau/2T_{\rm th}}$, where $\tau$ is the
duration of the single step of the noisy dynamics, and $T_{\rm th}$ is
the thermal relaxation time. The phase damping is modeled by the
channel with the Kraus operators $\sqrt{\lambda}\idty$ and
$\sqrt{1-\lambda}\pauliz$, where $\lambda = (1+e^{-\tau/T_{\rm
    ph}})/2$, $T_{\rm ph}$ being the phase damping time.  The
phase-damping channel is not strictly contractive because it has two
fixed points, $\ketbra{\psi_+}{\psi_+}$ and $\ketbra{\psi_-}{\psi_-}$,
where $\pauliz \psi_\pm = \pm \psi_\pm$.

Typically the phase damping time is much shorter than the thermal
relaxation time \cite{gc}, the value of the ratio $T_{\rm ph}/T_{\rm
  th}$ depending on the kinetics of the particular molecule.
Therefore the phase damping time sets a more stringent limitation on
the number of operations that can be carried out on an NMR quantum
computer, but, because the channel formed by composing the
thermalizing channel and the phase-damping channel is still strictly
contractive with $k = e^{-\tau/2 T_{\rm th}}$, only the thermal
relaxation time is relevant to our analysis.  With this in mind, we
can write down the following expression for the maximum number of
operations that can be carried out within the thermal relaxation time:
$$
n_{\rm max} = \frac{\log{(\epsilon/2)}}{\log{k}}
\sim \frac
	{ \log{(10^{-12}/2)} }
	{\log{e^{-\tau/2 T_{\rm th}}}}.
$$
The duration $\tau$ of the single step of the noisy dynamics is
comparable to the time it takes to execute a single unitary operation
\cite{gc,vsb}.  A single-spin unitary operation can be performed in
about 10 ms, whereas it may take roughly 100 ms to apply a two-spin
gate.  A conservative estimate for $\tau$ would therefore be around 45
ms, whence we obtain $n_{\rm max} = 1258.85 T_{\rm th}$, where the
thermal relaxation time $T_{\rm th}$ is measured in seconds.

Recently Vandersypen \etal \cite{vsb} implemented the simplest
nontrivial instance of Shor's quantum factoring algorithm (namely,
finding the prime factorization of 15) using a 7-spin NMR computer,
which required $\sim 300$ computational steps.  In the molecule they
used, the thermal relaxation times of the spins were as small as 2.8
seconds and as large as 45.4 seconds, which implies that the maximum
number of operations that could be carried out using a molecule with
spin relaxation times in this range is anywhere from 3,525 to
57,152. Because Shor's algorithm has cubic complexity, we may infer
that the NMR quantum computer utilizing such molecules would not be
scalable beyond 39 spins.  In general, the scalability would increase
with $T_{\rm th}$, so one of the ways to meet the scalability
challenge would be to engineer molecules with high thermal relaxation
times.  We mention in passing that, as pointed out by Schack and Caves
\cite{sc2}, there exists a purely classical model for NMR quantum
computation when the number $N$ of the spins per molecule is
sufficiently low (e.g., when the Boltzmann factor equals $2 \times
10^{-6}$, the NMR computer with $N < 16$ admits a classical model).

\subsection{Where do we go from here?}
\label{ssec:where}
 
The main lesson to be learned from the strictly contractive model of
decoherence is the following:  the longer the computation, the less
reliable its output.  The same problem arises in classical
circuit-based computation with noisy gates, and there are two ways to
handle it: (a) error-correcting codes, and (b) parallelization.  The
first of these techniques amounts to introducing a considerable amount
of redundancy into the network.  In particular, it was shown by
Dobrushin and Ortyukov \cite{do} (cf. also the more refined argument
by Pippenger, Stamoulis, and Tsitsiklis \cite{pst}) that, if a
noiseless network requires $N$ gates to compute a particular function,
then the noisy network would require at least $N \log{N}$ gates to
compute the same function reliably, provided that the error
probability per gate does not exceed 1/2.  The parallelization
technique, on the other hand, allows to reduce the computation time by
shrinking the circuit depth.

The problems that are efficiently parallelizable (i.e., can be
computed in polylogarithmic time using a polynomial number of
processors working in parallel) form the complexity class {\bf NC}
\cite[Ch. 15]{pap}.  The abbreviation {\bf NC} stands for \qt{Nick's
  class,} after Nicholas Pippenger who extensively studied this
complexity class.  Typical problems in {\bf NC} are, e.g., summing $m$
numbers [which can be done in time $O(\log{m})$], or copying the
contents of a particular memory cell into $n^{O(1)}$ memory cells
[which can be done in time $O(\log{n})$] \cite[p. 253 ff.]{as}.

Moore and Nilsson \cite{mn} recently introduced the quantum complexity
class {\bf QNC}. They showed that most circuits, including those for
performing error correction, can be parallelized to logarithmic
depth. A notable exception is the circuit for the quantum Fourier
transform (QFT), a crucial ingredient in Shor's factoring algorithm,
which can be parallelized only to linear depth. Moore and Nilsson
conjectured that it is impossible to parallelize the QFT circuit to
less than linear depth. In Section \ref{ssec:gencons} we provided some
numerical estimates for the threshold error rate in circuit-based
quantum computers as a function of the circuit depth. We saw that
circuits of logarithmic and linear depth turned out to be more robust
than circuits of polynomial and superpolynomial depth.  Hence, if one
does insist on implementing quantum computers using the circuit
paradigm, then it may be worthwhile to explore the class {\bf QNC} further.

A more radical solution is to abandon the quantum circuits in favor of
massively parallel systems of locally interacting particles (cellular
automata).  In a cellular automaton, the state of each particle at
some integer time $t+1$ is determined by its state, as well as by the
states of finitely many neighboring particles, at time $t$.  Classical
cellular automata, both deterministic \cite{cd,gar} and probabilistic
\cite{lms,lig}, model a rich variety of complex phenomena; in
particular, they can serve as a computational medium.  As was shown by
Toom \cite{too}, it is possible to store reliably a single bit of
information in a noisy two-dimensional cellular automaton.  The
approach of Toom was adopted by G\'acs \cite{gac} and by G\'acs and
Reif \cite{gr}, who have demonstrated that it is possible to perform
reliable computation in nosiy three-dimensional cellular automata.
The physical underpinning of reliable computation and information
storage in noisy cellular automata can be understood in terms of phase
transitions \cite{lms}.  The idea is to construct a {\em nonergodic}
cellular automaton, i.e., one that does not have a unique invariant
state which it would eventually reach irrespective of initial
conditions.

It would be interesting to see how much of this carries over to the
quantum domain.  There are many proposals for computation, both
classical and quantum, using quantum cellular automata (see, e.g.,
Briegel and Raussendorf \cite{br2}, Fussy \etal \cite{fgss}, Lent
\etal \cite{ltph}, Lloyd \cite{llo} or Meyer \cite{mey}).  The two
main attractions of quantum cellular automata are (a) massively
parallel structure, and (b) the possibility of a phase transition.  We
have already discussed massively parallel structure of quantum
cellular automata in Section \ref{ssec:approxec}; here we focus our
attention on phase transitions.  A necessary condition for the
existence of a phase transition in a cellular automaton is
nonergodicity.  Richter and Werner \cite{rw} gave an ergodicity
criterion for quantum cellular automata, formulated in terms
of the completely positve map that describes, in the Heisenberg
picture, the transition rule of the automaton.  Assuming that each
cell (site) of the automaton is under the influence of some strictly
contractive error channel $T$, an interesting problem would be to
devise such a transition rule that the automaton would be nonergodic.
In this respect we should mention that, even if $T$ is a strictly
contractive channel, it is not at all obvious whether $\tp{T}{T}$ is
strictly contractive as well:  it has a unique fixed point among the
product density operators, but there may also be another fixed point
of $\tp{T}{T}$ that is not a product density operator. 

%% file: chap4.tex
\chapter{Entropy-energy arguments}
\label{ch:enterg}

%\vskip 1in

The physics of relaxation processes is often understood, at least on a
heuristic level, through the consideration of the balance of energy
and entropy, as determined by the temperature.  There is a
thermodynamic function that relates energy, entropy, and temperature,
namely the {\em Helmholtz free energy} \cite[p. 98]{cal},
$$
F \defeq E - (1/\beta)S,
$$
where $E$ is the energy, $S$ is the entropy, and $\beta$ is the
inverse temperature. The second law of thermodynamics
\cite[p. 17]{dor} states that no energy-conserving process can
decrease the entropy.  Another way to state this is to say that, among
all the configurations of the system that have the same energy, the
ones with the largest entropy are \qt{thermodynamically favorable}
\cite[pp. 22-24]{fer}, by which we mean that the corresponding
configurations have very large probabilities.  We can also deal with
processes that do not conserve energy, in which case we are interested
in the incremental free energy, $\Delta F = \Delta E - (1/\beta)\Delta
S$.  We can consider a particular configuration stable if any local
modification of this configuration results in $\Delta F > 0$, i.e.,
the energy cost of the modification more than compensates for the
entropy gain.  On the other hand, if $\Delta F < 0$, then the energy
cost cannot offset the entropy gain, and the corresponding
configuration is unstable.

In this chapter we offer an interpretation of the relaxation dynamics
of noisy quantum computers in terms of the entropy-energy balance.  

\section{Definition and properties of entropy}
\label{sec:entropy}

We give a very brief overview of the concept of entropy.  Most of the
results are just stated without proofs.  The reader is encouraged to
consult the book by Gray \cite{gra} for the rigorous treatment of
entropy in the context of classical information theory; an excellent
survey of Wehrl \cite{weh} is devoted to the concept of entropy in
statistical physics.  For an abstract treatment of entropy in the
context of operator algebras, we recommend the book by Ohya and Petz
\cite{op}.

In statistical physics, the entropy of a system that can exist in $N$
possible configurations is given, up to a multiplicative constant, by
Boltzmann's formula
$$
S \defeq \ln{N}.
%\label{eq:maxent}
$$
This assumes, however, that all $N$ configurations of the system are
equiprobable.  When this is not the case, i.e., when the $i$th
configuration occurs with probability $w_i$, the entropy is defined as
\begin{equation}
S \defeq - \sum^N_{i=1} w_i \ln{w_i}.
\label{eq:entropy}
\end{equation}
Given the probability distribution $w = \set{w_i}$, we will denote the
corresponding entropy (\ref{eq:entropy}) by $S(w)$ or, when we want to
exhibit the probabilities explicitly, by $S(\set{w_i})$.  The entropy
$S(w)$ is referred to as the {\em Shannon entropy} of $w$.  It can be
shown that, for any probability distribution $w$ on an $N$-element
set, $0 \le S(w) \le \ln{N}$, where the lower bound is achieved if and
only if $w_i = \delta_{ik}$ for some $k \in \set{1,\ldots,N}$, and the
upper bound is achieved if and only if $w$ is the uniform
distribution, $w_i = 1/N$ for all $i$.  In view of this, it is natural
to regard the entropy as a measure of \qt{randomness} of a probability
distribution.  A crucial property of the entropy is {\em concavity}:
given any number $\lambda \in [0,1]$, we have $S(\lambda w +
(1-\lambda) w') \ge \lambda S(w) + (1-\lambda)S(w')$, where the convex
combination $\lambda w + (1-\lambda) w'$ of the probability
distributions $w = \set{w_i}$ and $w' = \set{w'_i}$ is the probability
distribution $\set{\lambda w_i + (1- \lambda) w'_i}$.

Now consider a classical system with the $N$-element configuration
space $\sX$ and the corresponding algebra of observables $\cC(\sX)$.
Then the one-to-one correspondence between the states over $\cC(\sX)$
and the probability measures on $\sX$ allows us to define the entropy
of the state $\omega$ as the entropy of the corresponding probability
distribution. We see that the only states with zero entropy are the
pure states; the mixed states all have strictly positive entropy.  The
concavity of the entropy then means that mixing leads to an increase
in entropy.  Because the normalized counting measure on $\sX$
corresponds to the unique state that maximizes the entropy, we will
refer to this state as {\em maximally mixed}.

Next we turn to the case of a quantum system, the corresponding
algebra of observables being the algebra $\cB(\sH)$ of bounded
operators on the Hilbert space $\sH$ associated with the system.  We
assume for simplicity that $\sH$ has finite dimension $N$.  Then there
is a one-to-one correspondence between the states over $\cB(\sH)$ and
the density operators in $\cB(\sH)$.  Given the density matrix $\rho$,
we define its {\em von Neumann entropy} as
\begin{equation}
S(\rho) \defeq - \tr{\rho \ln{\rho}}.
\label{eq:vnentropy}
\end{equation}
The eigenvalues $\lambda_i$ of $\rho$ form a probability distribution
on an $N$-element set, and it is clear from the definition
(\ref{eq:vnentropy}) that the von Neumann entropy of $\rho$ is
precisely the Shannon entropy of this probability distribution.  This
also implies that $S(\rho) = 0$ if and only if $\rho$ is a pure state,
and that the unique state that maximizes $S$ is the maximally mixed
state $\idty/N$.    The von Neumann entropy enjoys a concavity
property similar to that of the Shannon entropy.

The von Neumann entropy is a continuous functional on the state space
$\cS(\sH)$ when the latter is given the topology induced by the trace
norm.  More precisely, we have the following lemma \cite[p. 22]{op}.

\begin{lemma}{\bf (Fannes)}
Let $\rho,\sigma$ be two density operators on an $N$-dimensional
Hilbert space, and suppose that $\trnorm{\rho - \sigma} < 1/3$.  Then
\begin{equation}
\abs{S(\rho) - S(\sigma)} \le \ln{N} \cdot \trnorm{\rho - \sigma} -
\eta(\trnorm{\rho-\sigma}),
\label{eq:fannes}
\end{equation}
where $\eta(t) \defeq t \ln{t}$.
\label{lm:fannes}
\end{lemma}

Now suppose that we are given a channel $T$ with the following
properties:  (1) the $T$-invariant state $\rho_T$ is unique, and (2)
$S(T(\rho)) > S(\rho)$ unless $\rho = \rho_T$.  Then the Fannes
inequality (\ref{eq:fannes}) implies that the von Neumann entropy is a
strict Liapunov function for $T$. Theorem \ref{th:liapunov} can then
be used to establish the trace-norm convergence of the orbit
$\set{T^n(\rho)}$ to $\rho_T$ for any initial state $\rho$.

The entropy is an extensitve property (it scales with system size).
That is, if we consider two systems with the Hilbert spaces $\sH$ and
$\sK$, then for any $\rho \in \cS(\sH)$ and any $\sigma \in \cS(\sK)$
we have
$$
S(\tp{\rho}{\sigma}) = S(\rho) + S(\sigma).
$$
In fact, among all the density operators $\rho \in \cS(\tp{\sH}{\sK})$
with the same restrictions $\rho_\sH \defeq \ptr{\sK}{\rho}$ and
$\rho_\sK \defeq \ptr{\sH}{\rho}$, the product state
$\tp{\rho_\sH}{\rho_\sK}$ has the largest entropy.  This property is
referred to as the {\em subadditivity} of the entropy.  There is also
a property referred to as the {\em strong subadditivity}, which
consists in the following.  Let $\sH_1,\sH_2,\sH_3$ be Hilbert spaces;
we will use $\ptr{ij}{(\cdot)}$ to denote the partial trace over
$\tp{\sH_i}{\sH_j}$.  Given a density operator $\rho \in
\cS(\tp{\sH_1}{\sH_2}\tp{}{\sH_3})$, define the partial traces $\rho_1
\defeq \ptr{23}{\rho}$, $\rho_{12} \defeq \ptr{3}{\rho}$, and so
on. Then
$$
S(\rho) + S(\rho_2) \le S(\rho_{12}) + S(\rho_{23}).
$$
The proof of the strong subadditivity, which was first obtained by
Lieb and Ruskai \cite{lr}, is far from transparent, in stark contrast
to the fairly straightforward proof of the corresponding property of
the Shannon entropy. 

A useful quantity derived from entropy is the so-called {\em relative
  entropy}.  The classical definition, for a pair $w,w'$ of
probability distributions, is given by 
$$
S(w|w') \defeq \sum_i w_i \ln{\frac{w_i}{w'_i}}.
$$
It is easy to show that $S(w|w') \ge 0$ with equality if and only if
$w = w'$.  Sometimes the relative entropy is referred to as the {\em
  Kullback-Leibler distance,} but this is a misnomer because the
relative entropy is not symmetric and does not satisfy the triangle
inequality.  A more appropriate term is the Kullback-Leibler {\em
  divergence}.  The quantum relative entropy is defined, for two
density operators $\rho$ and $\sigma$, as 
$$
S(\rho | \sigma) \defeq
	\tr{(\rho \ln{\rho} - \rho \ln{\sigma})}.
$$
The quantum relative entropy has the same positivity property as the
corresponding classical quantity, namely $S(\rho | \sigma) \ge 0$ with
equality if and only if $\rho = \sigma$.  This can be proved using the
following lemma \cite{str2}, which also gives a handy lower bound on
$S(\cdot | \cdot)$.

\begin{lemma}
{\bf (Streater)} Let $\rho$ and $\sigma$ be two density operators.  Then
\begin{equation}
S(\rho | \sigma) \ge \frac{1}{2} \pnorm{\rho - \sigma}{2}^2,
\label{eq:streater}
\end{equation}
where $\pnorm{\cdot}{2}$ is the Hilbert-Schmidt norm.
\label{lm:streater}
\end{lemma}

\begin{proof}
Consider the function $\eta(x)$, defined as above, on the interval $I
= [0,1]$. For any pair $x,y \in I$ we have, by Taylor's theorem with
the Lagrange remainder,
$$
\eta(x) = \eta(y) + (x-y) \eta'(y) + \frac{1}{2}(x-y)^2 \eta''(t)
$$
for some $t \in I$.  Now $\eta''(t) = 1/t \ge 1$ for $t \in I$, which
leads to the estimate
\begin{equation}
\eta(x) - \eta(y) - (x-y) \eta'(y) - \frac{1}{2}(x-y)^2 \ge 0.
\label{eq:lagrange}
\end{equation}
Let $a_i$ and $\phi_i$ be the eigenvalues and the eigenvectors of
$\rho$, and let $b_i$ and $\psi_i$ denote the same objects for
$\sigma$.  Define $g_{ij} \defeq \braket{\phi_i}{\psi_j}$, so that
$\sum_j \abs{g_{ij}}^2 = 1$.  Then
\begin{eqnarray}
&& \braket{\phi_i}{[\eta(\rho) - \eta(\sigma) - (\rho - \sigma)
      \eta'(\sigma) - (\rho - \sigma)^2/2]\phi_i} \nonumber \\
&& \qquad = \sum_j \abs{g_{ij}}^2 [\eta(a_i) - \eta(b_i) - (a_i - b_i)
      \eta'(b_i) - (a_i - b_i)^2/2]. \nonumber
\end{eqnarray}
Summing over $i$ and using the estimate (\ref{eq:lagrange}), we get
$$
\tr{(\rho \ln{\rho} - \rho \ln{\sigma})} \ge \frac{1}{2} \tr{(\rho -
  \sigma)}^2 \equiv \frac{1}{2} \pnorm{\rho - \sigma}{2}^2,
$$
and the lemma is proved.
\end{proof}

In a limited sense, the relative entropy $S(\rho|\sigma)$ can be
thought of as a measure of closeness between $\rho$ and $\sigma$ [we
  say \qt{limited} because $S(\cdot |\cdot)$, just like its classical
  counterpart, fails to satisfy the triangle inequality].  In this
regard we mention the result of Lindblad \cite{lin3} that, for any
channel $\map{T}{\cS(\sH)}{\cS(\sH)}$ and for any pair $\rho,\sigma
\in \cS(\sH)$, we have $S(T(\rho)|T(\sigma)) \le S(\rho|\sigma)$.

\section{The Gibbs variational principle and thermodynamic stability}
\label{sec:gibbs}

In Section \ref{sec:chrelax} we have discussed the zeroth law of
thermodynamics, which essentially says that any macroscopic system
will generally be found in the state of equilibrium, characterized by
a few macroscopic parameters.  According to the well-known Gibbs
variational principle in statistical mechanics \cite[p. 348]{sim}, the
equilibrium states of a finite quantum system with Hamiltonian $H$ at
absolute temperature $T$ are precisely those states that minimize the
free-energy functional
\begin{equation}
F_\beta(\rho) \defeq \tr{(\rho H)} - \frac{1}{\beta} S(\rho)
\label{eq:gvp}
\end{equation}
(in case of an infinite system, one would instead minimize the
specific free-energy functional, i.e., free energy per \qt{particle}).

It is very easy to see that, when the system in question is finite,
the Gibbs state $\rho_\beta$ [cf. Eq.~(\ref{eq:gibbs})] is the unique
solution of the variational problem (\ref{eq:gvp}).  Let $\Phi(\beta)
\defeq -(1/\beta) \ln{Z_\beta}$, where $Z_\beta = \tr{e^{-\beta H}}$
is the canonical partition function. Then, for any density operator
$\rho$, we have
\begin{eqnarray}
F_\beta(\rho) - \Phi(\beta) &=& \tr{(\rho H)} +
\frac{1}{\beta}\left[\tr{(\rho \ln {\rho})} + \ln {Z_\beta} \right]
\nonumber \\
&=& -\frac{1}{\beta} \left[\tr{(\rho \ln{\rho_\beta})} +
  \ln{Z_\beta}\right] + \frac{1}{\beta} \left[\tr{(\rho \ln{\rho})} +
  \ln{Z_\beta}\right] \nonumber \\
&=& \tr{(\rho \ln{\rho} - \rho \ln{\rho_\beta})} \nonumber \\
&\equiv & S(\rho | \rho_\beta), \nonumber
\end{eqnarray}
which implies that $F_\beta(\rho_\beta) \equiv \Phi(\beta)$.  Now all
we need to show is that, for any $\rho \neq \rho_\beta$,
$F_\beta(\rho) > \Phi(\beta)$, but this follows immediately from Lemma
\ref{lm:streater}.  The uniqueness of the solution to (\ref{eq:gvp})
for finite systems makes them unsuitable for the study of macroscopic
degeneracy (i.e., when a given macroscopic system has multiple
equilibrium states at a given temperature) and phase transitions; it
is then necessary to pass to the so-called {\em thermodynamic limit}.

An alternative characterization of equilibrium states is developed
through the notions of global and local thermodynamic stability
\cite{sew}. Global thermodynamic stability is equivalent to the Gibbs
variational principle: states that are globally thermodynamically
stable (GTS) are precisely those that minimize the specific
free-energy functional.  On the other hand, a state $\rho$ is locally
thermodynamically stable (LTS) if the specific free energy of any
state $\sigma$, obtained by local perturbation of $\rho$, is greater
than that of $\rho$.  It is known that any GTS state is also LTS, but
the converse is not generally true for an infinite system
\cite[pp. 31-33]{sew}.  We thus obtain a useful device for showing
that a given state is not GTS: namely, showing that it is not LTS. An
argument of this kind is referred to as an \qt{entropy-energy
  argument} \cite{ss} since showing that the state $\rho$ is not LTS
amounts to showing that it is possible to perturb $\rho$ locally in
such a way that the resulting change in specific free energy is
negative, owing to the fact that the entropy gain due to the perturbation
overwhelms the corresponding energy shift.  Although the goal of an
entropy-energy argument is to show that a given infinite-volume state
is not GTS, it is often possible to consider only the finite-volume
scenario to show that the state is not LTS.

We illustrate a typical entropy-energy argument by giving a heuristic
description of the argument due to Thouless \cite{tho} concerning the
absence of ordering in a one-dimensional Ising system (spin chain)
with short-range interactions.  The original argument appeared in the
text by Landau and Lifshitz \cite[p. 537]{ll}; the version of Thouless
is a refinement of their reasoning.  Let us assume, to the contrary,
that ordering exists; that is, all of the spins in the chain point in the
same direction.  Then, if the ordered phase is stable, the
corresponding state must minimize the free-energy functional.  Now
suppose that we reverse all the spins in a segment of large (but finite)
size $N$.   Due to the short range of the interactions, the energy
cost of inserting this \qt{macroscopic droplet} is bounded above by a
constant.  Now, if we randomly insert this droplet in any one of $n$
contiguous segments, the entropy gain will be on the order of
$\ln{n}$, so the free-energy gain will be bounded from above by ${\rm
  const} - (1/\beta)\ln{n}$, which will be negative for large $n$.
Hence, by means of a local perturbation of the putative ordered phase,
we obtain a state of lower free energy, which implies that the ordered
phase is unstable.

Arguments of this kind depend crucially on both the dimension of the
model and on the range of the interactions.  For instance, they are
inapplicable whenever there exists a possibility that energy may
overwhelm entropy.  Consider, for instance, a $d$-dimensional Ising
model with short-range interactions, where $d \ge 2$.  Then, upon
being presented with the ordered phase, we flip all the spins in a
hypercube of volume $N$, chosen at random out of $n$ contiguous
hypercubes.  Again, this results in the entropy gain of $\ln{n}$.
However, the energy cost of flipping the spins in a hypercube of
volume $N$ will be on the order of its surface area, so that the
free-energy gain will be on the order of ${\rm const} \cdot
N^{(d-1)/d} - (1/\beta)\ln{n}$.  In this case it very well may happen
that the energy shift will offset the entropy change.

\section{Entropy-energy arguments and quantum information theory}
\label{sec:enterg}

Our goal in this chapter is to incorporate entropy-energy arguments
into the framework of quantum information science.  Our starting point
will be Streater's adaptation of an entropy-energy argument, described
in his monograph \cite{str} on nonequilibrium thermodynamics.  We
briefly illustrate his approach in order to set the stage for our own
investigation.

Consider a quantum system $\Sigma$ whose initial state is given by a
density operator $\rho$ and let $T$ be an irreversible discrete-time
dynamics constructed by mixing of reversible evolutions.  Then the von
Neumann entropy $S$ will increase monotonically along the orbit
$\set{T^n(\rho)}$.  Let $H$ be the bounded energy observable
(Hamiltonian) of $\Sigma$ with the property that its spectral
projections are left invariant by $T$.  Then the mean energy is
conserved along the orbit but, since the entropy keeps on growing, the
sequence of iterates $T^n(\rho)$ will eventually converge to the
mixture of microcanonical states on the eigenspaces (energy levels) of
$H$ for any choice of  the initial state $\rho$ (assuming that each
eigenvalue of $H$ has finite geometric multiplicity).  In case of an
irreversible quantum dynamics $T$ that does not conserve energy, the
same argument works as well because, as one can easily show, the
absolute value of the energy shift due to $T$ can be crudely bounded
from above by twice the operator norm of $H$ (i.e., by twice the
largest energy available to $\Sigma$).  

Now we present our twist on Streater's reasoning, as well as the
essence of our method.  While it is clear from the above discussion
that, for sufficiently high temperatures, entropy will eventually
overwhelm energy, we would like to obtain an estimate as to when that
will happen.  In order to do so, we appeal to the Gibbs variational
principle.  Let us, for simplicity, assume that the system $\Sigma$ is
finite, and is maintained at inverse temperature $\beta$.  As we have
already pointed out, the energy shift due to the iterated dynamics
$T^n$ is bounded from above by $2 \norm{H}$. Suppose, further, that we
have the lower bound on the entropy gain due to $T^n$ in the form
$\Delta S \ge f(n)$, where $f$ is an invertible function.  Then the
free energy will change by at most $2 \norm{H} - f(n)/\beta$.  This
number will, in turn, be negative when $f(n) > 2 \beta \norm{H}$.  If
$f$ is an increasing function, then so is the inverse function
$\inv{f}$.  As a consequence of this, entropy will exceed energy for
all $n \ge \inv{f}(2 \beta \norm{H}$).  Thus we see that, in order to
keep the system stable for a long time, we must either increase the
energy or lower the temperature (or do both).  The exact form of the
function $f$ will then allow us to appraise the energy-temperature
trade-off involved in keeping the system stable.

The dynamics of a noisy quantum computer can be pictured as the
competition between the entangling unitary transformations and the
localized errors which tend to destroy entanglement, thereby
increasing entropy \cite{aha}.  Because a large-scale quantum computer
is neither homogeneous in space nor homogeneous in time, there is no
straightforward way to tackle this problem using the formalism of
statistical mechanics of spin systems.  In particular, the usual
notion of the thermodynamic limit no longer applies. Recall, however,
our remark in the preceding section that it is possible to carry out
entropy-energy arguments without passing to the thermodynamic limit.
In our case we can reason as follows.  A large-scale quantum computer
is, for all practical purposes, a macroscopic system.  However, the
only part of this system of any interest to us is comprised by the
degrees of freedom directly involved in the actual computation; the
number of such degrees of freedom is ostensibly finite.  If we can
show that this finite system is not LTS, then the entire macroscopic
computer cannot be GTS.

There are two separate aspects of the thermodynamics of noisy quantum
computers --- the temporal and the spatial.  The temporal aspect
refers to the maximum number of computational operations that can be
carried out before being in any state, other than the microcanonical
(maximally mixed) state, becomes \qt{thermodynamically unfavorable}
for the computational degrees of freedom.  This is, of course, tied
closely to the relaxation time. The spatial aspect concerns the size
of the computational subsystem as measured in qubits --- on a
heuristic level, we can expect that, as the subsystem gets larger,
there is more room for \qt{randomness} in the locations of the errors, so that
it may be possible to show, using a typical entropy-energy argument a
l\'a Thouless, that the computer is not LTS.  We present analyses of
these two aspects in Sections \ref{sec:maxnum} and
\ref{sec:thermostab}, devoting the rest of this section to energy
shift estimates.

We agree at the outset to deal only with the circuit model of quantum
computation, in which case we adopt the model of noisy quantum
computation from Section \ref{sec:noisyqmem}.  Namely, each step of
the computation is the application of a unitarily implemented channel
(quantum gate), followed by an invocation of a fixed noisy channel
$T$.   We take $T$ to be strictly contractive and bistochastic.  In
fact, as we have shown in Section \ref{ssec:sccdense}, if the noise
modeled by $T$ is sufficiently weak (i.e., $\cbnorm{T-\id}< \epsilon$
with $\epsilon$ sufficiently small), then there exists a depolarizing
channel $D_\eta$ such that $\cbnorm{T-D_\eta} < \epsilon$ and $\eta <
\epsilon/{\rm const}$.  

Let us first consider the temporal aspect, in which case we are
interested in the estimate of the energy shift due to $n$ successive
invocations of the noisy channel $T$. In the case when the Hamiltonian
$H$ does not depend on time, the estimate is easy --- we have, for any
density operator $\rho$,
$$
\abs{\Delta E} = \abs{\tr{[T^n(\rho)H]} - \tr{(\rho H)}} \le
\norm{\hT^n(H) - H} \le 2\norm{H}.
$$
If we picture noiseless quantum computation as an evolution governed
by the Schr\"odinger equation, then the corresponding Hamiltonian is
manifestly time-dependent.  Noisy computation could then be described
by a Lindblad master equation \cite{lin}, but the Hamiltonian part of
the corresponding Liouvillian would still be time-dependent.  We can,
however, circumvent this issue for the following reason.  The goal of
\qt{temporal} entropy-energy arguments is to obtain an estimate of the
maximum number $n_{\rm max}$ of computational steps that can be
carried out before the entropy gain due to the repeated invocations of
$T$ overwhelms the energy cost of the computation, which may include
any energy resources required to perform error correction.
Specifically, we are interested in the expression for $n_{\rm max}$ in
terms of energy and temperature.  Therefore we {\em assume} that we
operate the computer under the maximum energy constraint, i.e., the
energy shift can be estimated as $\Delta E \le E_{\rm max}$ for some
given $E_{\rm max}$. Recalling the discussion above, we will then have
$n_{\rm max} \ge \inv{f}(\beta E_{\rm max})$, where $f$ is the
function that figures in the lower bound on the entropy gain due to
$T^n$.

Now we turn to the analysis of the spatial aspect.  Suppose that we
have a quantum computer comprised by a large number of qubits.  Let
$T$ be the channel that models the decoherence of a single qubit in the
computer.  Imagine picking, at random, one out of $n$ disjoint
$k$-qubit sets and applying the channel $T^{\tp{}{k}}$ to the qubits
in this set.  Suppose that the energy shift due to this local
perturbation is independent of $k$ and $n$.  Then we want to show that
if we take $k$ large enough, there will be some finite value of $n$
such that the corresponding entropy gain overwhelms the energy shift.
How can we show that the energy shift can, in fact, be bounded
independently of $k$ and $n$?  We reason as follows. Given an initial
state of $N$ qubits, consider a quantum circuit whose size is
polynomial in $N$.  Each gate in the circuit acts on at most $c$
qubits, where the number $c$ is independent of $N$.  Let $\rho_{s-1}$
be the state of the computer given by
$$
\rho_{s-1} =\left( \prod^{s-1}_{i=1}T \hU_i \right)(\rho_0),
$$
where $\rho_0$ is the input state, and $T$ is a fixed noisy channel.
Suppose that the $s$th gate has been applied, so we have the
transformation $\maps{\rho_{s-1}}{U_s \rho_{s-1}U^*_s}$.  Now, when we
invoke the channel $T$, the corresponding energy shift will be
determined by the Hamiltonian $H_s$, where $U_s = \exp{(-iH_s
  \tau/\hbar)}$ and $\tau$ is the time it takes to apply the $s$th
gate.  Because, by hypothesis, each gate acts on at most $c$ qubits, the
energy shift can be bounded from above by a function of $c$ alone.
This assumption can also be justified on the grounds of \qt{local
  reversibility} \cite{per} (cf. also the \qt{gearbox quantum
  computer} of DiVincenzo \cite{div}).

We can, in fact, put this energy shift estimate in a broader context
of simulation of quantum systems using quantum computers
\cite[pp. 204-212]{nc}.  Again, consider a system of $N$ qubits and
the Hamiltonian
$$
H = \sum^{P(N)}_{k=1} H_k,
$$
where $P$ is some polynomial, and each local interaction $H_k$
involves at most $c$ qubits, the number $c$ being, as before,
independent of $N$.  Then, assuming that we can implement the unitary
evolution generated by each term $H_k$ using a circuit whose size is
polynomial in $c$, we can simulate the unitary evolution generated by
$H$ using high-order approximations \cite{dnbt} of the Lie-Trotter
product formula
$$
e^{A+B} = \lim_{n \rightarrow \infty} \left(e^{A/n}e^{B/n}\right)^n,
$$
for any two matrices $A,B$ of the same shape. The main point is that,
at each step of the simulation, the number of interacting qubits is
bounded from above by a function of $c$ alone.

\section{Entropy-energy balance and the maximum number of operations}
\label{sec:maxnum}

In the preceding section we have argued that, as far as the temporal
entropy-energy arguments are concerned, we may assume that the energy
shift due to $n$ invocations of the noisy channel $T$ can be bounded
from above by some constant $E_{\rm max}$, which can be thought of as
the energy resources available for the computation.  In order to
proceed with the entropy-energy argument, we need the estimate of the
entropy gain due to $T^n$.  We state some preliminaries first.

Let $\map{T}{\cS(\sH)}{\cS(\sH)}$ be a channel.  Extending the map $T$
to all of $\cB(\sH)$ and treating the latter as a Hilbert space with the
Hilbert-Schmidt inner product $\inp{A}{B} \defeq \tr{(A^* B)}$, we see
that the Heisenberg-picture channel $\hT$ coincides with the adjoint
operator $T^*$. Indeed, let $\set{V_\alpha}$ be a Kraus decomposition
of $T$.  Then we have, for any $A,B \in \cB(\sH)$,
$$
\inp{A}{T(B)} = \sum_\alpha \inp{A}{V_\alpha B V^*_\alpha} =
\sum_\alpha \tr{(A^* V_\alpha B V^*_\alpha)} = \sum_\alpha
\tr{(V^*_\alpha A^* V_\alpha B)} = \inp{\hT(A)}{B},
$$
which shows that $T^* = \hT$.  Furthermore, if $T$ is bistochastic,
then the map $\hT T$ is also a bistochastic channel:  it is a composition
of two completely positive maps, and its Kraus decomposition
$\set{V^*_\alpha V_\beta}$ has the proper normalization,
\begin{eqnarray}
\sum_{\alpha,\beta}(V^*_\alpha V_\beta)^* V^*_\alpha V_\beta &=&
\sum_\beta V^*_\beta \left(\sum_\alpha V_\alpha V^*_\alpha \right)
V_\beta = \idty \nonumber \\
\sum_{\alpha,\beta} V^*_\alpha V_\beta (V^*_\alpha V_\beta)^* &=&
\sum_\alpha V^*_\alpha \left(\sum_\beta V_\beta V^*_\beta \right)
V_\alpha = \idty, \nonumber
\end{eqnarray}
where we have used the fact that $T$ is bistochastic. In addition, the
map $\hT T$ is self-adjoint in the sense that $\inp{A}{\hT T(B)} =
\inp{\hT T(A)}{B}$, and hence diagonalizable.  Furthermore, $\hT T$ is
a positive operator\footnote{Here we mean operator positivity in
  the usual Hilbert-space sense, not in the sense that $\hT T$ maps
  positive operators to positive operators, which it obviously does.}
because, for any $A \in \cB(\sH)$,
$$
\inp{A}{\hT T(A)} = \inp{T(A)}{T(A)} = \tr{\left[ T(A)^* T(A) \right]} \ge 0.
$$
Because the absolute values of the eigenvalues of any completely
positive map do not exceed unity \cite{tdv}, we conclude that the
spectrum of $\hT T$ is contained in the interval $[0,1]$ of the real
line.  We will need some additional ergodic and spectral properties
for the channel $T$, which we summarize in the following definition.

\begin{definition} \label{def:specgap}
Let $\map{T}{\cB(\sH)}{\cB(\sH)}$ be a bistochastic channel.  We say
that $T$ is {\em ergodic with spectral gap $\gamma$} if $\idty$ is the
only fixed point of $T$ in $\cB(\sH)$, and the spectrum of the channel
$\hT T$ is contained in the set $[0,1-\gamma] \cup \set{1}$.
\end{definition}
Our starting point will be the following entropy gain estimate due to
Streater \cite{str2}.

\begin{lemma}
{\bf (Streater)}
Let $\sH$ be a Hilbert space of finite dimension $d$.  If
$\map{T}{\cB(\sH)}{\cB(\sH)}$ is a bistochastic channel which is
ergodic and has spectral gap $\gamma$, then for any $\rho \in
\cS(\sH)$
\begin{equation}
S(T(\rho))-S(\rho) \ge \frac{\gamma}{2} \pnorm{\rho - \idty/d}{2}^2.
\label{eq:streater2}
\end{equation}
\label{lm:streater2}
\end{lemma}

\begin{proof}
Given a bistochastic channel $T$, a theorem of Alberti and Uhlmann
\cite{au} says that, for any $\rho$, there exist unitaries $U_\alpha$
and nonnegative numbers $p_\alpha$ with $\sum_\alpha p_\alpha = 1$
such that $T(\rho) = \sum_\alpha p_\alpha U_\alpha \rho U^*_\alpha$.
Define $\rho_\alpha \defeq U_\alpha \rho U^*_\alpha$. By Lemma
\ref{lm:streater} we have, for each $\alpha$,
$$
\tr{\left[\rho_\alpha \ln{\rho_\alpha} - \rho_\alpha
  \ln{T(\rho)}\right]} \ge \frac{1}{2}\pnorm{\rho_\alpha -
  T(\rho)}{2}^2.
$$
Becuase $\rho_\alpha$ and $\rho$ are unitarily equivalent, we have
$S(\rho_\alpha) = S(\rho)$ for all $\alpha$, and thus
$$
\sum_\alpha p_\alpha \tr{\left[\rho_\alpha \ln{\rho_\alpha} -
  \rho_\alpha \ln{T(\rho)}\right]} = S(T(\rho)) - S(\rho),
$$
which yields
\begin{equation}
S(T(\rho)) - S(\rho) \ge \frac{1}{2} \sum_\alpha p_\alpha
\pnorm{\rho_\alpha - T(\rho)}{2}^2.
\label{eq:strstep1}
\end{equation}
We can rewrite Eq.~(\ref{eq:strstep1}) as
\begin{eqnarray}
S(T(\rho)) - S(\rho)& \ge & \frac{1}{2} \sum_\alpha p_\alpha \left[
  \inp{\rho_\alpha}{\rho_\alpha} - \inp{\rho_\alpha}{T(\rho)} -
  \inp{T(\rho)}{\rho_\alpha} + \inp{T(\rho)}{T(\rho)} \right]
\nonumber \\
&=& \frac{1}{2} \left[ \inp{\rho}{\rho} -
  \inp{T(\rho)}{T(\rho)}\right] \nonumber \\
&=& \frac{1}{2} \inp{\rho}{(\id - \hT T)(\rho)}. \label{eq:strstep2}
\end{eqnarray}
Because 1 is a simple eigenvalue of $\hT T$, we can write $\rho$ as a
direct sum $\idty/d \oplus (\rho - \idty/d)$. Then
Eq.~(\ref{eq:strstep2}) becomes
\begin{eqnarray}
S(T(\rho)) - S(\rho) &\ge & \frac{1}{2} \inp{\rho - \idty/d}{(\id -
  \hT T)(\rho - \idty/d)} \nonumber \\
&\ge & \frac{\gamma}{2} \pnorm{\rho - \idty/d}{2}^2, \nonumber 
\end{eqnarray}
where the last inequality follows from the fact that $\gamma$ is the
smallest nonzero eigenvalue of $\hT T$.  We thus obtain
Eq.~(\ref{eq:streater2}), and the lemma is proved.
\end{proof}

\begin{rem}
Notice that the theorem of Alberti and Uhlmann cited in the proof
above does {\em not} imply that a channel is bistochastic if and only
if it is a convex combination of unitarily implemented channels.
While it is obvious that any convex combination of unitary
conjugations is a bistochastic channel, Landau and Streater \cite{ls}
showed that the converse is not true in general when the dimension of
the underlying Hilbert space is greater than 2.  The Alberti-Uhlmann
theorem only says that if $T$ is a bistochastic channel, then for each
$\rho \in \cS(\sH)$ there exist {\em $\rho$-dependent} unitaries
$U_\alpha$ and nonnegative weights $p_\alpha$ such that $T(\rho) =
\sum_\alpha p_\alpha U_\alpha \rho U^*_\alpha$.
\end{rem}

Lemma \ref{lm:streater2} says that the von Neumann entropy is a
strict Liapunov function for any bistochastic channel
which is ergodic with a spectral gap. Hence Theorem \ref{th:liapunov} can be
applied to show the trace-norm convergence of the orbit
$\set{T^n(\rho)}$ to the maximally mixed state $\idty/d$ for any $\rho
\in \cS(\sH)$; the rate of convergence is controlled by the spectral
gap. The next result shows the connection between ergodic bistochastic
channels on $\cM_2$ and strictly contractive channels.

\begin{theorem}
Let $\map{T}{\cM_2}{\cM_2}$ be a bistochastic channel.  If $T$ is
strictly contractive, then it is ergodic with spectral gap $\gamma = 1
- k^2$, where $k$ is the contractivity modulus.  Conversely, if $T$ is
ergodic with spectral gap $\gamma$, then it is strictly contractive
with $k = \sqrt{1-\gamma}$.
\label{th:specgap}
\end{theorem}

\begin{proof}
First we need a lemma.

\begin{lemma}
If $\map{T}{\cB(\sH)}{\cB(\sH)}$ is a bistochastic channel, then so is
the dual map $\map{\hT}{\cB(\sH)}{\cB(\sH)}$ in the sense that $\hT$
is a completely positive trace-preserving unital map.
\label{lm:superdual}
\end{lemma}

\begin{proof}
Let $\set{V_\alpha}$ be a Kraus decomposition of $T$, so that, for any
$X \in \cB(\sH)$, we have $T(X) = \sum_\alpha V_\alpha X V^*_\alpha$.
Then we have $\sum_\alpha V^*_\alpha V_\alpha = \idty$ because $T$ is
trace-preserving, and also $\sum_\alpha V_\alpha V^*_\alpha = \idty$
because $T$ is bistochastic.  Now, for any $A \in \cB(\sH)$, we have
$$
\tr{\hT(A)} = \tr{\left(\sum_\alpha V^*_\alpha A V_\alpha \right)} =
\tr{\left(A \sum_\alpha V_\alpha V^*_\alpha \right)} = \tr{A}.
$$
Hence $\hT$ is a completely positive trace-preserving unital map,
i.e., a bistochastic channel.
\end{proof}

\noindent{Now let $\map{T}{\cM_2}{\cM_2}$ be a strictly contractive
  bistochastic channel, and let $\cV$ be its interaction algebra
  (cf. Section \ref{ssec:nonns}). Theorem \ref{th:nns_bist} then says
  that $\cV' = \bbc \idty$, where $\cV'$ is the commutant of $\cV$.
  By Lemma \ref{lm:superdual}, the Heisenberg-picture channel $\hT$ is
  trace-preserving, and leaves invariant the maximally mixed state
  $\idty/2$, which is an invertible matrix. Then, according to the
  Fannes-Nachtergaele-Werner theorem (cf. Refs.~\cite{bjkw} and
  \cite{fnw} and the remark after Theorem \ref{th:nns_bist}), the set
  of operators in $\cB(\sH)$ left invariant by $T$ is precisely the
  commutant $\cV'$.  Thus the only operators in $\cM_2$ that are left
  invariant by $T$ are the complex multiples of the identity matrix,
  i.e., $T$ is ergodic. (In fact, the same argument can be used to
  show that any bistochastic strictly contractive channel is ergodic.)}

Now recall from Section \ref{ssec:sccqubit} that if $T$ is a
bistochastic channel on $\cM_2$, then there exists a real $3\times 3$
matrix $\tT$ such that, for any density operator
\begin{equation}
\rho = \frac{1}{2} (\idty + \mathbf{w}\cdot \bbsigma),
\label{eq:denop}
\end{equation}
we will have
$$
T(\rho) = \frac{1}{2} \left[\idty + (\tT \mathbf{w})\cdot \bbsigma\right].
$$
Furthermore, if $T$ is strictly contractive, then the contractivity
modulus $k$ is precisely the operator norm (the largest singular
value) $\norm{\tT}$ of $\tT$.  Now consider the channel $\hT T$, whose
action on the density operator (\ref{eq:denop}) is given by
$$
(\hT T)(\rho) = \frac{1}{2} \left[\idty + (\trn{\tT} \tT
\mathbf{w})\cdot \bbsigma \right].
$$
Since $T$ is strictly contractive, so is $\hT T$; this follows from
the fact that, for any two density operators $\rho,\rho'$, we have
$$
\trnorm{(\hT T)(\rho - \rho')} \le \trnorm{T(\rho - \rho')} \le k
\trnorm{\rho - \rho'}.
$$
Then the discussion above applies, and $\hT T$ is ergodic.
Furthermore, since $\hT T$ is self-adjoint, its second largest
eigenvalue equals $\norm{\trn{\tT}\tT}$.  But $\norm{\trn{\tT}\tT} =
\norm{\tT}^2 = k^2$, which implies that $1 - \gamma = k^2$.  Thus we
have shown that if $T$ is a bistochastic strictly contractive channel,
then it is ergodic with the spectral gap $\gamma = 1 - k^2$.  We skip
the proof of the converse statement, because it is quite similar to
this argument.
\end{proof}

\noindent{As far as the entropy gain estimate goes, Theorem
  \ref{th:specgap} has the following useful corollary.}

\begin{corollary}
Let $\sH \simeq (\bbc^2)^{\tp{}{N}}$, and consider the channel $T
\defeq R^{\tp{}{N}}$, where $\map{R}{\cM_2}{\cM_2}$ is a strictly
contractive bistochastic channel.  Then, for any $\rho \in \cS(\sH)$
and for any positive $n$, we have
\begin{equation}
S(T^n(\rho)) - S(\rho) \ge \frac{1 - k^{2n}}{2} \pnorm{\rho - \idty/2^N}{2}^2,
\label{eq:entgain}
\end{equation}
where $k$ is the contractivity modulus of $R$.
\label{cor:entgain}
\end{corollary}

\begin{proof}
Because $R$ is bistochastic and strictly contractive, so is $T$
(cf. Section \ref{ssec:sccqubit}). Hence, $T$ is ergodic.  Furthermore,
the contractivity modulus of $T$ equals that of $R$.  Therefore, in
order to prove Eq.~(\ref{eq:entgain}), all we need to do is to
estimate the spectral gap of $T^n$ and then apply Lemma
\ref{lm:streater2}.

If $k$ is the contractivity modulus of $R$, then the contractivity
modulus of $R^n$ (and hence of $T^n$) is at most $k^n$.  Now if
$1-\gamma$ is the second largest eigenvalue of $\hR^n R^n$, then
Theorem \ref{th:specgap} implies that $\gamma \ge 1- k^{2n}$. But
$$
\hT^n T^n = (\hR^{\tp{}{N}})^n (R^{\tp{}{N}})^n = (\hR^n)^{\tp{}{N}}
(R^n)^{\tp{}{N}} = (\hR^n R^n)^{\tp{}{N}},
$$
so the second largest eigenvalue of $\hT^n T^n$ equals that of $\hR^n
R^n$.  Thus the spectral gap of $T^n$ is at least $1-k^{2n}$, and the
corollary is proved.
\end{proof}

Now we are ready to proceed with our entropy-energy argument.  Suppose
that we have a quantum computer operating on $N$ qubits at the inverse
temperature $\beta$.  We assume that the input to the computer is
given by an effective pure state
\begin{equation}
\rho = (1-\epsilon) \idty/2^N + \epsilon \ketbra{\psi}{\psi}.
\label{eq:pureef}
\end{equation}
Let us adopt the typical model of {\em local stochastic noise}
\cite{aha}, meaning that the noisy channel $T$ has the form
$R^{\tp{}{N}}$, where $R$ is a bistochastic channel on $\cM_2$. We
will further assume that $R$ is strictly contractive. Given the
\qt{energy resources,} or maximum allowed energy cost, $E_{\rm max}$,
we are interested in the maximum number $n_{\rm max}$ of computational
steps that can be carried out before the balance of energy vs. entropy
tips in favor of the latter.  We claim that if $\beta$, $E_{\rm max}$,
$\epsilon$, and $N$ are such that
\begin{equation}
\beta E_{\rm max} < \frac{\epsilon^2(1-2^{-N})}{2},
\label{eq:entempcond}
\end{equation}
then
\begin{equation}
n_{\rm max} = \frac{\log{\left\{1-2 \beta E_{\rm max}/[\epsilon^2
      (1-2^{-N})]\right\}}}{2 \log{k}},
\label{eq:nmax}
\end{equation}
where $k$ is the contractivity modulus of $R$.  Indeed, we have
$$
\pnorm{\rho - 2^{-N}\idty}{2}^2 = \epsilon^2
\tr{\left(\ketbra{\psi}{\psi} - 2^{-N}\idty \right)^2} = \epsilon^2(1
- 2^{-N}).
$$
Using this result and Corollary \ref{cor:entgain}, we can bound the
entropy gain due to $T^n$ from below as follows:
$$
S(T^n(\rho)) - S(\rho) \ge \frac{\epsilon^2 (1-k^{2n})(1-2^{-N})}{2}.
$$
The corresponding change in the free energy is then
$$
2 \beta \Delta F \le 2 \beta E_{\rm max} - \epsilon^2 (1-k^{2n})(1-2^{-N}).
$$
The right-hand side of this expression will be negative
when $n > n_{\rm max}$ with $n_{\rm max}$ given by
Eq.~(\ref{eq:nmax}).    When the number $N$ of qubits in the computer
is very large, and when $\epsilon$ does not depend on $N$, the term
$2^{-N}$ can be neglected, so Eqs.~(\ref{eq:entempcond}) and
(\ref{eq:nmax}) become respectively
$$
\beta E_{\rm max} < \frac{\epsilon^2}{2}
$$
and
$$
n_{\rm max} = \frac{\log {(1 - 2 \beta E_{\rm max}/\epsilon^2)}}{2 \log{k}}.
$$

The condition (\ref{eq:entempcond}) restricts the range of
applicability of our entropy-energy argument to low-energy
quantum computers that are operated at high temperatures.  We can,
however, dispense with Eq.~(\ref{eq:entempcond}) altogether when the
noisy channel $T$ has the form
$$
T = (1-\delta)\id + \delta R^{\tp{}{N}}
$$
for some small positive $\delta$, where $R$ is a strictly contractive
bistochastic channel on $\cM_2$ with the contractivity modulus $k$.
Then we have
$$
T^n = \sum^n_{l=0} {n \choose l} (1-\delta)^{n-l}\delta^l (R^l)^{\tp{}{N}}.
$$
Let the input state be given by Eq.~(\ref{eq:pureef}) and suppose, as
before, that $N$ is sufficiently large and that $\epsilon$ does not
depend on $N$. Then, using the concavity of the von Neumann entropy as
well as Corollary \ref{cor:entgain}, we obtain the bound
\begin{eqnarray}
\Delta S \equiv  S(T^n(\rho)) - S(\rho) &\ge & \frac{\epsilon^2}{2}
\sum^n_{l=0} {n \choose l} (1-\delta)^{n-l} \delta^l (1-k^{2 l})
\nonumber \\
&= & \frac{\epsilon^2}{2} \left[ 1 - (1-\delta + \delta k^2)^n \right]
\nonumber \\
&=& \frac{\epsilon^2}{2} n\delta(1-k^2) + o(\delta), \nonumber
\end{eqnarray}
where $o(\delta)$ stands, as usual, for terms that go to zero faster
than $\delta$ as $\delta \rightarrow 0$.  By the assumed smallness of
$\delta$, we may neglect these terms.  Then the free-energy increment
can be calculated from
$$
2 \beta \Delta F \le 2 \beta E_{\rm max} - \epsilon^2 n \delta(1-k^2),
$$
so that we obtain
$$
n_{\rm max} = \frac{2 \beta E_{\rm max}}{\delta \epsilon^2 (1-k^2)}.
$$
This formula shows clearly that, in order to keep the computer stable,
we must either lower the temperature or raise the energy (or, perhaps,
do both).  The good news is that, when the noise is weak, the required
thermodynamic resources are polynomial in the number of operations.

\section{Thermodynamic stability of large-scale quantum computers}
\label{sec:thermostab}

In Section \ref{sec:maxnum} we used an entropy-energy argument to
estimate the maximum number of operations that can be carried out on a
noisy quantum computer before the cumulative entropy gain due to
decoherence overwhelms the energy cost of the computation, including
error correction.  Here we
consider the spatial aspect of decoherence in quantum computers and show,
using an entropy-energy argument, that there exists an upper bound on
the number of qubits that can be accommodated by circuit-based quantum
computers.

What we present here is an adaptation of the classical entropy-energy
argument (cf. Simon and Sokal \cite{ss}) that shows the absence of
ordering in a one-dimensional Ising ferromagnet with short-range
interactions.  The main ingredient of the Simon-Sokal argument is the
following entropy gain estimate.  Consider a finite portion of a chain
of classical Ising spins divided into $n$ disjoint segments, each
containing $k$ spins.  Now suppose that we choose, at random, a
$k$-spin segment and flip all of the spins in it.  The resulting
entropy gain is easily seen to be on the order of $\ln{n}$.  Simon and
Sokal then give a lower bound on the entropy gain when the segments
are \qt{almost disjoint,} in the sense that the corresponding
configurations can be associated with probability measures with
\qt{minimally overlapping} supports.  The following lemma and its proof are a
straightforward adaptation of a similar result due to Simon and Sokal
\cite{ss}.

\begin{lemma}
\label{lm:entmix}
{(\bf entropy of statistical mixtures)}. Let $\rho_i, i =1,\ldots,n,$
be $n$ mutually commuting density operators.  Suppose that there
exists a constant $\kappa \ge 0$ such that, for each $i$, we have
\begin{equation}
\tr{\sum_{j \neq i} \rho_j \rho_i} \le \kappa.
\label{eq:almostdisj}
\end{equation}
Let $\rho = \inv{n} \sum^n_{i=1}\rho_i$.  Then
\begin{equation}
S(\rho) \ge \inv{n} \sum^n_{i=1}S(\rho_i) + \ln n - 2 \kappa^{1/2}.
\label{eq:ssentgain}
\end{equation}
\end{lemma}

\begin{proof} Let $\set{q^{(i)}_\alpha}$ be the eigenvalues of
  $\rho_i$.  We have then
\begin{eqnarray}
S(\rho) &-& \inv{n} \sum_i S(\rho_i) - \ln{n} = - \inv{n} \sum_i \tr
{\rho_i \left( \ln{\sum_j \rho_j} - \ln{\rho_i} \right) } \nonumber \\
&=& - \inv{n} \sum_i \sum_\alpha q^{(i)}_\alpha
	\ln{
		\frac
			{\sum_j q^{(j)}_\alpha} {q^{(i)}_\alpha}
	} = - \inv{2n} \sum_i \sum_\alpha q^{(i)}_\alpha \ln{
	\left( \frac
			{\sum_j q^{(j)}_\alpha}
			{q^{(i)}_\alpha}\right)^{1/2}} \nonumber \\
& \ge & - \inv{2n} \sum_i \ln \sum_\alpha \left(\left(q^{(i)}_\alpha
	\right)^2 + \sum_{j \neq i} q^{(j)}_\alpha q^{(i)}_\alpha
	\right)^{1/2} \label{step1} \\
& \ge & - \inv{2n} \sum_i \ln \left(1 + \sum_\alpha \left(\sum_{j \neq
	  i} q^{(j)}_\alpha q^{(i)}_\alpha \right)^{1/2} \right)
	\label{step2} \\
& \ge & - \inv{2n} \sum_i \sum_\alpha \left(\sum_{j \neq i}
	q^{(j)}_\alpha q^{(i)}_\alpha \right)^{1/2} \label{step3} \\
& = & -\inv{2n} \sum_i \tr{\left( \sum_{j \neq i}\rho_j \rho_i
	\right)^{1/2}}, \label{finres}
\end{eqnarray}
where (\ref{step1}) is a consequence of Jensen's inequality \cite{hlp}
and the convexity of the function $\maps{x}{-\ln{x}}$, (\ref{step2})
uses $(a + b)^{1/2} \le a^{1/2} + b^{1/2}$, and (\ref{step3}) uses
$\ln{(x+1)} \le x$.  For $i$ fixed, we have, for the trace in
(\ref{finres}),
\begin{equation}
\tr{\left( \sum_{j \neq i}\rho_j \rho_i \right)^{1/2}} \le
\left(\tr{\sum_{j \neq i}\rho_j \rho_i }\right)^{1/2},
\label{finres2}
\end{equation}
which follows from the concavity of the function $\maps{x}{\sqrt{x}}$
and the self-adjointness of the operator $\sum_{j \neq i}\rho_j
\rho_i$ (the latter is a consequence of the fact that $\rho_i$ are
mutually commuting).  The right-hand side of (\ref{finres2}) can now
be bounded from above by $\kappa^{1/2}$, and the lemma is proved.
\end{proof}

\begin{rem}
The requirement that the density operators $\rho_i$ be
\qt{almost disjoint} is rigorously borne out by
Eq.~(\ref{eq:almostdisj}).
\end{rem}

Consider now a quantum computer operating on $N$ qubits.  In Section
\ref{sec:enterg} we have already noted that, during each computational
step, the number of interacting qubits is no greater than some fixed
constant $c$ that depends on the particular algorithm, but not on
$N$.  Hence it follows from the considerations of local reversibility
\cite{per} that the energy shift due to any noisy channel applied right
after the computational step can be bounded above by a function of $c$
alone.  Another important consequence of local reversibility is the
observation that we can partition the set $Q \defeq \set{1,\ldots,N}$
into disjoint sets $C_l$, $l \in \set{1,\ldots,L}$, such that, at some
stage of the computation, the overall state of the computer will be a
separable state of the form
\begin{equation}
\rho = \bigotimes^L_{l =1} \rho_l,
\label{eq:sepst}
\end{equation}
where $\rho_l$ is a state of the qubits indexed by elements of
$C_l$.  This observation can be justified as follows.  An input
state for a typical quantum network is the pure state
$$
\ket{\psi} = \frac{1}{\sqrt{2^N}} \sum^{2^N-1}_{m=0} \ket{m} =
\left(\frac{1}{\sqrt{2}}\ket{0} +
\frac{1}{\sqrt{2}}\ket{1}\right)^{\tp{}{N}},
$$
which is manifestly separable.  The quantum network responsible for
the computation consists of one- and two-qubit gates, which explains
the formation of disjoint clusters of qubits.  Furthermore, we can
picture the noisy computation as a competition between the entangling
gates, which tend to form clusters of qubits, and the errors, which
tend to detach qubits from clusters \cite{aha}.  We assume, for
simplicity, that the clusters all have the same size $d$, so that $N =
Ld$.

Now suppose that we are given two positive integers, $k$ and $n$, so
that $L \ge nk$.  Thus our computer is operating on at least $knd$
qubits.  Partition the set $\set{1,\ldots,nk} \subseteq
\set{1,\ldots,L}$ into $n$ disjoint $k$-element subsets $S_i, i =
1,\ldots,n$.  For some $\eta \in (0,1)$, let $D^{(l)}_\eta$ denote the
depolarizing channel
$$
D_\eta(A) \defeq (1-\eta)A + \eta 2^{-d} (\tr{A})\idty
$$
acting on the qubits in the cluster $C_l$. For each $i$, let $T_i$ be
the channel that acts as $\left(D^{(l)}_\eta\right)^{\tp{}{k}}$ on the
qubits in $\bigcup_{l \in S_i} C_l$, and is the identity channel on
the rest of the qubits.  For the state $\rho$ given by
Eq.~(\ref{eq:sepst}), define the density operators $\rho_i \defeq
T_i(\rho),1 \le i \le n$.  Then we have $[\rho_i, \rho_j]=0$ for all
$i,j$.  We also note the following elementary estimate:
$$
\sum_{j \neq i} \tr{(\rho_i \rho_j)} = \sum_{j \neq i}
\tr{\left[T_i(\rho) T_j(\rho)\right]} \le (n-1) \left(1-\eta +
\frac{\eta}{2^d}\right)^{2k}.
$$
Given $n$, $d$, and $\eta$, we can always find such a value of $k$ that
$$
\left(1 - \eta + \frac{\eta}{2^d}\right)^{2k} \le \frac{1}{n-1},
%\label{eq:kcond}
$$
so that the condition (\ref{eq:almostdisj}) of Lemma \ref{lm:entmix}
is satisfied by the $\rho_i$'s with $\kappa = 1$.  Hence the entropy
gain due to the channel $T \defeq n^{-1}\sum^n_{i=1}T_i$ can be
bounded from below as follows:
$$
S(T(\rho)) - S(\rho) \ge \frac{1}{n} \sum^n_{i=1} S(\rho_i) - S(\rho)
+ \ln{n} - 2 \ge \ln{n} - 2,
$$
where the last inequality follows from extensivity of the von Neumann
entropy and from the fact that we can write
$$
\rho = \bigotimes^L_{l =1}\rho_l = \bigotimes^n_{i=1}
\left(\bigotimes_{l \in S_i}\rho_l \right).
$$
The corresponding free-energy increment can then be bounded by
\begin{equation}
\beta \Delta F \le \beta E(c) - \ln n + 2,
\label{eq:fegain}
\end{equation}
where $E(c)$ is the $c$-dependent upper bound on the energy shift. We
can certainly pick $n$ so large that the right-hand side of
Eq.~(\ref{eq:fegain}) is negative.  With the appropriate choice of
$k$, we see that, if the computer operates on at least $nkd$ qubits,
then it is possible to \qt{depolarize} a randomly chosen set of qubit
clusters in such a way that the resulting entropy gain will overwhelm
the corresponding energy shift.\\

\begin{rem}
In order for our entropy-energy argument to work, it is crucial that
the energy shift be bounded independently of $n$ and $k$ (the latter
actually depends on $n$).  As we have argued above, this bound can be
justified for circuit-based quantum computation.  We also draw the
reader's attention to the following caveat.  The validity of the
entropy-energy argument just presented rests chiefly on the assumption
that, at some stage in the computation, the state of the computer can
be written in the form (\ref{eq:sepst}).  This is certainly true for
circuit-based computers, either because the initial state is
completely separable, or because an initially entangled state is
rendered separable by noise.  
\end{rem}

Notice that up to now we have left unspecified the value of the
depolarization rate $\eta$ without sacrificing much of our argument.
We can, however, pick $\eta$ in such a way that the state of the
computer will be separable with high probability.  This follows from
the work of Aharonov \cite{aha}, who showed that when the the computer
is realized as a circuit in $d+1$ dimensions, there exists a critical
value $\eta_c \in [1/3,1/2^{1/d}]$ such that, for all $\eta > \eta_c$,
the state of the computer will eventually end up in the form
(\ref{eq:sepst}) with a large value of $L$.  Our entropy-energy
argument can therefore be used to elucidate the thermodynamical
underpinnings of the process by which noisy quantum computers tend
towards essentially classical behavior.

\section{Putting it all in perspective}
\label{sec:eepersp}

The results reported in this chapter help shed some light on the
thermodynamics of noisy quantum computation.  We have seen, in
particular, that, when the noise affecting the system is modeled by a
bistochastic strictly contractive channel, there is an intimate
connection between the contraction rate and the rate of entropy
production.  This lends further support to our claim that strictly
contractive channels serve as a physically reasonable model of
relaxation processes in noisy quantum memories and computers.

We also showed that there is an upper bound on the number of qubits
that can be accommodated in a circuit-based quantum computer.  The
existence of this bound was shown under two crucial assumptions:  (a)
that the energy shift due to a single step of the noisy dynamics does
not depend on the number of qubits, and (b) that, at some point, the
state of the computer is separable.  As we have emphasized, both of
these assumptions are justified for the case of quantum circuits.
However, this limitation on the size of the computer is significant
only at high temperatures, as can be easily seen from
Eq.~(\ref{eq:fegain}).  On the other hand, Ozawa showed recently
\cite{oz2} that conservation laws impose a lower bound on the size of
quantum computers.  Unlike the competing upper bound implied by our
entropy-energy argument, Ozawa's bound is independent of temperature,
which suggests that operating quantum computers at low temperature may
go a long way towards curtailing the effects of decoherence.  

Another approach to stabilization of quantum memories and computers
would call for replacing circuit-based computation with the following
procedure \cite{br2,yue2}: upon preparing a multiparticle entangled
state, a suitable set of observables is measured on it, and the
measurement results are processed on a classical computer.  While any
computation performed using this technique is manifestly irreversible,
it has the advantage of being less susceptible to the effects of noise
by virtue of cutting down the computation time.

%% file: chap5.tex
\chapter{Information storage in quantum spin systems}
\label{ch:spinsys}

%\vskip 1in

In Chapter \ref{ch:enterg} we have considered the situation when the
effect of noise is such that the entropy produced exceeds the
resulting energy shift, at which point it becomes thermodynamically
unfavorable for the computer to be in any state other than the
maximally mixed (microcanonical) state, or a mixture of such states on
the energy eigenspaces.  In this chapter we briefly comment on the
possibility of reliable storage of information in quantum spin systems
in which there eixsts a possibility of a phase transition.  In this
case it can be shown, using the so-called Peierls argument
\cite{fer,gri,pei}, that the energy shift does, in fact, overwhelm the
entropy gain.

Our goal here is to reinterpret the results of rigorous perturbation theory
for quantum spin systems in the context of quantum information
processing. We hope that these preliminary findings might spur further
research into this topic both in the quantum information community and
in the statistical mechanics community.  The contents of this chapter follow
Ref.~\cite{rag3} essentially verbatim.

\section{Toric codes and error correction on the physical level}
\label{sec:torcodes}

Error correction is a key ingredient in any good recipe for quantum
information processing.  Many ingenious schemes have been invented to
that effect. A particularly interesting approach has been suggested by
Kitaev and colleagues in a series of beautiful papers
\cite{bk,dklp,kit}, namely the possibility of implementing quantum
error correction {\em on the physical level}.

\begin{figure}
\includegraphics{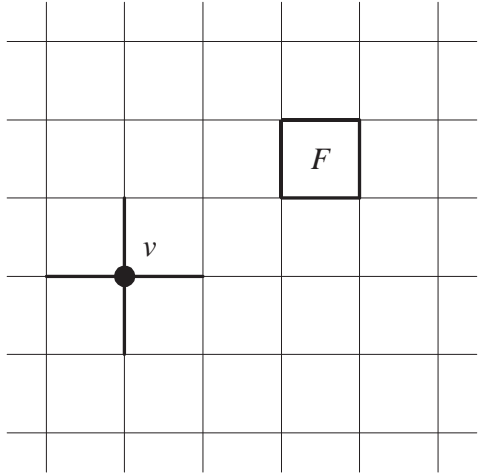}
\caption{Square lattice on a torus.}
\label{fig:toric}
\end{figure}

Consider a $k \times k$ square lattice $\Lambda$ on the torus
$\bbz^2/\bbz$.  Associate a qubit with each edge of $\Lambda$, for a
total of $n= 2 k^2$ qubits.  We can identify two kinds of geometric
objects on $\Lambda$,  {\em vertices} and {\em faces};
Fig.~\ref{fig:toric} shows a portion of the lattice together with a
vertex $v$, a face $F$, and the edges incident with $v$ and $F$.  It
is easy to see that there are $k^2$ vertices and $k^2$ faces.

Given a vertex $v$, we denote by $\Sigma(v)$ the set of all edges of
$\Lambda$ incident with $v$.  Similarly, given a face $F$, we
let $\partial F$ denote the boundary of $F$.  For any $v$, $\Sigma(v)$
contains exactly four edges; the same can be said of $\partial F$ for
any $F$.  Now define the {\em verification operators}
$$
A_v \defeq \bigotimes_{e \in \Sigma(v)} \paulix^{e},\qquad B_F \defeq
\bigotimes_{e \in \partial F} \pauliz^{e},
$$
where $\sigma^e_i$ denotes the Pauli matrix $\sigma_i$ acting on
the Hilbert space of the qubit associated to the edge $e$. It is easy to
see that all of these $n$ operators commute with each other, and are
self-adjoint with eigenvalues $\pm 1$.  

Let $\sH$ be the Hilbert space of the $n$ qubits on the lattice and
consider the {\em protected subspace}
$$
\sK \defeq \setcond{\psi \in \sH}{A_v \psi = \psi, B_F \psi = \psi
  \quad \forall v,F}.
$$
There are two relations connecting the operators $A_v$ and $B_F$,
namely $\prod_v A_v = \idty_\sH$ and $\prod_F B_F =
\idty_\sH$. Hence there are $m = n - 2$ independent verification
operators.  Using the theory of the so-called {\em stabilizer codes}
\cite{got}, it can be shown that the dimension of the protected
subspace is equal to $2^{n-m} = 4$.

In Ref.~\cite{kit1}, Kitaev proposed the following approach to quantum error
correction.  He considered the Hamiltonian
\begin{equation}
H_\Lambda \defeq -\sum_v A_v - \sum_F B_F,
\label{eq:kithamil}
\end{equation}
where the summations run over all the vertices and faces of $\Lambda$.
Note that this Hamiltonian is formed by 4-spin interactions, namely
the verification operators. The ground state of the Hamiltonian
(\ref{eq:kithamil}) is fourfold degenerate, and the corresponding
eigenspace is precisely the protected subspace $\sK$.  We can
therefore store a state of 2 qubits as a vector in $\sK$.  

Addition of a small local perturbation given by the sum of certain
single-spin terms and 2-spin interactions (cf. Ref.~\cite{kit1} for
details) modifies the Hamiltonian $H_\Lambda$ to $H_\Lambda(\epsilon)$, where
$\epsilon$ is the perturbation strength.  The effect of the
perturbation is to introduce an energy splitting between the
degenerate ground-state levels of the unperturbed Hamiltonian.  Kitaev
then argues that there exists a constant $\epsilon_0$ such that, at low
temperatures and for all $\abs{\epsilon} \le \epsilon_0$, the energy
splitting for sufficiently large $\Lambda$ is given by
$\exp{(-ck)}$ for some positive $c$. In other words, in the
thermodynamic limit $\Lambda \uparrow \bbz^2$, the ground state is
still fourfold degenerate, and any sufficiently weak perturbation is
\qt{washed out} by the system itself.

The four-body interactions comprising the Hamiltonian
(\ref{eq:kithamil}) were originally considered by Kitaev in
Ref.~\cite{kit} as a basis for the construction of a family $(2k^2,2)$
stabilizer codes, which he termed \qt{toric codes.} The remarkable
feature of toric codes is the fact that, despite their apparent
nonoptimality (in the sense of Calderbank and Shor \cite{cs}), they
require only local operations for their implementation and can correct
any number of errors (provided that the lattice is large enough). The
bulk of Kitaev's analysis of toric codes was concerned with their
properties as \qt{conventional} quantum error-correcting codes
\cite{kl} that require active intervention through frequent
measurements and other external processing. The issue of constructing
\qt{self-correcting} quantum spin systems on the basis of toric codes
has been taken up again only very recently by Dennis \etal
\cite{dklp}.  Their approach, however, is centered around the topological
features of toric codes and delves deep into such subjects as
nonabelian gauge theory \cite{bgkp,dklp}.

On the other hand, the very idea of physical error correction is so
tantalizing, both practically and conceptually, that one cannot help
but wonder: how generic are phenomena of this kind?  In this chapter
we show that a few results in statistical mechanics of quantum spin
systems point towards the conclusion that physical error correction is
fairly common in such systems, under quite reasonable conditions.

\section{Laying out the ingredients}

First of all, let us agree on the ingredients necessary for the analysis
of a self-correcting quantum spin system.  Let $\Lambda \subset
\bbz^\nu$, where $\nu \ge 2$, be a finite lattice. Let $\sH_0$ be the
$(2S+1)$-dimensional Hilbert space of a single particle of spin
$S$. Spins are situated on the lattice sites $l \in \Lambda$ (in
Kitaev's construction, spins were located on the lattice bonds).  In
order to retain a superficial analogy with stabilizer codes, we will
assume that the unperturbed Hamiltonian $H_\Lambda$ is {\em
  classical}, i.e., the interactions comprising it generate an abelian
subalgebra of the algebra $\cB(\sH_\Lambda)$ of all linear operators
on $\sH_\Lambda \defeq \bigotimes_{l \in \Lambda}\sH_l$, where $\sH_l$
is an isomorphic copy of $\sH_0$.  That is,
$$
H_\Lambda = \sum_{M \subset \Lambda}\Phi_M,
$$
where each $\Phi_M$ is a self-adjoint operator on $\sH_M \defeq
\bigotimes_{l \in M}\sH_l$, and $[\Phi_M,\Phi_N] = 0$. We assume
periodic boundary conditions [that is, the lattice $\Lambda$ is drawn
on the torus $(\bbz/k\bbz)^\nu$, where $k$ is the lattice size]. We
let $\set{\ket{\ul{\sg}}}$ be the orthonormal basis of $\sH_\Lambda$
in which $H_\Lambda$ is diagonal; the basis vectors are labelled by
classical spin configurations, $\ul{\sg} = \set{\sigma_l}_{l \in
  \Lambda}$ with $\sg_l \in \set{-S,-S+1,\ldots,S-1,S}$.  We also
assume that the smallest eigenvalue of $H_\Lambda$ is equal to zero,
and that its geometric multiplicity is $m \ge 2$.  We denote the
corresponding eigenspace by $\sH^g_\Lambda$.

The effect of errors is modeled by introducing an off-diagonal
perturbation term to the Hamiltonian:
$$
H_\Lambda(\epsilon) \defeq H_\Lambda + \epsilon P,
$$
where $\epsilon$ is a positive constant and $P$ is a self-adjoint
operator whose exact form is, for the moment, left unspecified.
Addition of the $\epsilon P$ term will perturb the eigenvalues of
$H_\Lambda$, resulting in energy splitting between orthogonal ground
states of the original (unperturbed) Hamiltonian.  Consequently, we
define
$$
\Delta E_\Lambda(\epsilon) \defeq \max_{ \ket {\ul{\sg}}  \in
  \sH^g_\Lambda} \braket{\ul{\sg}}{H_\Lambda(\epsilon)|\ul{\sg}}.
%\label{eq:esplit}
$$

Thus the basic idea behind a self-correcting quantum spin system boils
down to the following.  Information is stored in the ground-state
eigenspace of the unperturbed Hamiltonian $H_\Lambda$.  The
multiplicity $m$ is, obviously, dictated by the desired storage
capacity:  when $m = 2^k$, our \qt{ground-state memory cell} will hold
$k$ qubits.  Errors will cause some of the information to leak out
into excited states.  In order for error correction to take place, the
system should be able to recover its ground state from sufficiently
weak perturbations at sufficiently low temperatures (the fact that we
have to work with low temperatures is clear since we are dealing with the
ground state).  That is, we hope that there exists a threshold value
$\epsilon_0$ such that
\begin{equation}
\lim_{\Lambda \uparrow \bbz^\nu} \Delta E_\Lambda(\epsilon)=0,\qquad
\epsilon \le \epsilon_0.
\label{eq:esplit_tl}
\end{equation}
However, this condition is necessary but not sufficient for error
correction.  It may happen that the $m$-fold degeneracy of the ground
state does not survive in the thermodynamic limit [this possibility is
borne out by the off-diagonal matrix elements
$\braket{\ul{\sg}}{H_\Lambda(\epsilon)\ul{\sg'}}$, where
$\ket{\ul{\sg}},\ket{\ul{\sg'}} \in \sH^g_\Lambda$]. Therefore we
require that the ground state of the perturbed Hamiltonian remain
$m$-fold degenerate for all $\epsilon \le \epsilon_0$ in the
thermodynamic limit.  In the next section we elaborate further on
these requirements for self-correction and show that they are quite
easy to fulfill in a wide variety of quantum spin systems.

\section{Putting it together}

The main question is:  which restrictions ensue on the unperturbed
Hamiltonian $H_\Lambda$ and on the perturbation $P$?  It turns out
that this question can be answered using the same methods that are
employed for constructing low-temperature phase diagrams for classical
spin systems with quantum perturbations \cite{bku,dff,kt}.  Thus the
Hamiltonian $H_\Lambda$ can be comprised by $n$-spin interactions (for
fixed $n$) that satisfy the Peierls condition \cite{sla}:  the energy
cost of a local perturbation $\omega'$ of a translationally invariant
ground state $\omega$ is on the order of the surface area of the
region that encloses the part of the lattice on which $\omega$ and
$\omega'$ differ.  Additionally, the unperturbed Hamiltonian is
assumed to have a spectral gap $g > 0$ (i.e., its first nonzero
eigenvalue $\ge g$).  Admissible perturbations are formed by sums of
translates of an arbitrary self-adjoint operator $P_0$, whose support
(the set of sites on which the action of $P_0$ is nontrivial) is
finite and encloses the origin of the lattice.  Thus
$$
P = \sum_{l \in \Lambda} P_l,
$$
where $P_l = \gamma_l P_0$ with $\gamma_l$ being the automorphism
induced by the translation of the lattice $\Lambda$ that maps the
origin $0$ to the site $l$ and respects the periodic boundary
conditions.  Also, both the unperturbed and the perturbed Hamiltonians
are assumed to be invariant under unitary transformations induced by a
symmetry group acting transitively on the set $\set{\ket{\ul{\sg}}}
\cap \sH^g_\Lambda$.

Assuming these conditions are satisfied, we invoke a theorem of
Kennedy and Tasaki \cite{kt}, which says that there exists a constant
$\epsilon_0$ such that, for all $\epsilon \le \epsilon_0$, the
perturbed spin system has $m$ translationally invariant ground states
in the thermodynamic limit. Furthermore, if the $m$ translationally
invariant ground states of the unperturbed Hamiltonian are invariant
under some additional symmetries, these invariance properties carry
over to the ground states of the perturbed system.  The threshold
value $\epsilon_0$ of the parameter $\epsilon$ is determined by
developing a low-temperature expansion \cite{gin,kt} of the perturbed
partition function using a modified Lie-Trotter product formula,
$$
\tr{e^{-\beta H_\Lambda(\epsilon)}} = \lim_{N \rightarrow \infty} \tr
   {\left \{ \left[ \left(\idty - \frac{\epsilon P}{N} \right)
       e^{-(1/N)H_\Lambda} \right]^{N\beta} \right\} }.
$$
The trace is expanded in the basis $\set{ \ket{\ul{\sg}}}$, thus
allowing for combinatorial analysis on a \qt{space-time} grid, where
the space axis is labelled by the lattice sites $l$ and the time axis is
labelled by the values $0,(1/N)\beta,(2/N)\beta,\ldots,\beta$. The
perturbation theory is controlled by a suitable coarse-graining of the
time axis, which then allows to determine the threshold value
$\epsilon_0$ that will render the contributions of the perturbation
terms $P_l$ sufficiently small. In fact, the translation-invariance
requirements can be lifted, with the perturbation theory still going
through \cite{kt}.  A similar space-time analysis of the error rate
has been described heuristically by Dennis \etal \cite{dklp}.

Another important issue is the following:  while the infinite-volume
ground state of the perturbed system may retain the degeneracy of the
original (unperturbed) ground state, the degeneracy may be lost when
the lattice has finite size.  This phenomenon, referred to as
{\em obscured symmetry breaking} \cite{kt2}, is characterized by the
fact that the low-lying eigenstates of the finite-system Hamiltonian
converge to additional ground states in the thermodynamic limit.  In
this case we will have, for any finite $\Lambda$, 
$$
\Delta E_\Lambda(\epsilon) > 0.
$$
It is therefore important to obtain an estimate of the convergence
rate in (\ref{eq:esplit_tl}); namely, given some $\delta > 0$, find
$N_0$ such that
$$
\Delta E_\Lambda(\epsilon) < \delta,\qquad \abs{\Lambda} \ge N_0.
$$
Knowing the convergence rate allows us to appraise the resources
needed to implement error correction with the desired accuracy
$\delta$.  In this respect, an estimate of the form
\begin{equation}
\Delta E_\Lambda(\epsilon) = e^{-c \abs{\Lambda}},\qquad \epsilon \le
\epsilon_0,
\label{eq:expcon}
\end{equation}
where the constant $c$ depends on $\epsilon$, would be ideal --- an
exponential gain in error-correction accuracy could then be achieved
with polynomial resources.  This exponential convergence rate is, in
fact, one of the most attractive features of Kitaev's construction in
Ref.~\cite{kit1}.  

On the other hand, the rate at which $\Delta E_\Lambda(\epsilon)$
converges to zero is determined by the unperturbed Hamiltonian
$H_\Lambda$, the perturbation $P$, and the perturbation strength
$\epsilon$.  It is therefore important to know what we can expect in a
generic setting. Obviously, the exponential convergence, as in
Eq.~(\ref{eq:expcon}), is optimal, but it may as well turn out that
the particular implementation (e.g., with a different Hamiltonian)
does not allow for it. We can, however, hope for a slower (but still
quite decent) convergence rate.  According to a theorem of Horsch and
von der Linden \cite{hl}, certain quantum spin systems possess
low-lying eigenstates of the finite-lattice Hamiltonian with
$$
\Delta E_\Lambda(\epsilon) = c/\abs{\Lambda}.
$$
The conditions for this to hold are the following.  There has to exist
an {\em order observable} $O_\Lambda$ of the form
$$
O_\Lambda = \sum_{l \in \Lambda} O_l,
$$
where each $O_l$ is a self-adjoint operator such that $[O_l,
O_{l '}] = 0$.  Furthermore, for any interaction term $\Phi_{M
\subset \Lambda}$ in the perturbed Hamiltonian $H_\Lambda(\epsilon)$
(these also include the perturbation terms), we will have $[\Phi_M,O_l]
= 0$ unless $l \in M$.  The operators $\Phi_M$ and $O_l$ are
required to be uniformly bounded (in $M$ and $l$ respectively), and
the cardinality of the support set $M$ must not exceed some fixed
constant $C$ (the latter condition has also to be fulfilled for the
perturbation theory described above to converge). Finally, if
$\ket{\psi}$ is an eigenstate of $H_\Lambda(\epsilon)$, then we
must have $\braket{\psi}{O_\Lambda | \psi} = 0$, but
$\braket{\psi}{O^2_\Lambda | \psi} \ge \zeta \abs{\Lambda}^2$
(here the constant $\zeta$ depends on $O_l$).  The latter
conditions are taken as manifestations of obscured symmetry
breaking.  Examples of systems for which the Horsch-von der Linden
theorem holds include \cite{kt2} the Ising model in the transverse
magnetic field or the Heisenberg antiferromagnet with a N\'eel order.

\section{Summary}

Where does it all take us?  It appears, from the discussion in the
preceding section, that any quantum spin system, whose Hamiltonian is
formed by mutually commuting $n$-body interactions that satisfy the
Peierls condition, can recover from sufficiently weak quantum (i.e.,
off-diagonal) perturbations at low temperatures.  The admissible
perturbations can be either finite-range \cite{kt}, or exponentially
decaying \cite{bku}.  Under these (quite general) conditions, it
follows from rigorous perturbation theory for quantum spin systems
that there exists a critical perturbation strength $\epsilon_0$, such
that, for all $\epsilon < \epsilon_0$, the degeneracy and the symmetry
properties of the ground state of the original (unperturbed) system
survive in the thermodynamic limit.  Furthermore, even if ground-state
degeneracy is removed by perturbation of the finite-size system, the
effect of the error (perturbation) is effectively \qt{washed out} in
the thermodynamic limit, as the low-lying excited states of the
perturbed system converge to additional ground states.

However, the systems we have considered were assumed to have classical
Hamiltonians and discrete symmetries.  What about truly quantum
Hamiltonians and continuous symmetry (e.g., the quantum Heisenberg
model)?  The situation here is not so easy.  For instance, it is
apparent from our discussion that, in order to be self-correcting, the
perturbed system must exhibit an order-disorder transition as the
parameter $\epsilon$ is tuned: in the \qt{ordered phase,} error
correction is possible; in the \qt{disordered phase,} occurrence of
errors results in irrevocable loss of information.  (This has already
been noted by Dennis \etal \cite{dklp}.)  Since we require the ground
state of the perturbed Hamiltonian to exhibit the same degeneracy as
the corresponding state of the unperturbed Hamiltonian, it makes sense
to talk about spontaneously broken symmetry in the ordered phase
(i.e., when $\epsilon < \epsilon_0$).  However, according to the
so-called Goldstone theorem \cite{lfw}, symmetry cannot be broken in a
system with continuous symmetry and a gap.  It would certainly be
worthwhile to explore physical error correction in systems with
continuous symmetries as well, but the models in which it can work
will not be as easy to find.

%% file: chap6.tex
\chapter{Conclusion}
\label{ch:concl}

%\pagestyle{intro}

%\vskip 1in

The now-fashionable field of \qt{physics of information and
  computation} is at least 73 years old if we count from 1929, the year
when Leo Szilard published his seminal paper \cite{szi} on the
Maxwell's demon.  Today, the amount of published work in this area is
astounding; to be sure, we have gained quite a bit (pun intended) of
new knowledge and new insights, owing to the continuous
cross-fertilization between the disciplines of mathematics, physics,
and computer science.  However, we hardly made a dent in the Big
Problem of harnessing the enormous information-processing potential of
quantum-mechanical systems.  Reliable storage of quantum information
still remains the paramount challenge.

In this dissertation, we presented a systematic study of the dynamical
aspects of information storage in quantum-mechanical systems.  We have
already outlined in Chapter \ref{ch:intro} the way in which dynamics
(statistical dynamics, in particular) relates to our investigation of
information storage in noisy quantum registers and computers.  Let us
therefore elaborate on the \qt{big picture.}

First of all, what do we mean by \qt{information?}  We take a very
simple approach:  we refer to any assignment of an initial state
(density operator) as the information stored in the register (or
supplied to the computer).  Sstatistical dynamics then comes in useful
as we attempt to follow the orbit traced over time by this initial
information in the state space of the register (computer).  Now, given
any pair of different initial states, the states along the
corresponding orbits in the noiseless case will be distinguishable
from one another exactly to the same extent as the input states.
This, in general, will not be the case when noise is present.  It is
precisely this feature that is central in our analysis of the noisy
dynamics.

In a sense, quantum information science is a nonequilibrium theory:
it deals with systems whose quantum-mechanical states are not
necessarily the thermodynamically favorable ones.  We believe,
therefore, that an important aspect of the noisy dynamics of quantum
registers and computers is the tendency towards equilibrium.  Since
the rate of convergence to equilibrium is expected to be quite rapid
(and this is precisely the behavior we have shown strictly contractive
dynamics to exhibit), any active intervention (such as error
correction) would have to take place extremely often.  In fact, if
the computational network is very elaborate, we may expect most of it
to be taken up by the degrees of freedom that are responsible for
keeping the computation stable.  However, because these degrees of
freedom are also susceptible to the noisy dynamics, the overall trend
toward equilibrium will still be present.

In this dissertation, the noise afflicting the quantum register
(computer) was modeled by a strictly contractive channel
\cite{rag2,rag5}.  This model is justified for several reasons, the
main one being the rapid convergence of disjoint orbits toward each
other.  This behavior naturally leads towards ergodicity and mixing,
two important ingredients in the theory of approach to equilibrium.
Furthermore, strictly contractive channels give us a way to incorporate
the crucial assumption of {\em finite precision} of any experimentally
available apparatus into the mathematical model of a physically
realizable (i.e., nonideal) quantum computer.  We have shown, in
particular, that no two states of such a computer can be distinguished
from one another with absolute certainty, even if they are maximally
distinguishable in the noiseless case.  Finally we have shown show
that, given any channel $T$, there will always be a strictly
contractive channel $T'$ in any neighborhood of $T$ in the cb-norm
topology.  Using the fidelity measure we have developed for quantum
channels \cite{rag1}, we showed that, for any channel $T$, there
always exists a strictly contractive channel $T'$ that cannot be
distinguished from $T$ by any {\em experimental} means.  We then went
on to demonstrate that, in the absence of error correction, the
sensitivity of quantum memories and computers to strictly contractive
errors would grow exponentially with storage time and computation time
respectively, and would depend only on the contraction rate and on the
measurement precision.  We proved that strict contractivity rules
out the possibility of perfect error correction, and gave an
argument that approximate error correction, which covers previous
work on fault-tolerant quantum computation as a special case, is
possible.  

We have then applied our model to the problem of determining the
threshold error rate for noisy quantum computation.  If the noise is
sufficiently weak, we may model the decoherence mechanism by a
depolarizing channel, the error rate being precisely the rate of
depolarization.  We would like to emphasize that we did not make any
assumptions about the specific procedure employed for error
correction, nor did we appeal to combinatorial considerations.  The
threshold error rate was shown to depend on the measurement precision
and on the physical circuit depth.  We presented some numerical estimates
for the threshold error rate for the case when the measurement
precision is on the order of the standard quantum limit, and found
that, even with such ridiculously precise measurements, the maximum
tolerable error rate would drop to zero extremely rapidly for circuits
of polynomial and superpolynomial physical depth.

After having described the general properties of strictly contractive
channels, along with implications for quantum information processing,
we took up the following question \cite{rag4}.  How does strict
contractivity relate to the balance of energy and entropy in a noisy
quantum register (computer)?  We found that there is a close
connection between the contraction rate of a channel and the rate of entropy
production in a noisy quantum computer.  We adapted the so-called
\qt{entropy-energy arguments} in order to determine the maximum number
of operations that can be carried out reliably on a noisy quantum
computer in terms of energy and temperature, thus enabling us to judge
the {\em thermodynamic cost} of keeping the computer stable.  Ideally
we would like to do error correction as infrequently as possible; the
longer the relaxation time, the closer we will be to this goal.  We
also proved that, under certain conditions, there exists an upper
bound on the number of qubits in a circuit-based quantum computer.

Finally we looked into the possibility of using quantum spin systems
with phase transitions for reliable storage of quantum information
\cite{rag3}.  Our inspiration came from Kitaev's idea \cite{kit1} to
store quantum information in the degenerate ground state of a system
of interacting anyons on a periodic lattice.  Proper treatment of the
ground states calls for the analysis of {\em
  low-temperature} behavior of quantum spin systems, where  the main
issue is not entropy, but rather the energy fluctuations above
the ground-state level, caused by quantum perturbations.  We have, in
particular, addressed the following question:  what kinds of
interactions are admissible for constructing such quantum memory
devices, and what are the perturbations against which these memories
will be stable?  We indicated that a few results in rigorous
statistical mechanics of quantum spin systems \cite{rue} point toward
the conclusion that such \qt{self-correction} is fairly common in
quantum spin systems with Hamiltonians that are comprised by
interactions satisfying the so-called Peierls condition (the standard
example being an Ising-type Hamiltonian), the admissible perturbations
being either finite-range or exponentially decaying.

Most of the results we have presented in this dissertation are of a
somewhat negative nature. The implications, however, are more of a
blessing than a curse for the future of quantum information
processing. We believe that the successful solution of problems faced
by researchers in this field will require models of computers far more
ingenious than networks of one- and two-qubit gates.  As we have mentioned
in Chapter \ref{ch:scc}, massively parallel systems of interacting
particles (quantum cellular automata) may well prove to be a viable
medium for the experimental realization of large-scale quantum
computers.

%% file: app.tex
\renewcommand{\thechapter}{\Alph{chapter}}

\setcounter{chapter}{0}
\chapter*{Appendix A\\
Mathematical background}

\addcontentsline{toc}{chapter}{Appendix A:  Mathematical background}

\renewcommand{\thesection}{A.\arabic{section}}
\renewcommand{\theequation}{A.\arabic{equation}}
\renewcommand{\thetheorem}{\thesection.\arabic{theorem}}
\setcounter{equation}{0}
\setcounter{section}{0}
\setcounter{theorem}{0}

\section{C*-algebras}
\label{sec:cstar}

We summarize here the absolute minimum of the C*-algebra theory.  For
the mathematical treatment, the reader is referred to the books by
Bratteli and Robinson \cite{br}, Conway \cite{con}, and Davidson
\cite{dav1}, and for the C*-algebras in the context of quantum theory
and statistical mechanics to the books by Emch \cite{emc}, Haag
\cite{haa}, and Streater \cite{str}.

First we give a few definitions.

\begin{definition}
An {\em algebra} is a complex linear space $\cA$, equipped with the
product operation $(A,B) \in \cA \times \cA \mapsto AB \in \cA$ such
that, for all $A,B,C \in \cA$ and all $\alpha,\beta \in \bbc$, (1)
$A(BC) = (AB)C$, (2) $A(B+C) = AB + AC$, (3) $(\alpha \beta) (AB) =
(\alpha A)(\beta B)$. The product operation does not have to be
commutative; an algebra with the commutative product is called {\em
abelian} or simply {\em commutative}.  
\label{def:alg}
\end{definition}

\begin{definition}
An {\em algebra with identity} or a {\em unital algebra} is an algebra
$\cA$ with the unique element $\idty \in \cA$ such that $A\idty =
\idty A = A$ for any $A \in \cA$.
\label{def:algid}
\end{definition}

\begin{definition}
An {\em involution} on an algebra $\cA$ is a mapping $A \in \cA
\mapsto A^* \in \cA$ such that, for all $A,B \in \cA$ and all
$\alpha,\beta \in \bbc$, (1) $(A^*)^* = A$, (2) $(AB)^* = B^* A^*$,
(3) $(\alpha A + \beta B)^* = \bar{\alpha} A^* + \bar{\beta} B^*$. An
element $A \in \cA$ with $A = A^*$ is called {\em self-adjoint}. An
algebra with an involution is referred to as a {\em $*$-algebra}.
\label{def:invol}
\end{definition}

\begin{definition}
An algebra $\cA$ is a {\em normed algebra} if it is equipped with a
norm $\norm{\cdot}$ which is, in addition to the usual properties of
the norm, {\em submultiplicative}, i.e., for any $A,B \in \cA$,
$\norm{AB} \le \norm{A}\norm{B}$.  If a normed algebra $\cA$ is a
complete normed space, and if the norm has the property $\norm{A^*} =
\norm{A}$ for any $A \in \cA$, then $\cA$ is called a {\em Banach
$*$-algebra}.
\label{def:normedalg}
\end{definition}

\begin{definition}
A Banach $*$-algebra $\cA$ is called a {\em C*-algebra} if its norm
has the {\em C*-norm property} $\norm{A^*A} = \norm{A}^2$.
\label{def:cstar}
\end{definition}

In this case, the requirement that the involution on $\cA$ be
isometric with respect to the norm is redundant, since this property
follows from the C*-norm property and the submultiplicativity of the
norm.

There are two classic examples of C*-algebras.  (1) Let $\sX$ be a
compact Hausdorff space, and let $\cC(\sX)$ be the set of all
continuous complex-valued functions on $\sX$.  Then $\cC(\sX)$ is a
C*-algebra with the operations defined pointwise, $(f+g)(x) \defeq
f(x)+g(x)$, $(fg)(x) \defeq f(x)g(x)$, and $f^*(x) \defeq
\overline{f(x)}$, and the norm $\norm{f} \defeq \sup_{x \in
\sX}\abs{f(x)}$.  The C*-algebra $\cC(\sX)$ is an abelian algebra. (2)
Let $\sH$ be a Hilbert space, and let $\cB(\sH)$ be the set of all
bounded operators acting on $\sH$.  Then $\cB(\sH)$ is a C*-algebra
with the usual sum and product operations, and the involution given by
the Hilbert-space (Hermitian) adjoint.  The norm is the operator norm
$\norm{A} \defeq \sup_{\psi \in \sH; \norm{\psi}=1}\norm{A \psi}$.
The C*-algebra $\cB(\sH)$ is a noncommutative algebra.  Both of these
algebras are algebras with identity; in the first case, the identity
is the constant function 1, and, in the second case, the identity is
the identity operator, $\idty \psi = \psi$ for all $\psi \in \sH$.

It turns out that these two examples are already exhaustive in the
following sense.  A theorem of Gelfand and Naimark asserts that, for
any C*-algebra $\cA$, there exists a Hilbert space $\sH$ such that
$\cA$ is isomorphic to $\cB(\sH)$.  Furthermore, according to a
theorem of Gelfand, any commutative C*-algebra $\cA$ is isomorphic to
the algebra $\cC_0(\sX)$ of complex-valued continuous functions that
vanish at infinity on some locally compact Hausdorff space $\sX$.
[This algebra is a more general object than $\cC(\sX)$ defined above,
but, whenever $\cA$ has an identity, the space $\sX$ will
automatically be compact.]  From now on we assume that all the
C*-algebras, with which we are dealing, have an identity.

\section{States, representations, and the GNS construction}
\label{sec:gns}

\begin{definition}
An element $A$ of a C*-algebra $\cA$ is called {\em invertible} if
there exists an element $\inv{A}$, called the {\em inverse of $A$},
such that $A \inv{A} = \inv{A} A = \idty$. The {\em resolvent set of
$A$}, denoted by $r(A)$, is the subset of $\bbc$ consisting of all
complex numbers $\lambda$ such that $A - \lambda \idty$ has an
inverse.  The {\em spectrum of $A$}, denoted by $\sigma(A)$, is the
complement of $r(A)$ in $\bbc$, $\sigma(A) \defeq \bbc \backslash
r(A)$.
\label{def:spectral}
\end{definition}

It is a celebrated result in spectral analysis that the spectrum of
any $A \in \cA$ is a nonempty compact set.  In particular, the {\em
spectral radius $r_A$ of $A$}, defined as $r_A \defeq \sup_{\lambda
\in \sigma(A)} \abs{\lambda}$, does not exceed the norm of $A$.

In case of a self-adjoint $A \in \cA$, the spectrum $\sigma(A)$ is
contained within the interval $[-\norm{A},\norm{A}]$ of the real line,
and $r_A \equiv \norm{A}$.  A self-adjoint element $A$ of a
C*-algebra $\cA$ is called {\em positive} (this is denoted by $A \ge
0$) if $\sigma(A) \subseteq [0,\norm{A}]$.  An element $A \in \cA$ is
positive if and only if there exists some $B \in \cA$ such that $A =
B^*B$.

After these preliminaries, we can define a state over a C*-algebra. 

\begin{definition}
A {\em state} over a C*-algebra $\cA$ is a normalized positive linear
functional $\omega$ over $\cA$, i.e., $\omega(\idty) = 1$ and
$\omega(A^*A) \ge 0$ for any $A \in \cA$.  The set $\cS(\cA)$ of all
states over a C*-algebra $\cA$ is a convex set, and its extreme points
are referred to as {\em pure states}.
\label{def:cstarstate}
\end{definition}

A canonical example of a state over a C*-algebra is furnished by
considering the algebra $\cB(\sH)$ of bounded operators on some
Hilbert space $\sH$.  Let $\psi \in \sH$ be a unit vector, and define
the linear functional $\omega_\psi(A) \defeq \braket{\psi}{A\psi}$.
It is quite easy to see that $\omega_\psi$ is a state.  A state
defined in this way is called a {\em vector state}.  It is the gist of
the famous Gelfand-Naimark-Segal (GNS) construction that any state
over a C*-algebra $\cA$ has the form of a vector state over a
C*-subalgebra of $\cB(\sH)$ for some Hilbert space $\sH$.  For this
reason, the GNS construction is central to the C*-algebraic quantum
theory.  However, in order to state it properly, we first have to
introduce some additional machinery.

\begin{definition}
A {\em $*$-homomorphism} between C*-algebras $\cA$ and $\cB$ is a
mapping $\map{\pi}{\cA}{\cB}$ such that, for all $A,B,C \in \cA$ and
all $\alpha,\beta \in \bbc$, (1) $\pi(\alpha A + \beta B) = \alpha
\pi(A) + \beta \pi(B)$, (2) $\pi(AB) = \pi(A)\pi(B)$, (3) $\pi(A^*) =
\pi(A)^*$. In other words, a $*$-homomorphism between two C*-algebras
is a mapping that preserves the C*-algebraic structure.  A bijective
$*$-homomorphism is referred to as a {\em $*$-isomorphism}.\label{def:starhomo}
\end{definition}
Any $*$-homomorphism maps positive elements to positive elements
because $\pi(A^*A) = \pi(A^*)\pi(A) = \pi(A)^* \pi(A) \ge 0$, and is
also continuous:  $\norm{\pi(A)} \le \norm{A}$.

\begin{definition}
A {\em representation} of a C*-algebra $\cA$ is a pair $(\sH,\pi)$,
where $\sH$ is a Hilbert space and $\pi$ is a $*$-homomorphism of
$\cA$ into $\cB(\sH)$.  The representation $(\sH,\pi)$ of $\cA$ is
called {\em faithful} if $\ker{\pi} \defeq \setcond{A \in \cA}{\pi(A)
= 0}$ is trivial.
\label{eq:cstarrep}
\end{definition}
From now on, when we talk about representations, we will omit the
mention of the Hilbert space $\sH$ whenever it is clear from the
context which Hilbert space we are talking about.

\begin{definition}
A vector $\Omega \in \sH$ is called the {\em cyclic vector} for the
representation $(\sH,\pi)$ of a C*-algebra $\cA$ if the set
$\setcond{\pi(A)\Omega}{A \in \cA}$ is dense in $\sH$, i.e., if for
any $\phi \in \sH$ and any $\epsilon > 0$, there exists some $A \in
\cA$ such that $\norm{\phi - \pi(A)\Omega} < \epsilon$.  The triple
$(\sH,\pi,\Omega)$ is called the {\em cyclic representation} of
$\cA$. The representation $(\sH,\pi)$ is called {\em irreducible} if
every vector $\psi \in \sH$ is cyclic for $\pi$ or, equivalently, if
the only invariant subspaces of the set $\pi(\cA) \defeq
\setcond{\pi(A)}{A \in \cA}$ are $\set{0}$ and $\sH$.  Otherwise, the
representation is called {\em reducible}.
\label{def:cyclic}
\end{definition}
It can be shown that any representation of a C*-algebra as an algebra
of operators over a Hilbert space can be decomposed into a direct sum
of irreducible representations.  In general, a set $\cM$ of bounded
operators on a Hilbert space $\sH$ is called irreducible if it has no
nontrivial invariant subspaces.  Thus we can say that the
representation $(\sH,\pi)$ of a C*-algebra $\cA$ is irreducible if and
only if the set $\pi(\cA)$ is irreducible.

A useful irreducibility criterion is provided by Schur's lemma,
which states that a set $\cM \subseteq \cB(\sH)$ which is
self-adjoint (i.e., closed under the operation of taking the adjoint),
is irreducible if and only if the {\em commutant}
of $\cM$, i.e., the set $\cM' \defeq \setcond{X \in \cB(\sH)}{[X,M] =
0,\forall M \in \cM}$, consists only of complex multiples of the
identity operator (this is written as $\cM' = \bbc \idty$).  Thus the
representation $(\sH,\pi)$ of a C*-algebra $\cA$ is irreducible if and
only if $\pi(\cA)' = \bbc \idty$.

Now we are ready to state the theorem which is the essence of the GNS
construction.

\begin{theorem}
{\bf (Gelfand-Naimark-Segal)} Let $\omega$ be a state over the
C*-algebra $\cA$.  Then there exists a cyclic representation
$(\sH,\pi,\Omega)$ of $\cA$ such that
$$
\omega(A) = \braket{\Omega}{\pi(A)\Omega}
$$
for all $A \in \cA$, and $\Omega$ is a unit vector.  This
representation, to which we will refer as the {\em GNS representation of
$\cA$ associated with $\omega$}, is unique up to unitary
equivalence.\label{th:gns}
\end{theorem}
Given a C*-algebra $\cA$ and a state $\omega$, the corresponding GNS
representation is irreducible if and only if $\omega$ is a pure state.
This result has an interesting consequence for pure states over
abelian C*-algebras.  Namely, $\omega$ is a pure state over an abelian
C*-algebra $\cA$ if and only if $\omega(AB)=\omega(A)\omega(B)$ for
all $A,B \in \cA$.  Indeed, let $(\sH,\pi)$ be the corresponding GNS
representation.  Since $\omega$ is a pure state, the representation is
irreducible, and therefore $\pi(\cA)' = \bbc \idty$.  But, because
$\cA$ is abelian, we have $\pi(\cA) \subseteq \pi(\cA)'$, which means
that $\pi$ is irreducible if and only if $\sH$ is one-dimensional,
i.e., isomorphic to $\bbc$.  The factorization property of $\omega$ is
now apparent.

\section{Trace ideals of $\cB(\sH)$}
\label{sec:trideals}

\begin{definition}
Let $\cA$ be an algebra.  A subset $\cI$ of $\cA$ is called a {\em
two-sided ideal} (or simply an {\em ideal}) of $\cA$ if, for any $I
\in \cI$ and any $A \in \cA$, the elements $AI$ and $IA$ are also in $\cI$.
\label{def:ideal}
\end{definition}
In this section we will give a brief description of a class of ideals
of the algebra $\cB(\sH)$, the so-called {\em trace ideals}. For a
well-written exposition of the theory of trace ideals, as well as its
applications to mathematical physics, consult the text by Simon
\cite{sim2}; another good source is the classic monograph by Schatten
\cite{scha}.  

The starting point in the theory of trace ideals is the concept of a
{\em compact operator}.

\begin{definition}
A bounded operator $A \in \cB(\sH)$ is called a {\em finite-rank
operator} if it has finite-dimensional range.  A bounded operator is
called {\em compact} if it is a norm limit of finite-rank operators.
\label{def:compactop}
\end{definition}

Let $\cC(\sH)$ denote the set of all compact operators on $\sH$.  It
is easy to see that $\cC(\sH)$ is a two-sided ideal of $\sH$.  Indeed,
if $A$ is a finite-rank operator and $B$ is a bounded operator, then
$AB$ and $BA$ are both finite-rank operators.  Because $\cC(\sH)$ is a
norm closure of the set of all finite-rank operators, we see that $AB$
and $BA$ are compact whenever $A$ is compact and $B$ is bounded.  In
fact, any two-sided ideal of $\cB(\sH)$ is a subset of
$\cC(\sH)$. Furthermore, we have the following key theorem.

\begin{theorem}
Let $A$ be a compact operator.  Then $A$ has a norm-convergent {\em
canonical expansion}
$$
A = \sum^N_{n=1}\mu_n(A) \ketbra{\psi_n}{\phi_n},
%\label{eq:svd}
$$
where $N$ is a nonnegative integer or infinity, each $\mu_n(A) > 0$
with $\mu_1(A) \ge \mu_2(A) \ge \ldots,$ and $\set{\psi_n}$ and
$\set{\phi_n}$ are (not necessarily complete) orthonormal sets.
Moreover, the numbers $\mu_n(A)$, called the {\em singular values of
$A$}, are the nonzero eigenvalues of $\abs{A} \defeq (A^*A)^{1/2}$,
arranged in descending order.
\label{th:svd}
\end{theorem}
Now suppose that we are given a compact operator $A$.  For $p =
1,2,\ldots,$ define the {\em Schatten $p$-norm} of $A$ as
$$
\pnorm{A}{p} \defeq \left(\sum_n \mu_n(A)^p \right)^{1/p} \equiv
(\tr{\abs{A}^p})^{1/p}
%\label{eq:pnormdef}
$$
Then $A$ is said to belong to {\em Schatten $p$-class} if
$\pnorm{A}{p}^p$ is finite; in this case we will write $A \in
\cT_p(\sH)$.  It can be shown that, for any $p$, the Schatten
$p$-class $\cT_p(\sH)$ is a two-sided ideal of $\cB(\sH)$;
alternatively, $\cT_p(\sH)$ is the closure of the finite-rank
operators in the Schatten $p$-norm.

We are mainly interested in the cases $p=1$ and $p=2$.  Let us look at
the first case, where we have the norm
$$
\trnorm{A} \defeq \tr{\abs{A}},
%\label{eq:trnorm}
$$
referred to as the {\em trace norm}.  The Schatten 1-class
$\cT_1(\sH)$ is also referred to as the {\em trace class}, and any
operator $A \in \cT_1(\sH)$ is called a {\em trace-class operator}.
For a self-adjoint trace-class operator $A$ we also have $\abs{\tr{A}}
\le \trnorm{A}$.  Furthermore, for any trace-class operator $A$ and
any bounded operator $B$, we have the inequalities
\begin{eqnarray}
\trnorm{AB} &\le & \norm{B}\trnorm{A} \nonumber \\
\trnorm{BA} &\le & \norm{B}\trnorm{A} \nonumber,
\end{eqnarray}
which can, of course, be taken as an indication that the trace-class
is a two-sided ideal of $\cB(\sH)$.

When $p=2$, the corresponding $p$-norm,
$$
\pnorm{A}{2} \defeq (\tr{\abs{A}^2})^{1/2} \equiv [\tr{(A^*A)}]^{1/2},
%\label{eq:hsnorm}
$$
is called the {\em Hilbert-Schmidt norm}, and the space $\cT_2(\sH)$
is called the space of the Hilbert-Schmidt operators.  The
Hilbert-Schmidt norm is the norm induced by the inner product
$\tr{(A^*B)}$, and it can be shown that the space of the
Hilbert-Schmidt operators is a Hilbert space.  The product of a
Hilbert-Schmidt operator $A$ and a bounded operator $B$, in any order,
is again a Hilbert-Schmidt operator with
$$
\pnorm{AB}{2}, \pnorm{BA}{2} \le \norm{B}\pnorm{A}{2}.
$$
This shows that the Hilbert-Schmidt space $\cT_2(\sH)$ is a two-sided
ideal of $\cB(\sH)$.

\section{Fixed-point theorems}
\label{sec:fpt}

Let $\sX$ be a metric space with the metric $d(\cdot,\cdot)$.
An operator $\map{A}{\sX}{\sX}$ is called a {\em contraction} if,
for any $x,y \in \sX$, $d(Ax,Ay) \le d(x,y)$, and a {\em strict
contraction} if there exists some $k \in [0,1)$ such that
$d(Ax,Ay) \le kd(x,y)$.  If $\sX$ is a complete metric space, then
the contraction mapping principle \cite{rs} states that any
strict contraction $A$ on $\sX$ has a unique fixed point.  In
other words, the problem $Ax = x$ has a unique solution on $\sX$.
If $\sY$ is a closed subset of $\sX$, then it follows that any
strict contraction $\map{A}{\sY}{\sY}$ has a unique fixed point on
$\sY$.

Strict contractivity is a remarkably strong property.  Indeed, if
we pick any $y \in \sY$, then the sequence of iterates $A^ny$
converges to the fixed point $y_0$ of $A$ exponentially fast,
because
\begin{equation}
d(A^ny,y_0) = d(A^ny,A^ny_0) \le k^nd(y,y_0).
\label{eq:expconv}
\end{equation}
This fact is of tremendous use in numerical analysis when one
wants to solve the fixed-point problem $Ay=y$ by the iteration method
with some initial guess $\hat{y}$.  If the operator $A$ is a
strict contraction on a closed subset of a complete metric space,
then, for any choice of $\hat{y}$, the iteration method is
guaranteed to zero in on the solution in $O(\log \epsilon^{-1})$
steps, where $\epsilon$ is the desired precision.

It should be noted that existence and uniqueness of a fixed point
of some operator $A$ are, by themselves, not sufficient to
guarantee convergence of the sequence of iterates $A^ny$ for any
point $y$ in the domain of $A$. Indeed, according to the
Leray-Schauder-Tychonoff theorem \cite{rs}, any continuous
map on a compact convex subset of a locally convex space $\sX$ has
at least one fixed point.  Furthermore, any {\em weak contraction}
on a compact subset $C$ of a Banach space, i.e., a map
$\map{W}{C}{C}$ with the property $\norm{Wx-Wy} < \norm{x-y}$
for any $x,y \in C$, has a unique fixed point \cite{sta}.
The key to the rapid convergence in Eq.~(\ref{eq:expconv}) is the
fact that a strict contraction $\map{A}{\sY}{\sY}$ shrinks
distances between points of $\sY$ {\em uniformly}.

\chapter*{Appendix B\\
List of Symbols}
\addcontentsline{toc}{chapter}{Appendix B:  List of symbols}

\renewcommand{\thesection}{B.\arabic{section}}
\setcounter{section}{0}

\begin{tabular}{|r|l|}

\hline
$\cA \simeq \cB$ & $\cA$ is isomorphic to $\cB$\\
\hline
$\cA,\cB$ & operator algebras \\
\hline
$\cB(\sH)$ & the algebra of bounded operators on the Hilbert space $\sH$ \\
\hline
$\bbc \idty$ & the set of the complex multiples of the identity operator \\
\hline
$\sH,\sK,\sE,\sF,\sG$ & Hilbert spaces \\
\hline
$\idty$ & identity operator \\
\hline
$\id$ & identity mapping \\
\hline
$\id_n$ & identity mapping on $\cM_n$ \\
\hline
$\cM_n$ & the algebra of $n \times n$ complex matrices \\
\hline
$\trn{M}$ & matrix transpose of $M$ \\
\hline
$\rho,\sigma$ & density operators \\
\hline
$\cS(\cA)$ & the state space of the C*-algebra $\cA$ \\
\hline
$\cS(\sH)$ & the set of density operators on $\sH$ \\
\hline
$\paulix,\pauliy,\pauliz$ & Pauli matrices \\
\hline
$\cT_p(\sH)$ & Schatten $p$-class on $\sH$ \\
\hline
$\abs{X}$ & cardinality of the set $X$ \\
\hline
$X^*$ & operator adjoint to the operator $X$ \\
\hline
$\sX,\sY$ & general (e.g., topological or measurable) spaces \\
\hline
$\bar{z}$ & complex conjugate of $z \in \bbc$ \\
\hline

\end{tabular}

%% file: vita.tex
\chapter*{Vita}
\label{ch:vita}
\addcontentsline{toc}{chapter}{Vita}

\noindent{Maxim Raginsky was born in 1977 in Vladimir, Russia.  His family
immigrated to the United States of America in 1992.   In 2000 he
obtained his B.S. and M.S. degrees, both in electrical engineering,
from Northwestern University. Maxim's doctoral research, carried out
in the Department of Electrical and Computer Engineering at
Northwestern under the supervision of Prof. Horace P. Yuen, was
concerned with the problem of information storage in
quantum-mechanical systems.}\\

\noindent{{\bf Publications}
\begin{enumerate}
\item[$\bullet$] (with P. Kumar) \qt{Generation and manipulation of
squeezed states of light in optical networks for quantum communication
and computation,} J. Opt. B: Quantum Semiclass. Opt. {\bf 3}, L1
(2000).
\item[$\bullet$] \qt{A fidelity measure for quantum channels,}
Phys. Lett. A {\bf 290}, 11 (2001).
\item[$\bullet$] \qt{Strictly contractive quantum channels and physically
realizable quantum computers,} Phys. Rev. A {\bf 65}, 032306 (2002).
\item[$\bullet$] \qt{Almost any quantum spin system with short-range
interactions can support toric codes,} Phys. Lett. A {\bf 294}, 153
(2002).
\end{enumerate}
}